\documentclass[longauth]{aa}  
\usepackage{graphicx}
\usepackage{xcolor}
\usepackage{txfonts}

\usepackage{hyperref}
\hypersetup{
     breaklinks,
     colorlinks   = true,
     citecolor    = blue,
     urlcolor     = blue,
     linkcolor    = blue
}

\newcommand{\sd}{\texttt{4MIDABLE-HR}}

\usepackage{array}
\newcolumntype{L}[1]{>{\raggedright\arraybackslash}p{#1}}

\begin{document}

\title{4MIDABLE-HR: Science motivation and overview of the 4MOST Milky Way disk and bulge high-resolution survey}
\subtitle{Stellar physics, star-planet connection, and Galactic archaeology}

\titlerunning{4MIDABLE-HR: Science motivation and overview}
\authorrunning{M. Bergemann, T. Bensby, et al.}
\author{
M.~Bergemann\inst{mpia}
\and
T.~Bensby\inst{lunda}
%
%
\and
J.~Gerber\inst{jmg}
\and
A.~Serenelli\inst{as}
\and
G.~Traven\inst{gt}
\and
C.~Schneider\inst{ham}
%
%
\and
R.~Andrae\inst{mpia}
\and
V.~D'Orazi\inst{vo,gb}
\and
H.~Ernandes\inst{lunde}
\and
S.~Feltzing\inst{lunde}
\and
A.~Perdomo Garcia \inst{mpia}
\and
N.~Storm\inst{mpia}
%
%
%
\and
R.~Albarracin\inst{mpia}
\and
G.~Bono\inst{vo, gb, infn}
\and
B.~Bitsch\inst{bb}
\and
R.~Church\inst{lunda}
\and
R.A.~Garc\'{i}a \inst{rag}
\and
O.~Gerhard\inst{mpe}
\and
G.~Guiglion\inst{mpia}
\and
V.~F.~Ksoll\inst{ita}
\and
G.~Kordopatis\inst{oca}
\and
B.~Liagre \inst{rag}
\and 
K.~Lind\inst{kl}
\and
T.~Merle\inst{tm}
\and
S.L.~Martell\inst{unsw}
\and
S.~Mathur\inst{sm}
\and
I.~Minchev\inst{im}
\and
H.-W.~Rix\inst{mpia}
\and
\'{A}.~Sk\'{u}ladóttir \inst{uf, oaa}
\and
M. Valentini\inst{im}
%
%
%
%
\and
O.~Agertz\inst{lunda}
\and
S.~Antony\inst{lsw}
\and
J.~Bland-Hawthorn\inst{syd}
\and
C.~Chiappini\inst{im}
\and
A.J.~Dimoff\inst{mpia}
\and
D.K.~Feuillet\inst{uu}
\and
S.~Khoperskov\inst{sk}
\and
H.~Korhonen\inst{mpia}
\and
J.~Kos\inst{gt}
\and
D.~Massari\inst{oas}
\and
A.~Plotnikova\inst{lunde}
\and 
Z.~Prudil\inst{zp}
\and
A.~Quirrenbach\inst{lsw}
\and 
V.~Ripepi\inst{vr}
\and
N.~Ryde\inst{lunda}
\and
R.~Sch\"onrich\inst{rs}
\and 
T.~Sicignano\inst{zp,ts,vr}
\and
R. Singh\inst{uf}
\and
N.~Vernekar\inst{OU}
\and
N.J.~Wright\inst{kuk}
}

\institute{(mpia) Max Planck Institute for Astronomy, Koenigstuhl 17, 69117, Heidelberg \\
\email{bergemann@mpia-hd.mpg.de}
\and
(lunda) Lund Observatory, Division of Astrophysics, Department of Physics, Lund University, Box 118, SE-221\,00 Lund, Sweden 
\and 
(lunde) Lund Observatory, Department of Earth and Environmental Sciences, S\"olvegatan 12, SE-223\,62 Lund, Sweden 
\and 
(as) Institute of Space Sciences (ICE, CSIC), C/Can Magrans S/N,  08193, Bellaterra, Spain 
\and
(as) Institut d'Estudis Espacials de Catalunya (IEEC), C/Esteve Terradas 1, Edifici RDIT, Castelldefels, 08860, Spain
\and
(ucc) Department of Physics, University College Cork, Cork, Ireland
\and
(tm) ULB, Institut d'Astronomie et d'Astrophysique, Universit\'e Libre de Bruxelles, CP 226, Bvd du Triomphe, B-1050 Brussels
\and 
Sydney Institute for Astronomy, University of Sydney, NSW 2008, Australia
\and 
(unsw) School of Physics, University of New South Wales, Sydney, NSW 2052, Australia
\and 
(vo, gb) Dept. of Physics, Universit\'{a} di Roma Tor Vergata, Via della Ricerca Scientifica, 1 , Roma, 00133, Italy
\and 
(vo, gb) INAF Osservatorio Astronomico di Roma, via Frascati 33, Monte Porzio Catone, Italy
\and
(infn) INFN Sezione di Roma2, Universit\'{a} di Roma Tor Vergata, Via della Ricerca Scientifica, 1 , Roma, 00133, Italy
\and 
(im) Leibniz-Institut f\"{u}r Astrophysik Potsdam (AIP), An der Sternwarte 16, D-14482 Potsdam, Germany
\and 
(gt) Faculty of Mathematics and Physics, University of Ljubljana, Jadranska 19, 1000 Ljubljana, Slovenia
\and 
(rag) Universit\'e Paris-Saclay, Universit\'e Paris Cit\'e, CEA, CNRS, AIM, 91191, Gif-sur-Yvette, France
\and
(sm) Instituto de Astrof\'isica de Canarias (IAC), E-38205 La Laguna, Tenerife, Spain
\and
(sm) Universidad de La Laguna (ULL), Departamento de Astrof\'isica, E-38206 La Laguna, Tenerife, Spain
\and 
(jmg) Department of Physics and Astronomy, Purdue University, 525 Northwestern Avenue, West Lafayette, Indiana, USA
\and 
(kl) Department of Astronomy, Stockholm University, AlbaNova University Centre, SE-106 91, Stockholm, Sweden
\and
(oas) Osservatorio di Astrofisica e Scienza dello Spazio di Bologna, INAF, Via Gobetti 93/3, Bologna, I-40129, Italy
\and
(zp) European Southern Observatory, Karl-Schwarzschild-Strasse 2, 85748 Garching bei M\"{u}nchen, Germany
\and
(lsw) Zentrum f\"{u}r Astronomie der Universit\"{a}t Heidelberg, Landessternwarte, K\"{o}nigstuhl 12, 69117 Heidelberg, Germany  
\and 
(ts) Scuola Superiore Meridionale, Largo San Marcellino10 I-80138 Napoli, Italy Napoli, Italy
\and
(vr) NAF-OACN Osservatorio Astronomico di Capodimonte, Salita Moiariello 16, 80131, Napoli (Italy)
\and
(uu) Observational Astrophysics, Department of Physics and Astronomy, Uppsala University, Box 524, SE-751 20 Uppsala, Sweden
\and
(uf) Universit\'{a} degli Studi di Firenze, Via G. Sansone 1, I-50019 Sesto Fiorentino, Italy.
\and
(oaa) INAF/Osservatorio Astrofisico di Arcetri, Largo E. Fermi 5, I-50125 Firenze, Italy 
\and 
(kuk) Astrophysics Research Centre, Keele University, Keele, ST5 5BG, UK.
\and
(mpe) Max-Planck-Institute for Ex. Physics, Giessenbachstrasse, D-85748 Garching, Germany
\and
(ita) Universit\"{a}t Heidelberg, Zentrum f\"{u}r Astronomie, Institut für Theoretische Astrophysik, Albert-\"{U}berle-Str. 2, D-69120 Heidelberg, Germany
\and
(OU) University of Oklahoma, 440W Brooks St., Norman, USA
\and
(oca) Observatoire de la Côte d’Azur,Laboratoire Lagrange, CNRS UMR 7293, CS 34229, 06304 Nice Cedex 04, FRANCE
\and
(syd) Sydney Institute for Astronomy, School of Physics, Faculty of Science, Sydney Astrophotonic Instrumentation Labs, the University of Sydney, Australia
\and
(ham) Universit\"{a}t Hamburg, Hamburger Sternwarte, Hamburg, Germany
}
\date{Received 19.05.2026; accepted 22.10.2026}

\abstract
{The Milky Way galaxy is home to hundreds of billions of stars, and the vast majority of them are in its disk and in the central region - the bulge. To probe the formation and evolution of these two components, it is essential to have large observational datasets of precise stellar chemical abundances, ages, positions, and 3D-space velocities probing stellar populations of all ages.}
{This paper presents the motivation, basic observing strategy, and the main planned scientific projects of the fourth 4MOST consortium survey (S4): the Milky Way disk and bulge high-resolution (\sd) survey. The primary goal is to explore and constrain the morphology, evolution, and chemical enrichment of the Milky Way disk and bulge. The 4MOST survey operations started in November 2026 and will proceed for five years.}
{We describe the selection of stellar samples, analyse them in the context of previous Galactic observational programmes, and define the Galactic components to be studied with the 4MOST data. We provide the first \textit{Gaia}-based characterisation of the input catalogue. We also describe the main sub-surveys and their content along with additional sub-surveys that complement other ground- and space-based missions, such as eROSITA, OGLE, K2, TESS, and PLATO.}
{The input survey catalogue comprises 3.6 million targets in the \textit{Gaia} magnitude range $7.5\lesssim G \lesssim17.5$. About $67$ thousand stars have G $> 16$, and $\sim 100\,000$ stars have [Fe/H] $\lesssim -2$. With its uniform spatial coverage of the entire southern sky ($-80^\circ < \mathrm{dec} < +5^\circ$ and the entire range of RA, spanning over 15 kpc radially and 5 kpc vertically from the Galactic mid-plane), the \sd\ survey will deliver the abundances of more than 15 chemical elements (namely, Li, C, O, Na, Mg, Al, Si, Ca, Ti, Cr, Mn, Co, Ni, Sr, Y, Ba, and Eu), stellar ages, and precise radial velocities, enabling studies of radial, vertical, and temporal variability of chemical gradients in the Galactic disk and bulge. We discuss how the precision of abundances and kinematics obtained by \sd\ will enable advances in studies of stellar physics and multiplicity, the origins of chemical elements, and the evolution of Galactic stellar populations.}
 {}
%
%


\maketitle

%

\section{Introduction}

\begin{figure*}
\centering
\resizebox{\hsize}{!}{
\includegraphics[width=0.55\hsize,angle=0]{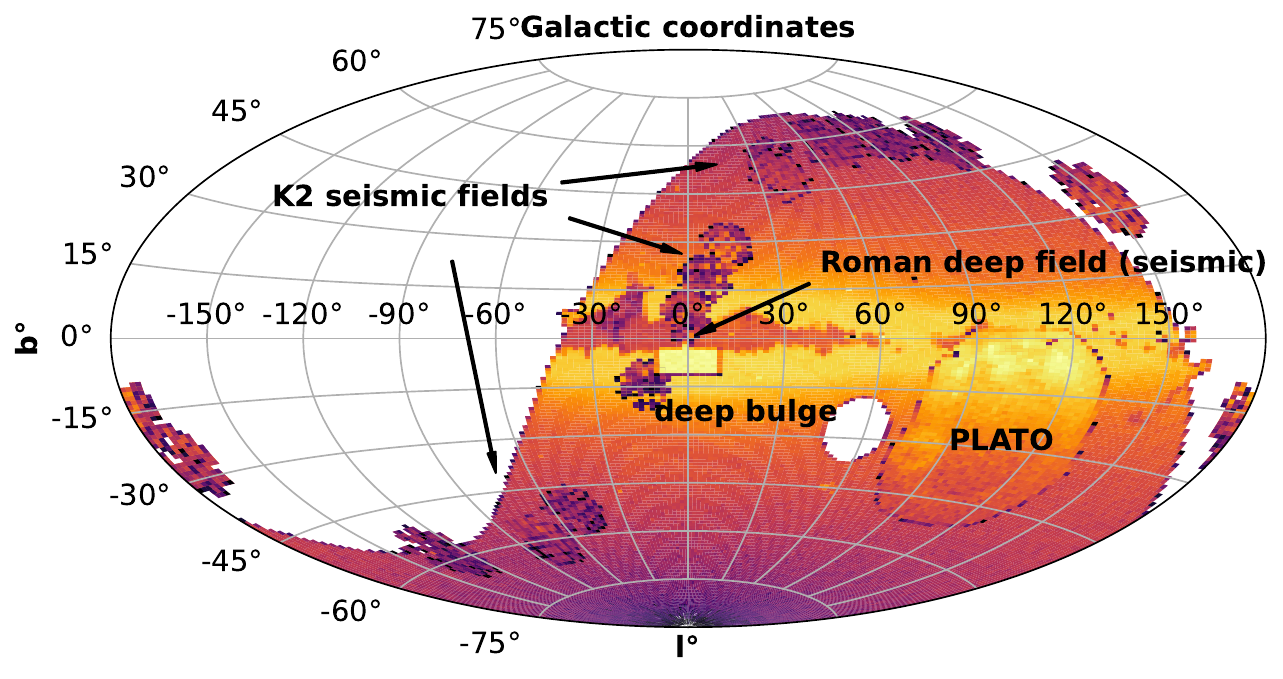}
\includegraphics[width=0.45\hsize,angle=0]{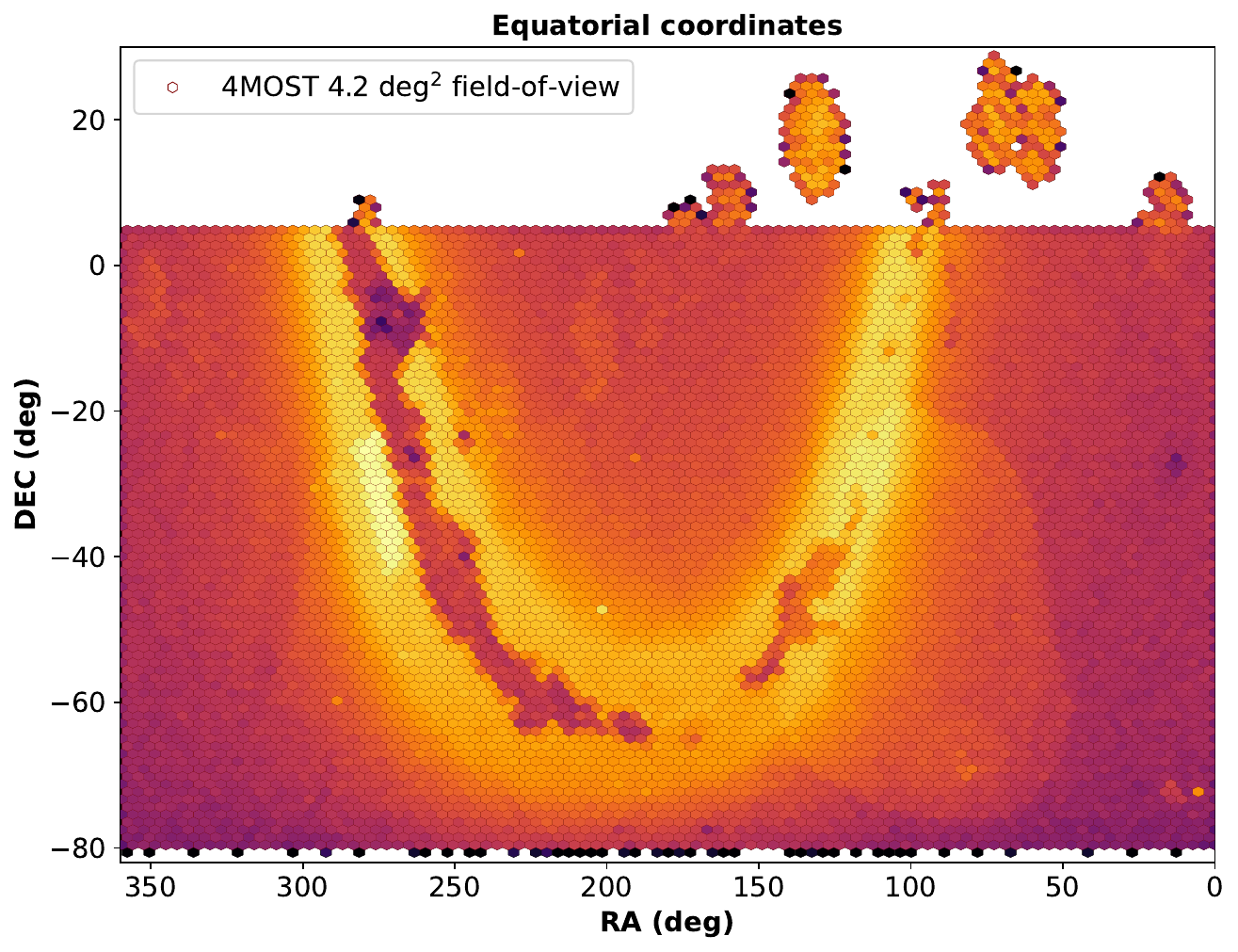}}
\caption{Sky distribution of targets in the \sd\ survey. {\it Left:} Galactic longitude and latitude in the Aitoff projection. Important fields in the survey are marked. {\it Right:} Space of RA-Dec. The grid shows the hexagons, and each represents the size of the 4MOST field of view (4.2 sq.deg.). The colours show the target density in our catalogue; however, given the fact that different sub-surveys have vastly different target densities, here we show each sub-survey with its own density scale (see Table 1 for more information). Some of the over-densities covered by individual scientific sub-surveys are also distinguishable, such as the PLATO LOPS2 (Long-Pointing Southern Field) at RA$\sim 100$ deg and the K2 fields (seen as small mosaics) as well as our deep Bulge field and the Roman asteroseismology field (towards the Galactic centre).}
\label{fig:radec}
\end{figure*}

Large-scale Galactic surveys hold tremendous potential for detailed studies of the formation and evolution of the Milky Way and its constituents, ranging from major stellar populations to smaller gravitationally bound systems such as star clusters and even individual stellar systems \citep{Edvardsson1993, Freeman2002, Feltzing2013}. Spectroscopic surveys such as SDSS-V\citep{Kollmeier2026}, RAVE \citep{Steinmetz2006}, \textit{Gaia}-ESO \citep{Gilmore2022, Randich2022}, LAMOST \citep{zhao2012}, APOGEE \citep{Majewski2017, Abdurrouf2022}, GALAH \citep{DeSilva2015, Buder2025}, and \textit{Gaia} \citep{Gaia-Collaboration2020} have demonstrated this potential impressively. With homogeneous all-sky spectroscopy of $10^5$ to $10^6$ individual stars, these programmes have achieved an extraordinary scientific return. 

However, over the past decade, new and broader research themes have emerged that position the Milky Way as a time-evolving and structurally complex benchmark system in both galactic and cosmological contexts \citep{Kewley2008, Lian2023, Vislosky24, Monty2025, McCluskey2025, Zasowski2025}. This perspective has gained further relevance in light of the properties of the high-redshift galaxies that are now becoming accessible through facilities such as the \textit{James Webb} Space Telescope (JWST; \citealt{Ferreira2022,Ferreira2023,Yan2024,Tohill2024}) and the Atacama Large Millimetre/submillimeter Array (ALMA; \citealt{Smit2018,Rizzo2021,Roman-Oliveira2023}). Many of these themes focus on the Galactic disk and bulge, and they benefit enormously from the order-of-magnitude improvement in the precision and volume of \textit{Gaia} astrometry and photometry \citep{GaiaDR3} compared to its predecessor \textsc{Hipparcos} \citep{Perryman1997} as well as from major advances in the spatial resolution and sub-grid physics of numerical simulations of Milky Way-like galaxies \citep[e.g.][]{grand16, Minchev2014, pillepich24, Buck2020}. These simulations now allow testable evolutionary scenarios of these two components to be formulated and their distinct observable signatures to be predicted.

Key open questions that remain include characterisation of the early phases and the formation of the primitive Galactic disk \citep{Semenov2025, Sotillo-Ramos2023, vanDonkelaar2025}; secular processes in disk evolution, such as perturbations induced by satellite galaxies \citep{Belokurov2018, Laporte2019, Bland-Hawthorn2019, Khoperskov2023-hestia1}; dynamical heating and scattering by transient spiral arms and the Galactic bar \citep{Schoenrich2009, Minchev2014, Khoperskov2020-bar}; the role of extreme events such as compact binary mergers in Galactic chemical evolution \citep{Cote2019, Cowan2021, Storm2025}; and the origin and star formation history of the bulge \citep{Barbuy2018} and ultimately the first stars associated with it \citep{Klessen2023, Caffau2024b, Bonifacio2025}. Addressing these topics require significantly larger number statistics, completeness, uniformity, and depth in spatial, kinematic, chemical, and in photometric domains. 

In this paper, we present and describe the \sd\ survey on the ESO 4-Meter multi-Object Spectroscopic Telescope (4MOST) at the ESO Paranal observatory in Chile. We clarify that \sd\ is an acronym for 4MOST MIlky way Disk And BuLgE High-Resolution survey. This will be the largest optical high-resolution (resolving power of $R=\lambda/\Delta \lambda \approx 20\,000$) stellar survey to date in terms of number of targets, photometric depth, and spatial area. At present, $66$ scientists from $14$ countries\footnote{Sweden, Germany, Spain, Switzerland, Chile, United Kingdom, Austria, Australia, Italy, France, Slovenia, Israel, Belgium, and New Zealand} have joined the survey, and diverse collaborations with other 4MOST community and consortium surveys are planned. The timeline of the survey operations and data release is defined by that of the 4MOST Data Release Plan.\footnote{\url{https://www.eso.org/sci/observing/PublicSurveys/docs/4MOSTnew/VIS-PLA-4MOST-47110-9210-0003_3_00_Data_Release_Plan_signed.pdf}} With the survey operations starting in November 2026, it is planned that the public Data Release 0 (DR0) will occur after 7 months, and it will contain the key stellar parameters obtained on the best effort. The DR1 will follow after 2.25 years after the start of survey operations and will contain all basic pipeline parameters, including detailed chemical abundances (as defined for the $\sd$ survey in the survey management plan\footnote{\url{https://www.eso.org/sci/observing/PublicSurveys/4MOSTsmp.html}}), whereas DR2, DR3, and DR4 will take place after 3.25, 4.75, and 6.75 years, respectively.
%
%

\begin{figure*}[ht]
\centering
\includegraphics[width=1.6\columnwidth, angle=0]{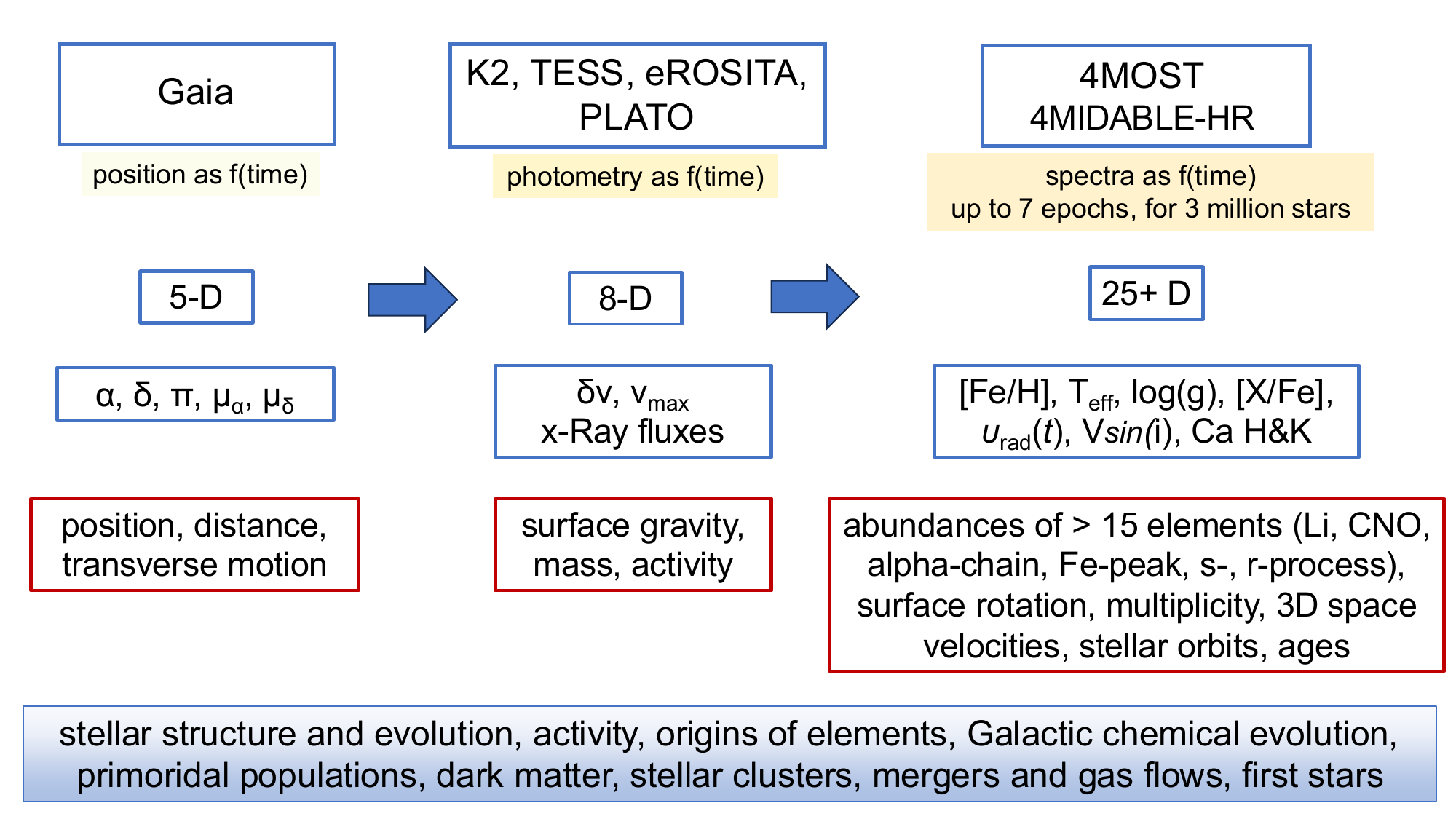}
\caption{Scientific data to be produced and topics to be addressed by the \sd\ survey. Figure inspired by \citet{Gilmore2012}.} 
\label{fig:sciencedata}
\end{figure*}

\begin{figure*}[ht]
\centering
\includegraphics[width=1.5\columnwidth, angle=0]{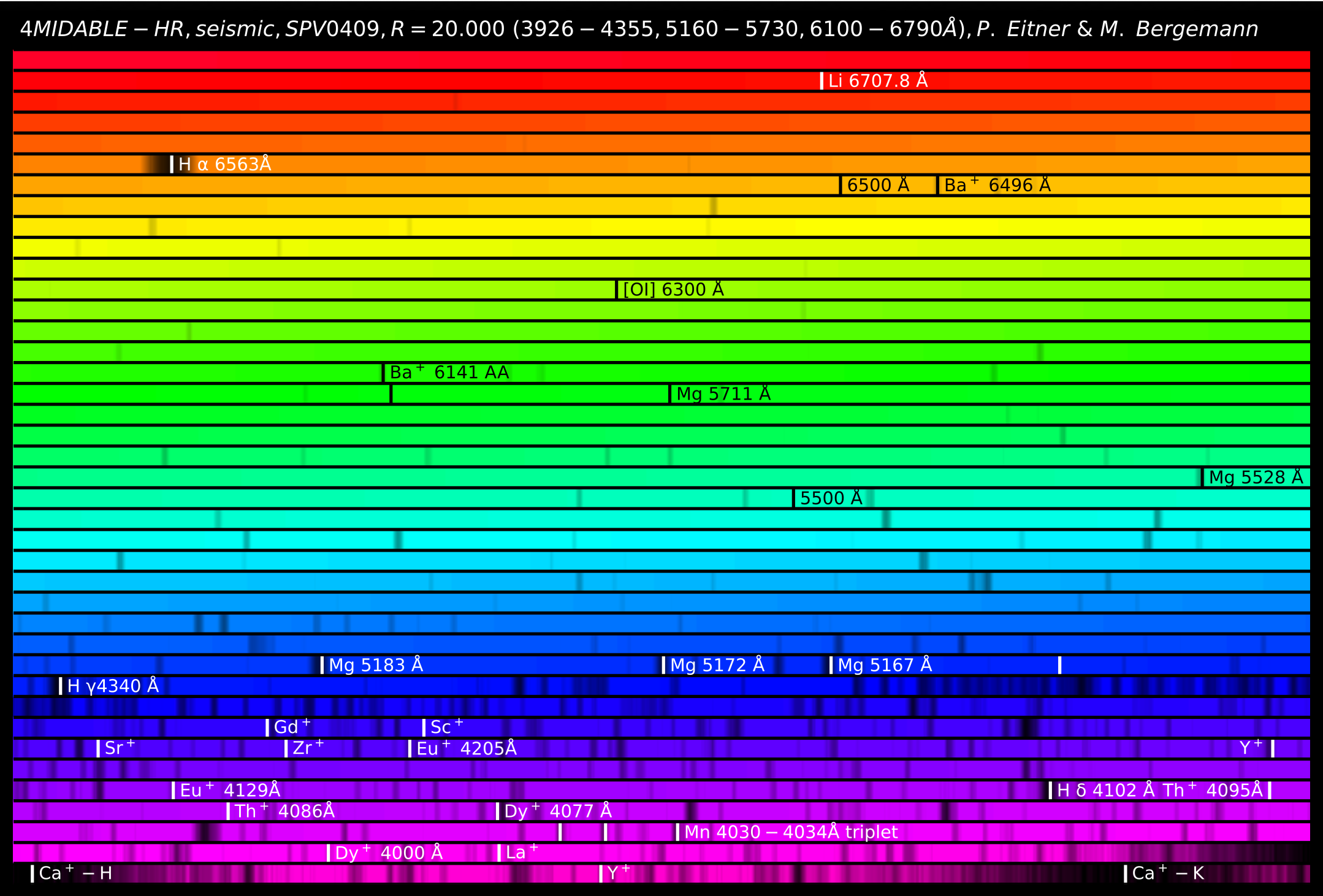}
\caption{Visualisation of a selected \sd\ stellar spectrum of a red giant taken during the SPV program.} 
\label{fig:visdata}
\end{figure*}

The baseline selection of the \sd\ stellar sample is simple. It is fully based on \textit{Gaia} DR3 positions and photometry \citep{GaiaDR3}; covers the spatial area of $28\,000$ square degrees,\footnote{The 4MOST field of view is 4.2 deg.$^2$, for comparison the size of the Moon is 0.2 deg.$^2$. Thus, 4MOST covers an area of 20 Moons in one pointing, which is large by astronomy standards. But as seen in Fig. 1, the largest field of view in astronomy is that of PLATO, as visible in this figure by the large area of $\sim 2200$ deg.$^2$ around RA$\sim 100$ deg; see \url{https://www.cosmos.esa.int/web/plato/first-observation-field}.} with $-80^\circ < \mathrm{dec} < +5^\circ$ and $0^\circ < \mathrm{ra} < 360^\circ$  (Fig.~\ref{fig:radec}); and includes stars within the {\it Gaia} apparent magnitude range of $7.5 \lesssim G \lesssim 17.5$. The \sd\ survey does not pre-select on colour or in the $T_{\rm eff}-\log g$ plane; therefore, stars of different spectral types are present in the input catalogue. However, given the volume coverage and magnitude limits, the vast majority of targets are late A-, F-, G-, K-, and M-type stars of all evolutionary phases and ages. 
The wavelength range of the 4MOST high-resolution spectra (three windows: $3926 - 4355$\,\AA, $5160 - 5730$\,\AA, and $6100 - 6790$\,\AA) covers key lines of chemical elements tracing major nucleosynthesis channels, from the lightest, such as Li, CNO, $\alpha$-chain elements (O, Mg, Si, Ca), to odd-Z elements (Na, Al) through to the Fe-group (Ti, Mn, Co, Ni) and to neutron-capture elements, such as Ba, Nd, and Eu. These spectra (Fig. \ref{fig:visdata}) will also enable measurement of stellar surface rotation velocities with the uncertainty of $3-5$\,km\,s$^{-1}$ and chromospheric activity (based on the near-UV H \& K lines of Ca\,{\sc ii} and H$\alpha$ for a subset of stars) as well as the detection of binaries, yielding information on stellar multiplicity and evolution of stellar rotation rates. Figure \ref{fig:sciencedata} shows the scientific data to be produced and the topics to be addressed by the \sd\ survey.

Moreover, the \sd\ survey will complement all other 4MOST Galactic surveys, such as the 4MIDABLE-LR (4MOST MIlky way Disk And BuLgE Low-Resolution survey) disk and bulge survey  \citep{Chiappini2019},\footnote{We note that the scope of science in both 4MIDABLE surveys is similar, which is why the same acronym is used. However, the top-level questions are different and they will be described in detail in the publications accompanying the 4MIDABLE-LR survey.} the 4MOST LR and HR halo surveys \citep{Christlieb2019, Helmi2019},  the Magellanic Clouds survey \citep{Cioni2019}, and the Milky Way dwarf galaxy survey \citep{Skuladottir2023}. For completeness, we also note that the ecological footprint of 4MOST, and specifically that of \sd\ survey, which occupies about 10~\% of time on the telescope, is very modest. According to our conservative estimates, the total emission equivalent will be of the order of two to five hundred tons CO$_2$e$/$year, thus about a few tens of tons of CO$_2$e$/$year for \sd.\footnote{We computed the number as follows. According to this ESO report:  https://www.eso.org/public/germany/about-eso/sustainability/green/
Cerro Paranal and Cerro Amazones together have the footprint of $4600$ tons CO2$_2$e/year, where CO$_2$e per year stands for tonnes of carbon dioxide equivalent per year. Most energy is likely spent in Paranal, so $4000$ tons CO$_2$ per year (Paranal only). Then 10 telescopes on Cerro Paranal (8 VLT units, 2 survey telescopes), https://www.eso.org/sci/facilities/paranal.html we get for 4MOST $2300/10 = 230$ tons CO$_2$e per year} Considering the scope of \sd\ (comparable to the \textit{Gaia}-ESO survey), this compares very favorably to other astronomical facilities, even to {\it Gaia} \citep{Knodlseder2022}.
%
%
%
\begin{figure*}
\hbox{
\includegraphics[width=0.33\textwidth]{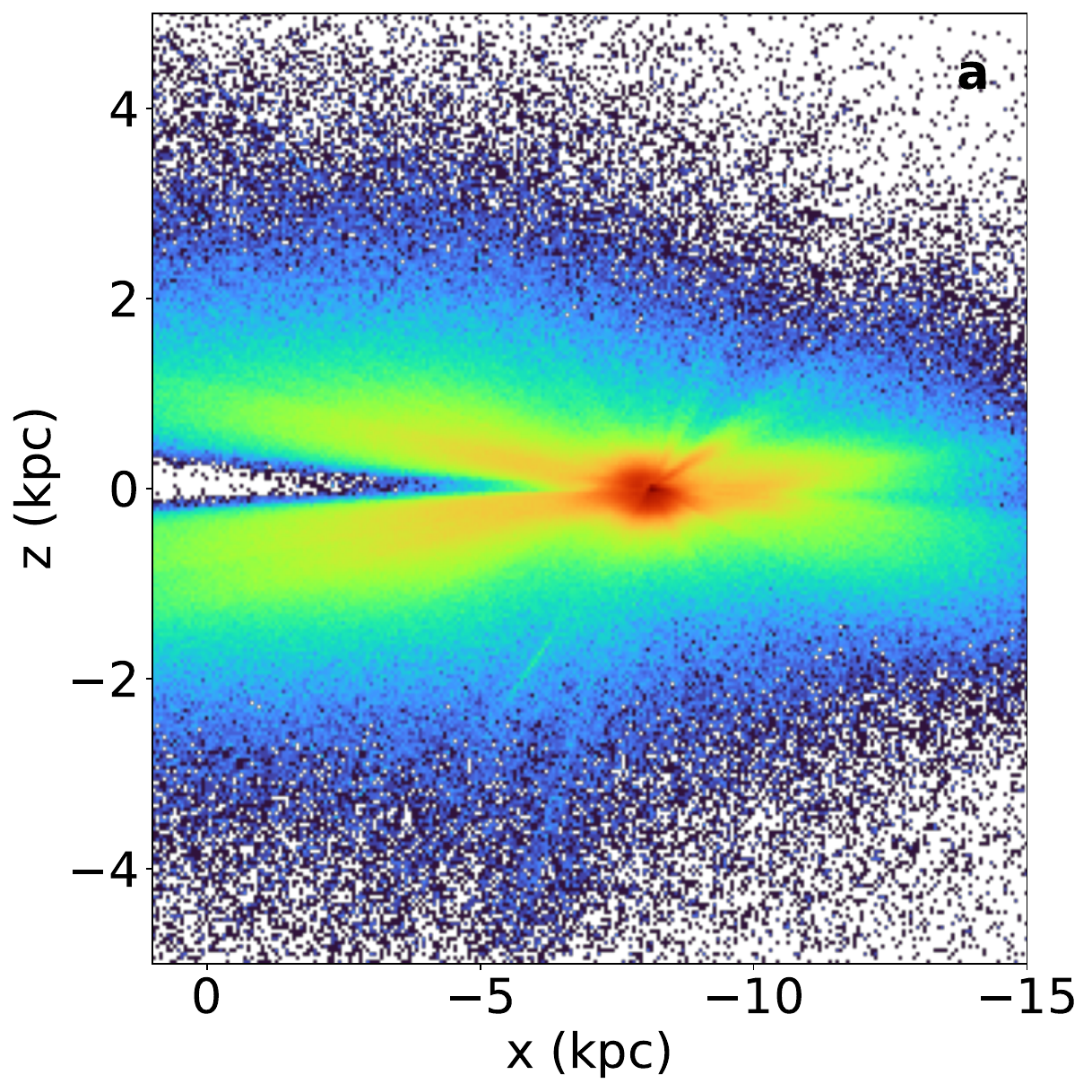}
\includegraphics[width=0.33\textwidth]{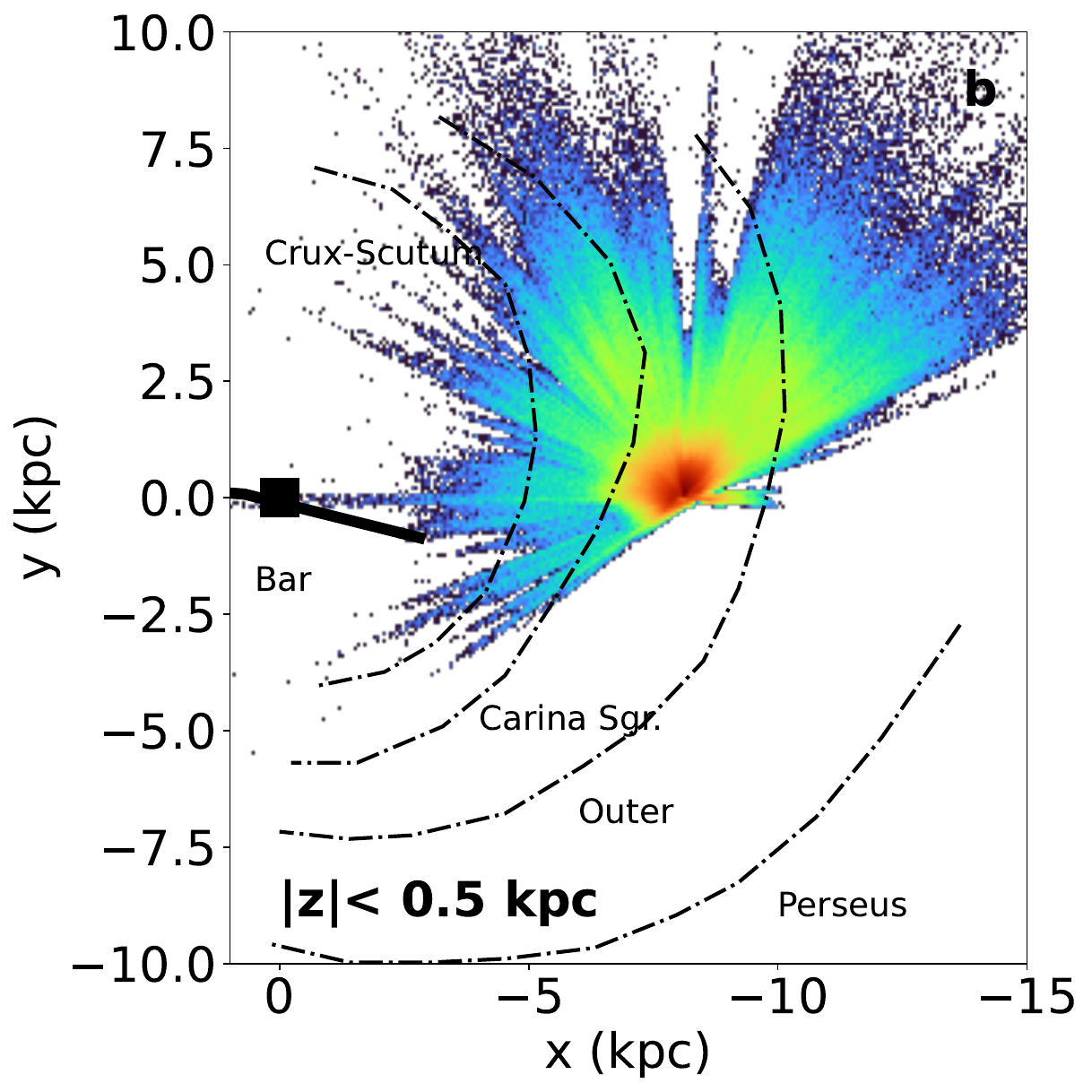}
\includegraphics[width=0.33\textwidth]{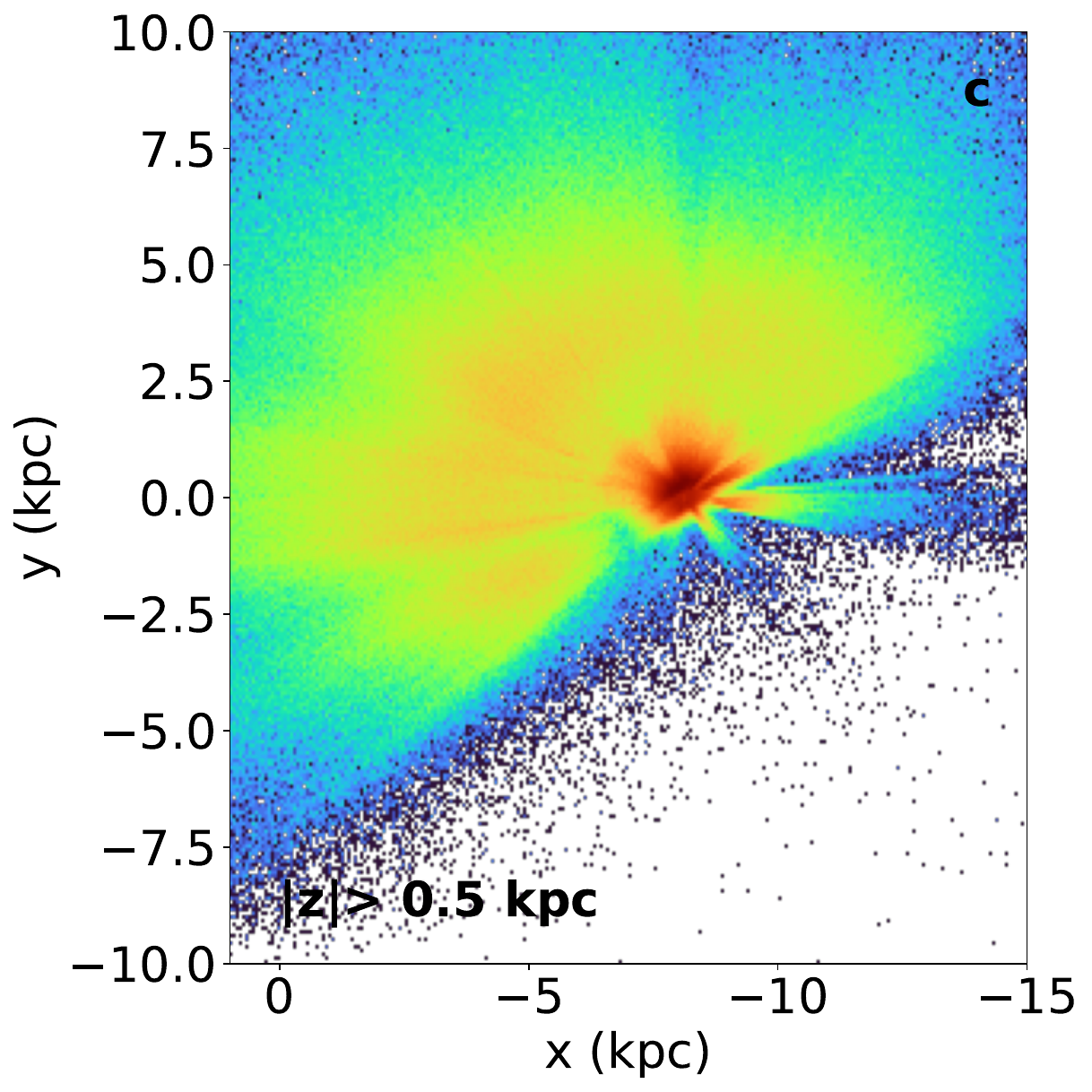}
}
\caption{Distribution of \sd\ targets in the Galactic Cartesian coordinate planes $X-Z$ (\textit{left}) and $X-Y$ (\textit{middle} and \textit{right}). The locations of the Galactic bar and spiral arms are overplotted (data adopted from \citealt{Momany2006} and \citealt{Bland-Hawthorn2016}).}
\label{fig:summary3d}
\end{figure*}

The paper is organised as follows. In Sect.~\ref{sec:1} we outline the main motivation and goals of \sd\,, including Galactic, stellar, and planetary science projects envisioned by our team with \sd\ data. Section~\ref{sec:2} provides a working definition of Galactic populations to enable decomposition and identification of stars in the \sd\ samples according to their parent Galactic components. Section~\ref{sec:3} presents a brief overview of the 4MOST telescope, the observational strategy, and the astrophysical analysis of the \sd\ spectra. In Sect.~\ref{sec:4} we describe the individual \sd\ sub-surveys and identify the scientific synergies with other facilities, including eROSITA, TESS, and PLATO. In Sect.~\ref{sec:5} we review auxiliary programs, including multiplicity, supplementary science programs to be carried out on 4MOST in collaboration with \sd\,, and the science verification experiments. Lastly, the paper closes with a summary, conclusions, and our outlook in Sect.~\ref{sec:6}.
\section{Baseline considerations about the sample size}

The conceptual design of the \sd\ survey is driven by the goal to study stellar populations  using their chemical abundances, kinematics, and ages. The 4MOST facility enables this to be done over three quadrants of the Galactic disk, and the sensitivity of the instrument enables probing out to several kiloparsecs vertically and 10\,kpc radially from the Sun. For the \sd\ survey to address its core science questions, we require a stellar sample of more than three million stars with elemental abundances that are accurate to better than $0.1$\,dex.

We estimate that an order of $100$ stars are needed to obtain a statistically robust mean down to 10\,\% of sigma ($\sqrt{N}$)  in order to characterise the moments of the kinematic distributions of a single population at a given location. The expected precision in elemental abundances allows us to cut the sample in 0.1\,dex width in abundance space. Sampling the abundance plane in [$\alpha$/Fe] versus [Fe/H] meaningfully would thus require on the order of 50 abundance groups times 100 stars. The total sample size in this survey thus allows us to map out these populations at about 400 locations. This will enable us to distinguish the separation of elemental abundance trends between the $\alpha$-poor and $\alpha$-rich disks, and between the halo and the metal-poor disk, and to trace the position of the $\rm [\alpha/Fe]-[Fe/H]$ `knee' from the solar neighbourhood, through the inner disk, and into the bulge \citep{hayden2011}.

To resolve kinematic perturbations in the Galactic disk \citep{antoja2018}, one needs to resolve the mean velocity of the stars to better than 0.3\,km\,s$^{-1}$, which, at the velocity dispersion of 50\,km\,s$^{-1}$, requires $20\,000$ stars for each data point and would allow us to resolve about 100 different points in angular momentum space or physical position space. We also want to map the age-velocity dispersion relation at different points in elemental abundance space and physical position space to shed light on the heating mechanisms in the Galactic disk in order to differentiate between possible high velocity dispersion at birth for older stars, secular heating by disk structure (for example, the Galactic bar, or spiral arms) or giant molecular clouds, and merger induced heating. To detect a merger, we can roughly estimate that we need to precisely track an increase of 5\,km\,s$^{-1}$ in the velocity dispersion over the course of roughly one billion years. This would require $5\,000$ stars in each age bin (about one billion years wide) or about $50\,000$ stars in total. These need to be sampled at 10 to 20 different Galactocentric radii and altitudes dissected in metallicity to pinpoint the radial pattern of heating. 
 
\begin{figure*}
\includegraphics[width=1.99\columnwidth]{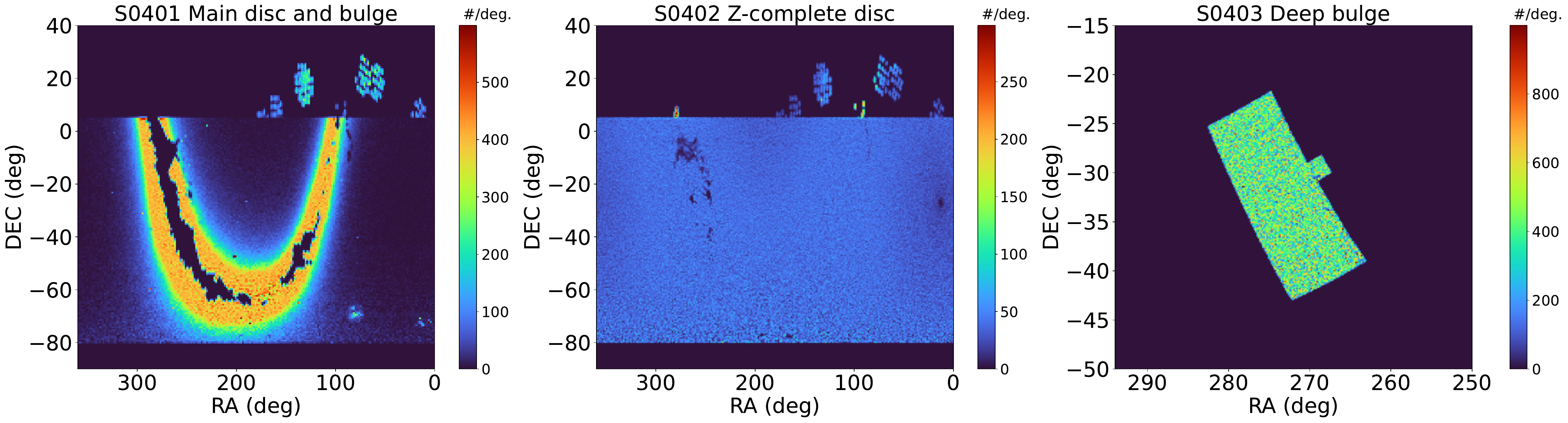}
\includegraphics[width=1.99\columnwidth]{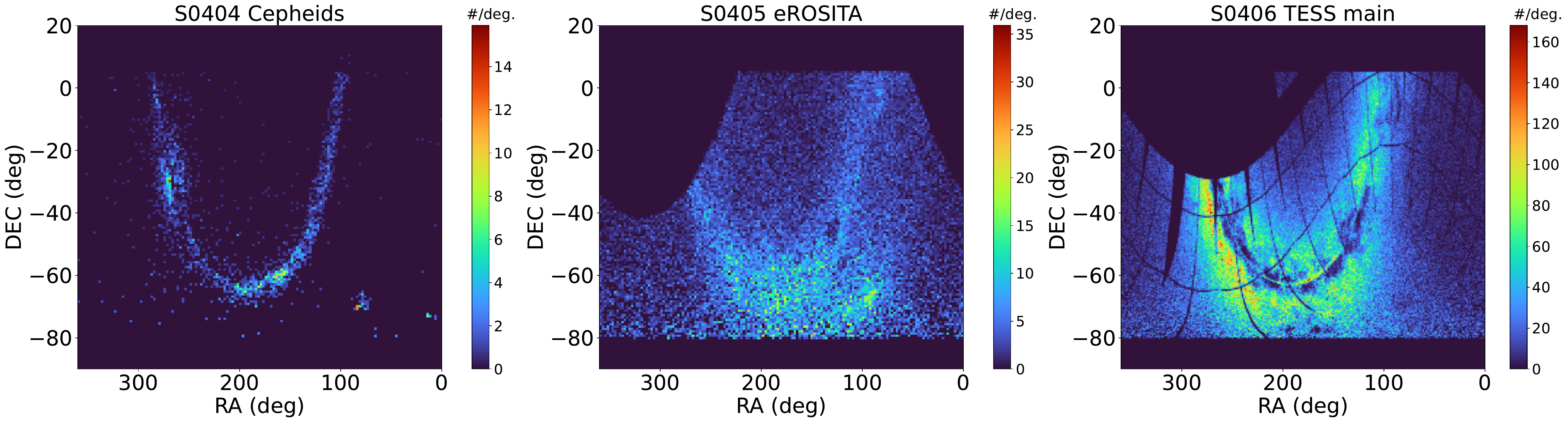}
\includegraphics[width=1.99\columnwidth]{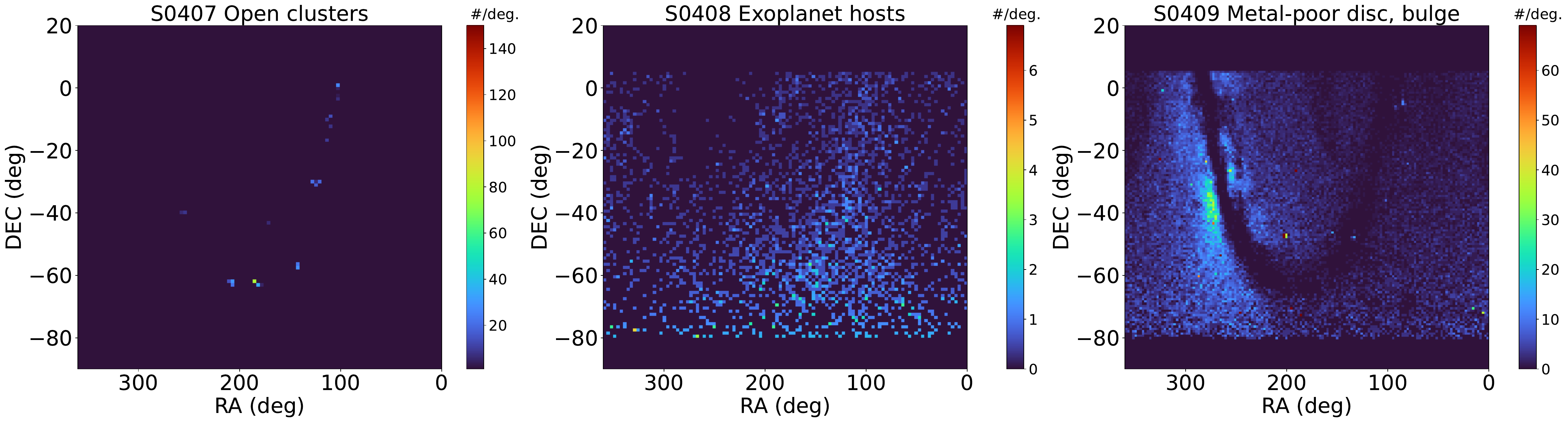}
\includegraphics[width=1.99\columnwidth]{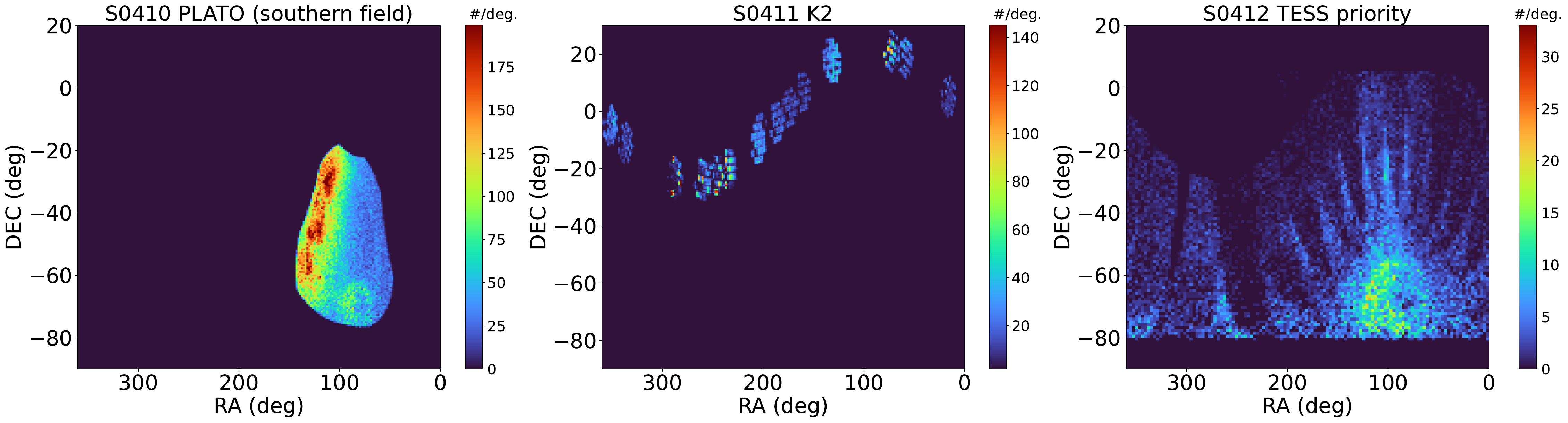}
\caption{RA-Dec distribution of targets in individual \sd\ sub-surveys. The colours show the stellar density per degree; note the difference in scales of individual sub-surveys.}
\label{fig:targetsRADEC}
\end{figure*}

With these estimates, we need on the order of four million stars to comprehensively address the formation of the Galactic components, which motivates the baselines sample size and selection of the \sd\ survey. Figure \ref{fig:summary3d} shows the distribution of our entire  stellar sample in the Galactic coordinate plane. In table \ref{tab:main} we provide a list of all sub-surveys that are part of the \sd\ program, and these will be used to address the scientific questions described in the next sections. The RA-Dec distributions of these 12 sub-surveys are shown in Fig. \ref{fig:targetsRADEC}. More detailed information about these sub-surveys is provided in Sect. \ref{sec:4}.
%
%
%

%
%
%
\section{Definitions and Galactic population identifiers}\label{sec:2}

The design of the \sd\ survey is driven by the goal to address the structure, formation, and evolution of the Galactic bulge and disk. The survey's sample selection $3\,638\,780$ targets broadly covering the entire southern hemisphere, and it requires an operative definition of the components in order to distinguish stellar populations - also from those of other 4MOST Galactic surveys (such as the halo surveys, \citealt{Christlieb2019, Helmi2019}) - and to ensure comprehensive and robust scientific exploitation of the observed data.

Historically, a set of definitions and conventions has been developed to describe Galactic populations, some based on chemistry \citep{Fuhrmann1998,Gehren2004, Bensby2011, Adibekyan2012, Nissen2010} and ages \citep{Bensby2014,Xiang2022}, other based on geometry, kinematics \citep{Bensby2003, Fuhrmann2004, Bensby2014} and/or orbits \citep{Khoperskov2025-disc}, and stellar densities \citep{Gilmore1983} and further approaches making use of some combination of these \citep{Ruchti2011,Chandra2024}. Generally, these descriptions are limited, as they rely on empirical evidence of multi-modalities in chemical space \citep[e.g.]{Adibekyan2012,Bensby2014,Hayden2015,Gent2024}, or assumptions such as the Gaussianity of velocity dispersions in stellar populations \citep{Bensby2003}. They also consider results from analytical or cosmological galaxy formation models tailored to the Milky Way \citep{Buck2020, Sotillo-Ramos2023, Semenov2025}, and the latter suggest that these populations are not isolated components, but occupy different regimes of a continuous orbit-stellar-population distribution \citep{Khoperskov2025-disc}. Overall, however, it has been recognised that formation and dynamical histories of  bulk of star in these components (disk, bar, bulge, and halo), are different, although likely influencing each other's evolution. These differences, when mapped into the chemo-dynamics space offer a utility in our definition of Galactic components, as described below.

For the Milky Way stars, for which properties of orbits are generally well-understood quantitatively, definitions of components rests upon the following considerations. The classical approach to isolate stars by kinematics \citep[e.g.][]{Fuhrmann2002,Bensby2003,Nissen2004,Ruchti2011} commonly assumes that components have a Gaussian distribution with local characteristic velocity distributions \citep{Beers1995}. Specifically, the total velocity of $\sqrt(V_{\rm LSR}^2 + V_{\rm r} + V_{\rm z}) < 70$ km/s, $80 - 180$, and $>200$ km/s is frequently used to identify possible thin, thick disk, and halo stars \citep{Nissen2004,Bensby2014,Kordopatis2011}. By this definition, our survey comprises 5\,\% halo, 19\,\% thick disk, and 74\,\% thin disk stars. 
%
\begin{figure}
\centering
\includegraphics[width=0.8\columnwidth]{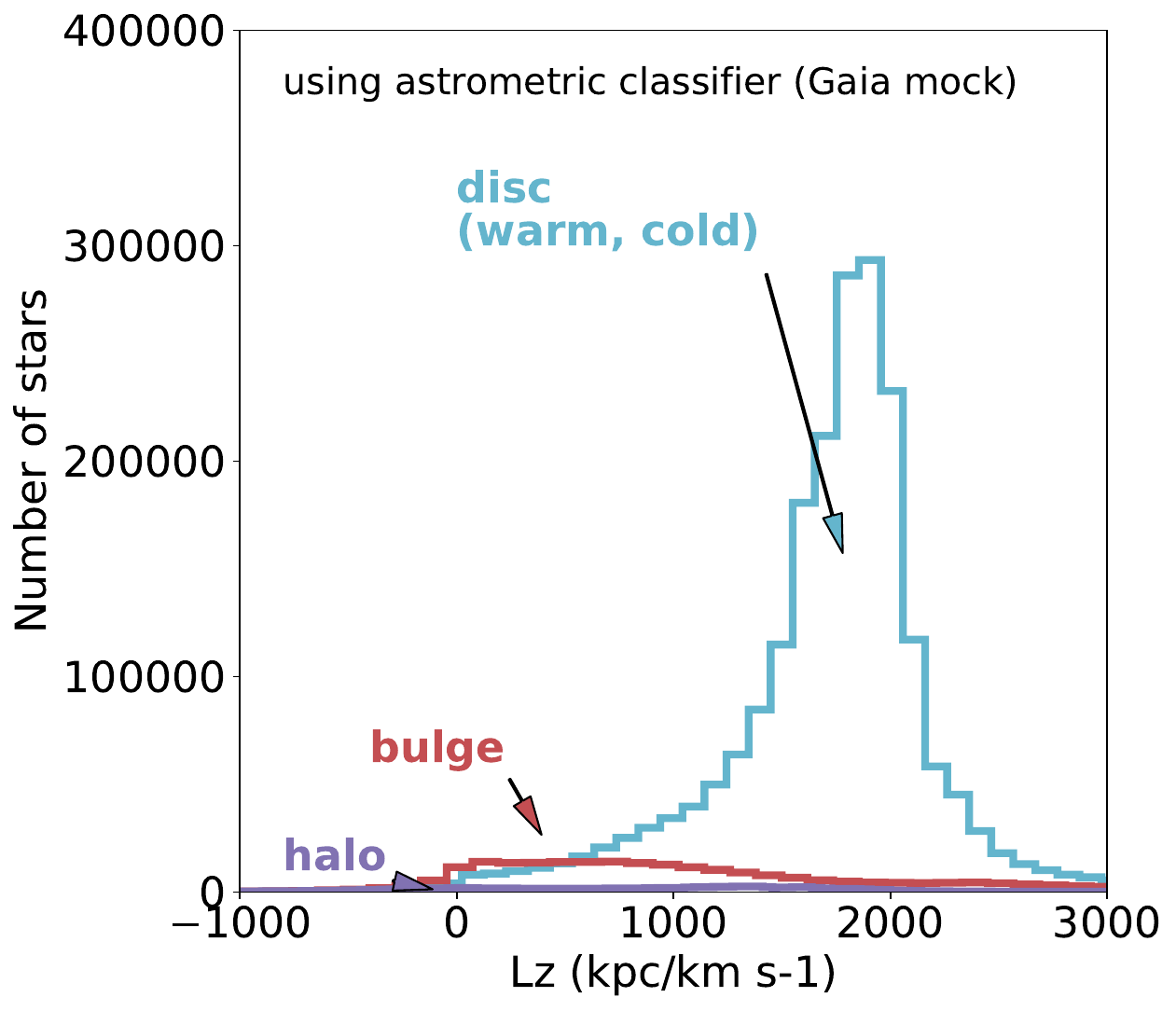}
\includegraphics[width=0.8\columnwidth]{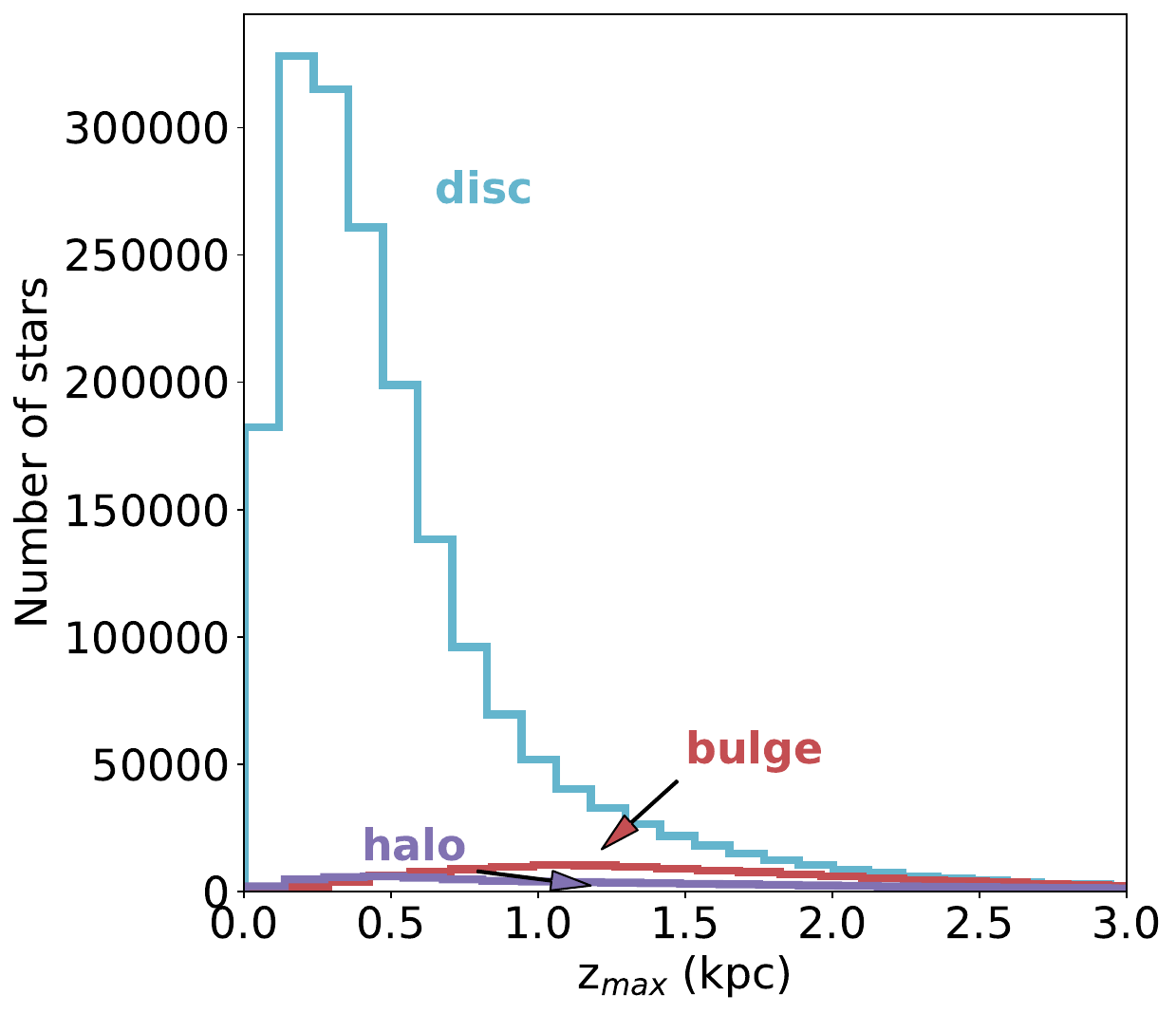}
\caption{Histograms of individuals components (disk, bulge, and halo) of the \sd\ survey as a function of vertical angular momentum L$_z$ and $z_{\rm max}$.}
\label{fig:lzzmax}
\end{figure}
%
%
Another approach is quantify the components via orbital quantities, such as the integrals of motion (energy, angular momentum), parameters like eccentricity and inclination, and actions \citep{Hunt2025}, because in dynamically quiescent phases these quantities remain invariant along an orbit \citep{Simpson2021,Feuillet2021,NAidu2021,Chandra2024,Bonaca2025}. The distributions of our targets in the parameter space of these quantities is demonstrated in Fig. \ref{fig:targetsup}.

For the majority of \sd\ targets ($2\,430\,540$, that is, over $65$ \% of all \sd\ stars), radial velocities are available from \textit{Gaia} RVS DR3, enabling orbital analysis. Assuming disk stars as those confined to the disk plane, such as J$_z$ $< 30$ km/s/kpc, where J$_z$ is the vertical action, with the bulge being mostly represented by stars in the inner Galaxy, $r_{\rm apo} <$ R$_{\rm Sun}$. So, we may define the halo as a population of stars whose orbits do not confine them close to the disk plane and not to the inner part of the Galaxy, thus J$_z$ $> 30$ km/s/kpc and $r_{\rm apo} >$ R$_{\rm Sun}$, although one has to keep in mind that also halo stars at suitable orbit phases may be found near the Galactic plane or close to the Galactic center. Using this definition, the \sd\ survey has: Disk/(Bar)/Bulge 93\%, Halo 7\%. These main populations in the \sd\ catalogue are demonstrated in Fig. \ref{fig:lzzmax}, where the detailed procedure to assign the membership is described in the appendix. We note that whereas the main focus of the \sd\ survey is the disk and the bulge, probing halo-disk interface requires stars of both components to be observed homogeneously and with the same quality of parameters and abundances. Therefore a small fraction, at the level of 5-10\,\%, of halo stars (currently, kinematically tagged) is allowed for in our catalogues.

We emphasise that these definitions are adopted for clarity, and they may be revised taking into account chemistry and/or ages of stars, which will be self-consistently obtained within the \sd\ survey. We note that with 4MOST HR spectra, the RV precision is expected to be better than $\sim 0.5 - 1$\,km\,s$^{-1}$. 
\section{4MOST instrument and observed spectra}\label{sec:3}
The 4MOST facility is described in detail in \cite{deJong2019} and we refer the reader to that paper for a detailed description. Here we only provide a brief summary of the most relevant information. 

\begin{figure}[!ht]
\hbox{
\includegraphics[width=\columnwidth]{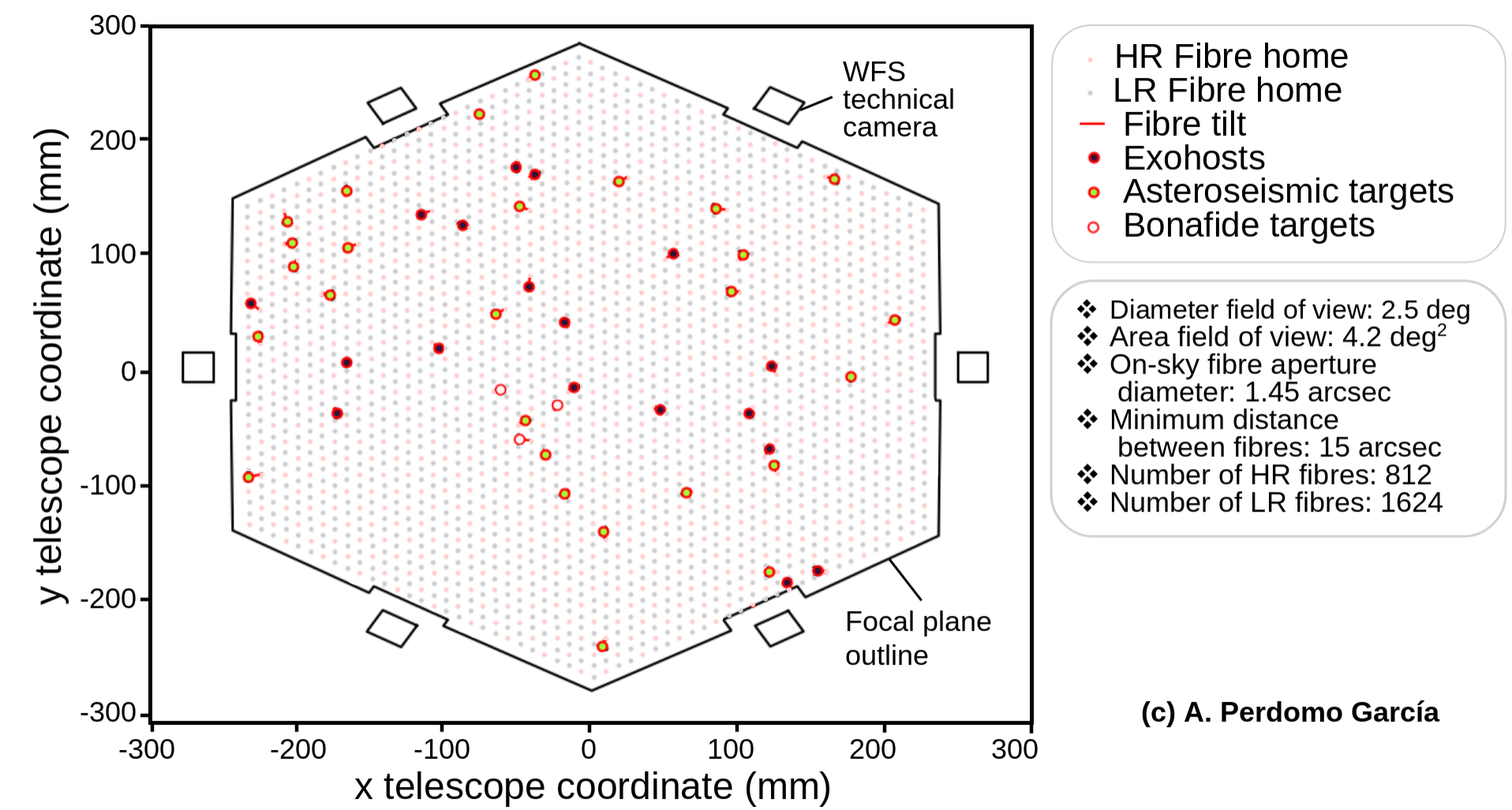}}
\caption{One example of fibre target allocation during the creation of an OB. The image shows the Cassegrain focal plane of the VISTA telescope (projected area on the sky of the field of view of 4.2\,deg$^2$; diameter of 2.5\,deg) at $\rm RA, Dec=(0,0)$ at 20:18:38 UTC on 17 February 2026. The fibres in the example are placed on 366 HR and 647 LR targets.} 
\label{fig:fop}
\end{figure}

As a highly multiplexed facility, 4MOST with its hexagonal $4$ square degree field of view and AESOP fibre positioner \citep{Brzeski2018} has the capability to simultaneously observe $2\,400$ targets. The facility is therefore ideally suited for parallel observations of targets from the $25$ Galactic and extragalactic 4MOST surveys \citep{deJong2019,Walcher2019}. $812$ the fibres deliver light to the high-resolution spectrograph (HRS; $R= \lambda/d \lambda \approx 20\,000$) and $1624$ fibers feed two identical low-resolution spectrographs (LRSs; $R=6\,500$). At high resolution, the resolving power varies between $19\,000$ at the short-wavelength edge of each channel to $21\,000$ at the long-wavelength end. The fibre diameter is $1.45$\,arcsec, and the system allows for a flexible fibre allocation with each object typically  accessible by two fibres and there is a minimum distance between adjacent targets of about $15$\,arcsec. The standard exposure time is 20 minutes, but the exposure time varies depending on the sky conditions: $\sim$ 3 - 5 minutes (super-bright), 6-8 (bright), 15-19 (grey), 20-21 minutes (dark).\footnote{Based on the 4MOST Visit planner simulator, e.g. 778 and 803.} 
Targets can be observed in multiple visits\footnote{A visit is also known internally as an Observation Block, OB}, and the data can be co-added to build up to the required signal-to-noise ratio \citep{Guiglion2019a}. One example of this OB allocation is shown in Fig. \ref{fig:fop}. Due to the saturation limit of the instrument, the brightest subset of our targets (those with $G < 10.5$) will be observed only during the 4MOST Poor Conditions Programme (PCP). We do not expect this to have any adverse impact on data quality.

The sensitivity requirement for the HRS are as follows. One hour of exposure time gives a spectrum with a signal-to-noise ratio $S/N\approx100$ per {\AA}ngstrom in the blue window for a main-sequence star with $G=14$ (three pixels per resolution element) and the same $S/N$ in the red is reached at $G\approx15.5$. The spectra will cover three spectral windows: $3926 - 4355$\,\AA, $5160 - 5730$\,\AA, and $6100 - 6790$\,\AA. These specific windows were chosen to include key diagnostic absorption lines of major chemical elements representing all main nucleosynthesis channels \citep{Hansen2015,Kordopatis2023}.\footnote{\url{https://line-detector.oca.eu}} Most of the elements will be targeted in the neutral phase, but for some species like Ca, Fe, and Ti, diagnostic lines of both neutral and singly ionised species are available. In addition, the key diagnostic lines for stellar parameter determination are covered, including the gravity-sensitive \ion{Mg}{i} triplet lines at $5170$\,\AA~\citep{Fuhrmann1998, Bergemann2017a} and the effective temperature-sensitive H Balmer lines, including H$_{\alpha}$ at $6563$\,\AA~and H$_{\beta}$ at $4863$\,\AA.

\begin{figure}
\hbox{
\includegraphics[width=\columnwidth]{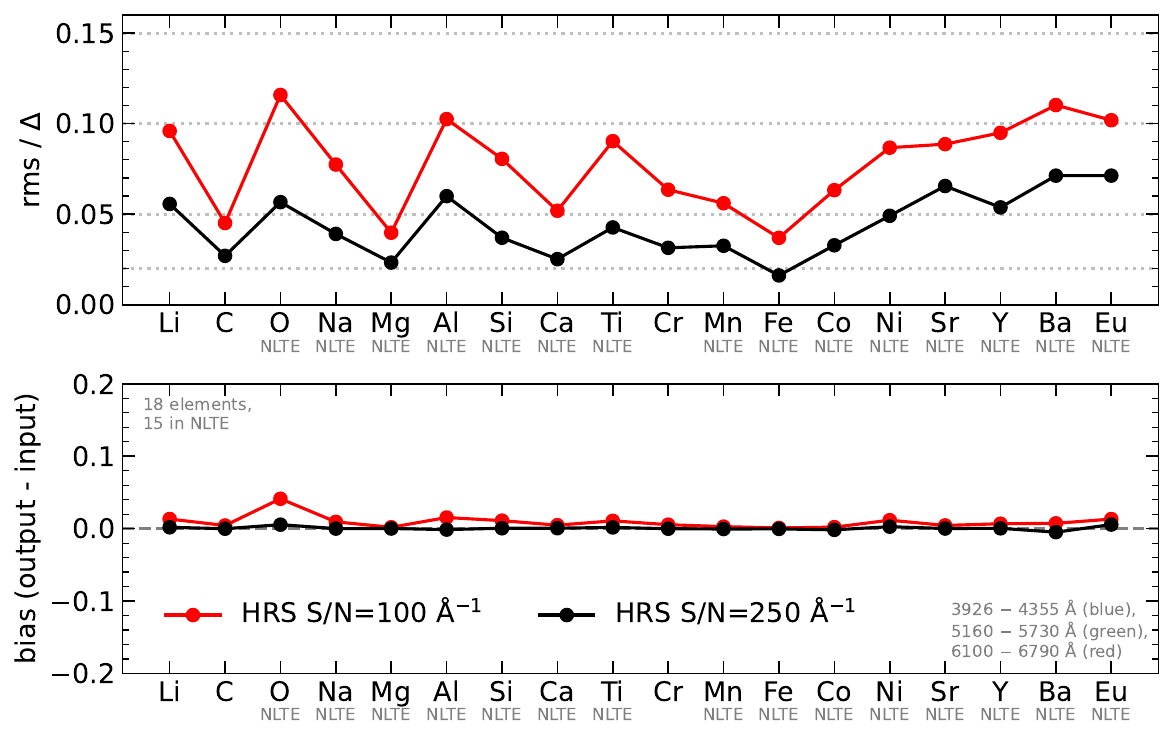}}
\caption{Characteristic scatter (\textit{top}) and bias of elemental abundances expected for the \sd\ survey data. The values were computed using the Payne NLTE code \citep{Storm2025b} and 4MOSTified synthetic templates.}
\label{fig:precision}
\end{figure}

The abundance analysis of the \sd\ data will be carried out using established codes within the IWG7 4GP (4MOST Galactic Pipeline). These modules will determine abundances of more than $15$ elements with an expected precision better than 0.05\,dex and a systematic uncertainty of $\sim 0.1$\,dex, as demonstrated in \cite{Gent2022, Storm2025b}. Figure~\ref{fig:precision} shows the achievable precision of chemical abundances in terms of bias and scatter for two values of $S/N$ (100 and 250 per \AA) adopted for \sd\ sub-surveys.

Among the pipelines that will be used for the analysis of the \sd\ stars, we have identified three key codes. These include the SAPP pipeline \citep{Gent2022,Storm2025b}, the NLTE-cINN  \citep{Ksoll2026}, and the Cepheids$/$SAPP code \citep{Albarracin2026}. The codes are based on NLTE synthetic grids of stellar spectra computed with the solar chemical composition from \citet{Magg2022} and NLTE line formation is included for 16 elements: H, O, Na, Mg, Al, Si, Ca, Ti, Mn, Fe, Co, Ni, Sr, Y, Ba, and Eu. The SAPP code and its NLTE Payne module, in particular, has been extensively validated on archival spectra of \textit{Gaia}-ESO and PLATO benchmark stars, including main-sequence, subgiants, and red giants with asteroseismic $\log g$ estimates, and can be readily used for chemical abundance analyses of large stellar samples. As discussed in \citet{Storm2025b}, the SAPP is the first single-network NLTE spectroscopic analysis code that self-consistently provides NLTE stellar parameters and NLTE abundances of over 15 elements, which is essential for accurate and precise studies of chemical composition of FGKM-type stars \citep{Asplund2005,Lind2024,Bergemann2025}. For cross-verification, our interface of the SAPP / NLTE Payne is available for direct analyses of 4MOST data \footnote{\url{https://nlte.mpia.de}}. 

The analysis of stellar masses, ages, and radii will be carried out using the SAPP code, using the GARSTEC evolutionary tracks (see \citealt{Gent2022} for a validation on benchmark stars). Also asteroseismic parameters - global oscillation parameters - will be used, where available, following the approach described and validated in that study. Also, since the 4MOST HR blue and green windows cover the H$_{\alpha}$ line, the CH G-band, and the violet CN band (B2SIG$+$ - X2SIG$+$) around 4125\,\AA~\citep{Villanova2025}, spectroscopic age tracers \citep{Bergemann2016}, and C/N abundances \citep{Ness2016} will be explored to constrain masses of these stars. 
\section{Sub-surveys and observing plans}\label{sec:4}
%

\begin{figure}
\sidecaption
\includegraphics[width=7.1cm]{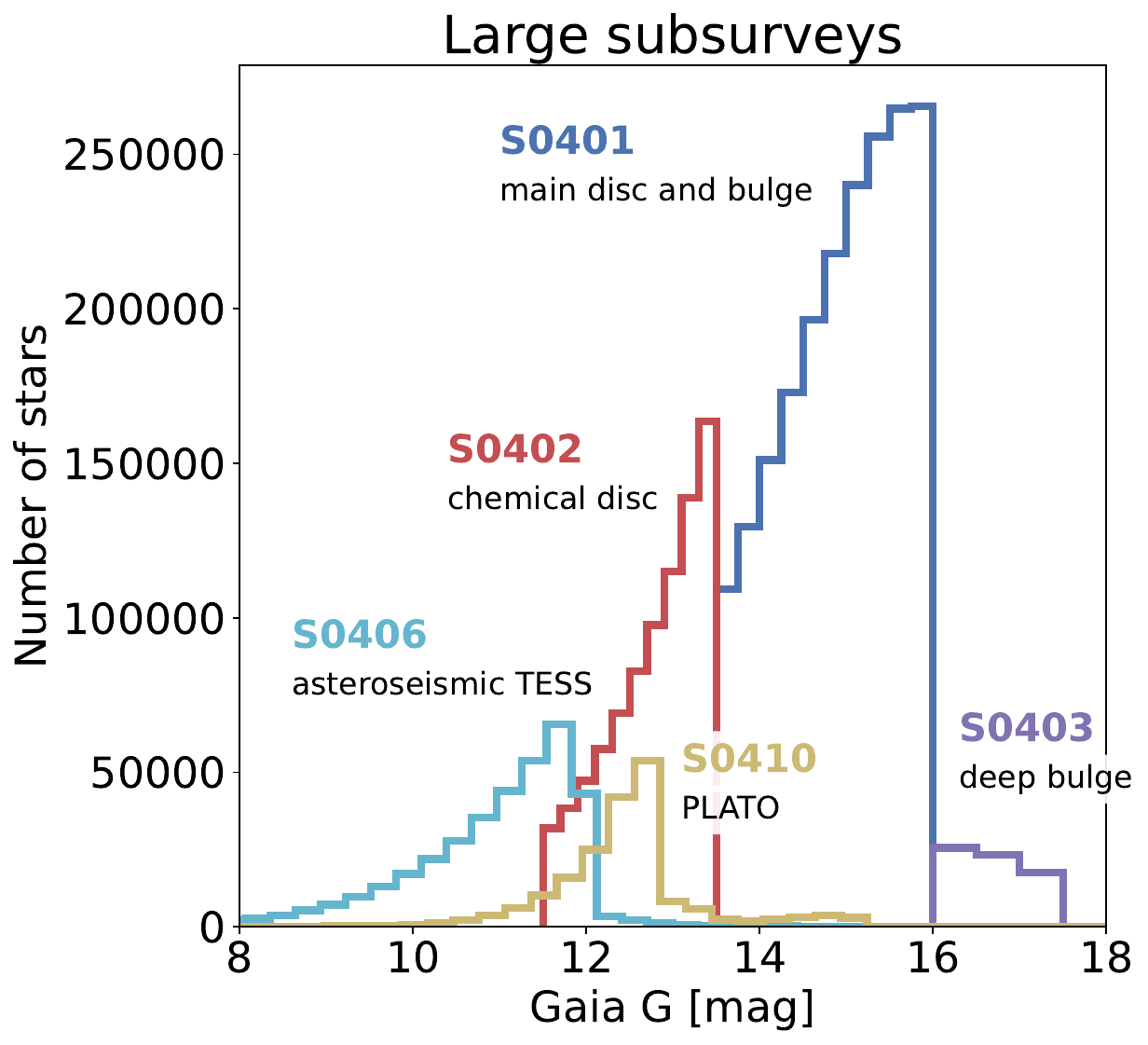}
\caption{Distribution of targets in large sub-catalogues of the \sd\ survey. See Sect. \ref{sec:4}.}
\label{fig:targetsmag1}
\end{figure}

\begin{figure}
\sidecaption
\includegraphics[width=6.7cm]{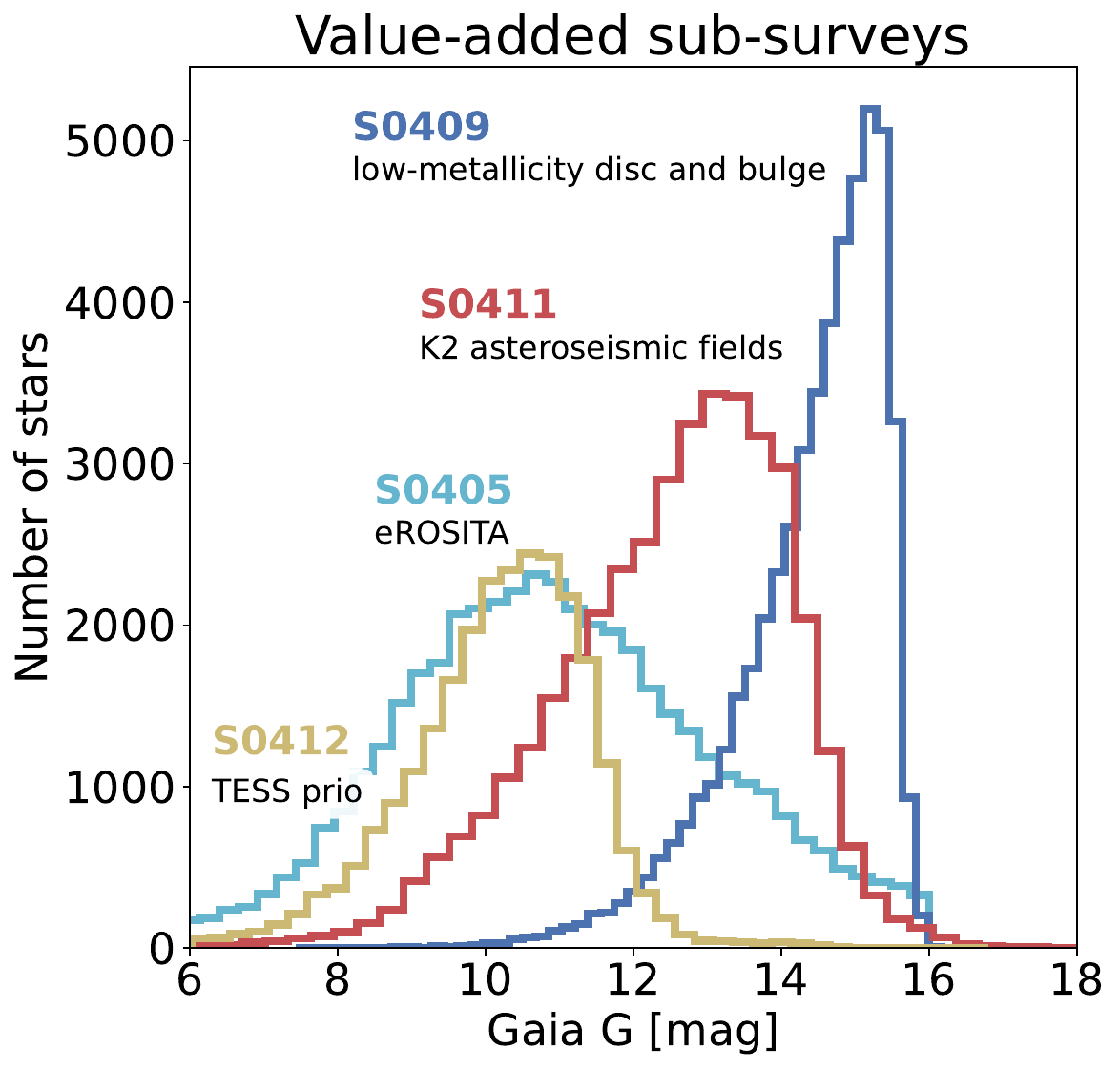}
\caption{Same as Fig. \ref{fig:targetsmag1} but for value-added sub-surveys.}
\label{fig:targetsmag2}
\end{figure}

\begin{figure}
\centering
\includegraphics[width=0.85\columnwidth]{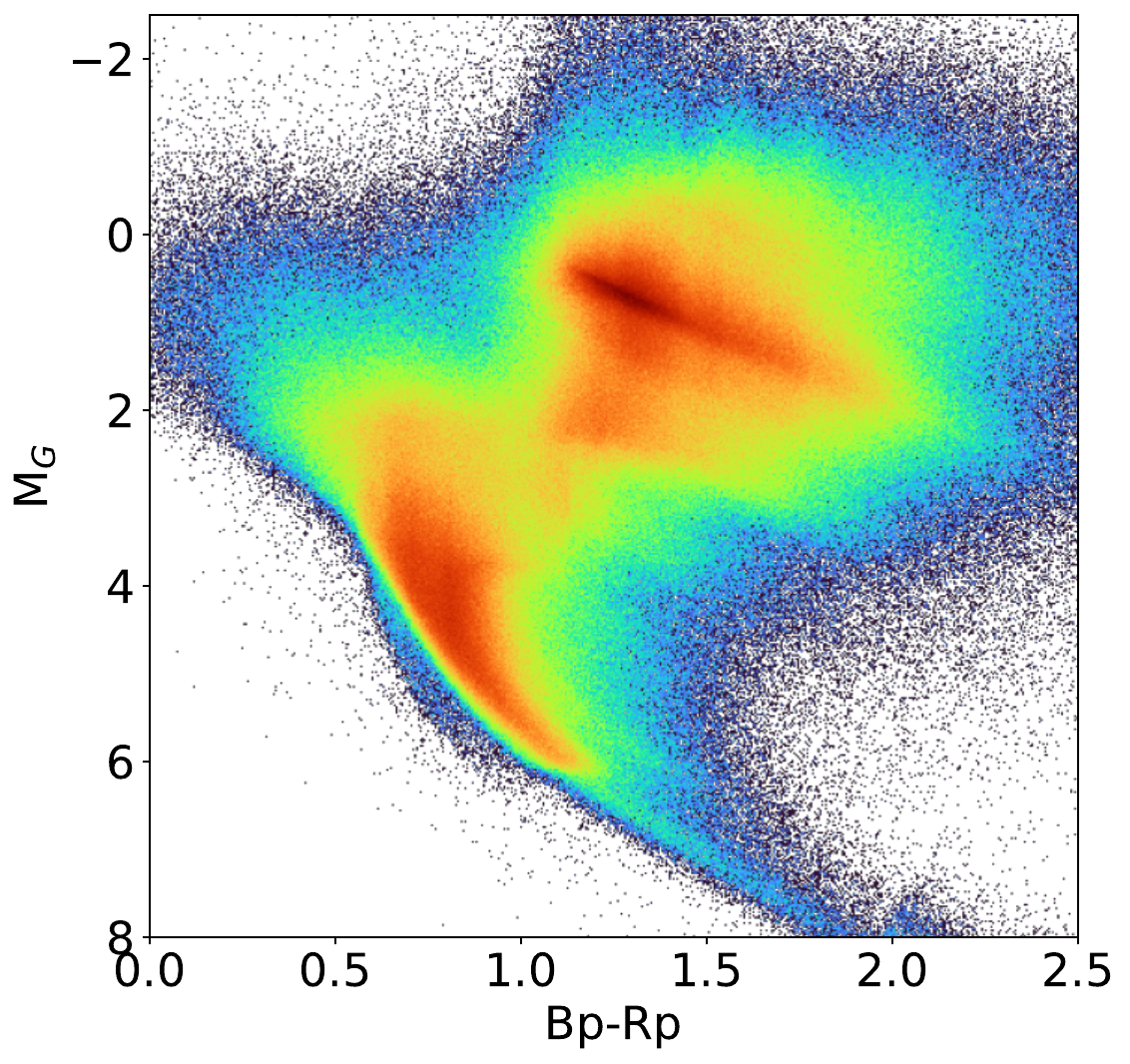}
\includegraphics[width=0.8\columnwidth]{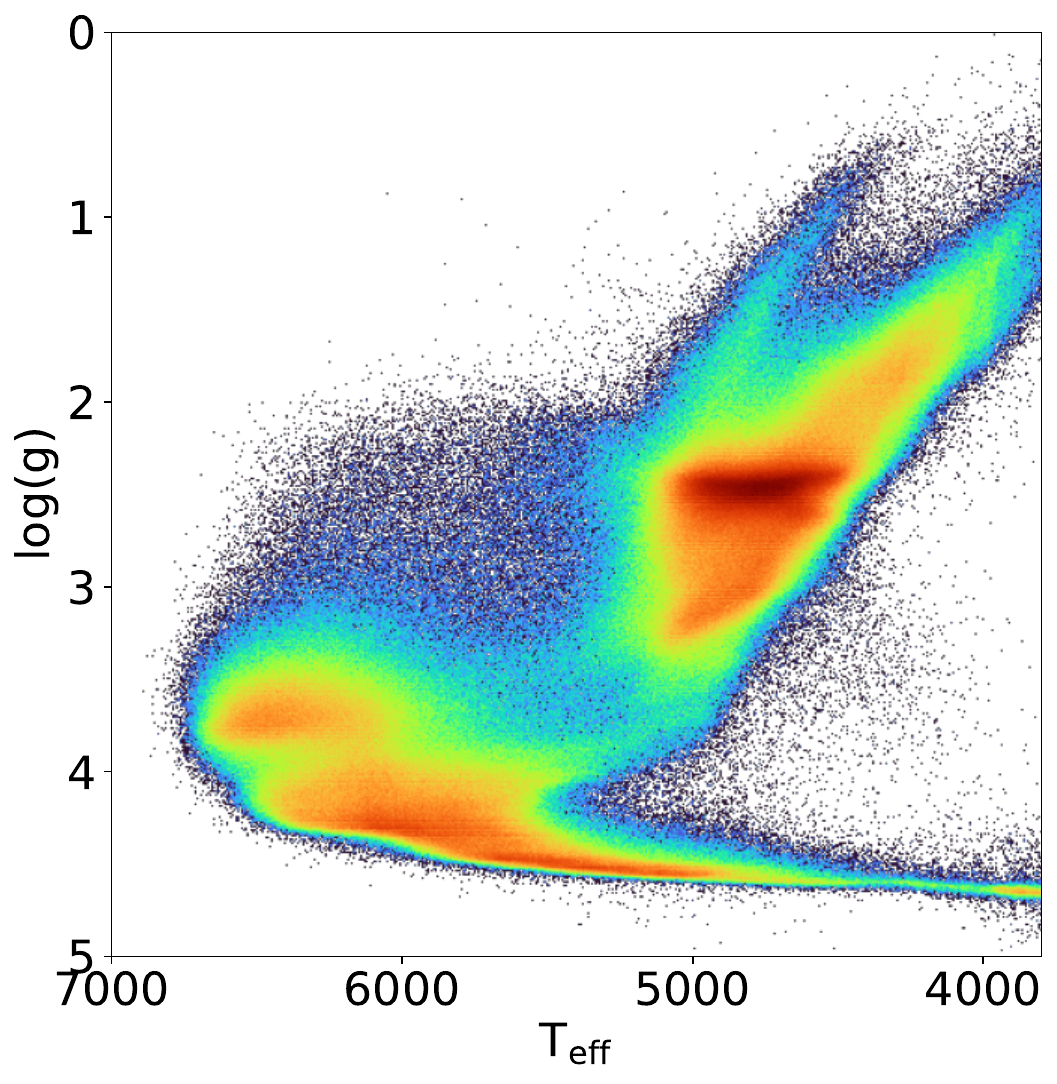}
\caption{Our \sd\ stellar sample using \textit{Gaia} photometry (\textit{top}) and the XGBoost parameters (\textit{bottom}).}
\label{fig:stelparam}
\end{figure}
%
%
%
\begin{table*}[ht] 
\centering
\setlength{\tabcolsep}{2mm}
\caption{Basic properties of the \sd\ sub-surveys.}
\label{tab:main}
\begin{tabular}{l l  r  r  r  L{3.1cm}  p{2.5cm}  p{2cm}}
\hline
\hline
\noalign{\smallskip}
SS & Name &  mag limits$^*$ & Area &  N$_{\rm stars}$ &  Cadence  & Selection & SNR criteria \\
\noalign{\smallskip}
  &  & \textit{Gaia} $G$  & sq.deg. & & &  \\ 
\noalign{\smallskip}
\hline
\noalign{\smallskip}
S0401  & Main disk and bulge & 13.5$-$16.0 & 18\,061 & 2\,002\,924 & B(1,0,0): 4\,024, C(1,5,1): 467, D(0,0,340): 1301515 & M$_G < 2$, (x,y,z) sel. & 100 green\\ 
\noalign{\smallskip}
S0402  & Z-complete disk     & 11.5$-$13.5 & 19\,273 & 842\,159    & B(1,0,0): 31, C(1,5,1): 299, D(0,0,340): 569\,761 & $2 <$ M$_G < 6$, \newline$\rm E(B-V)<0.6$ & 250 blue \\ 
\noalign{\smallskip}
S0403  & Deep bulge          & 16.0$-$17.5 & 228     & 66\,468     & B(1,0,0): 45    & --  & 100 green \\ 
\noalign{\smallskip}
S0404  & Cepheids            &  10.8$-$16.4 & 2\,854  & 2\,687      & B(1,0,0): 1\,742, C(1,9,60): 112    & -- & 100 green, 250 blue (brightest)\\ 
\noalign{\smallskip}
S0405  & eROSITA             &  6.3$-$15.2 & 13\,616 & 45\,696     &  B(1,0,0): 2, C(1,5,1): 306  &--  & 100 green \\ 
\noalign{\smallskip}
S0406  & TESS main           &  8.4$-$12.1 & 14\,533 & 364\,794    &  B(1,0,0): 3   &--  & 100 green \\ 
\noalign{\smallskip}
S0407  & Open clusters       &  10.5$-$16.2 & 77      & 1\,039      &     & -- & 100 green \\ 
\noalign{\smallskip}
S0408  & Exoplanet hosts          &  8.5$-$14.7 & 5\,908  & 3\,216      &     & -- & 100 green \\ 
\noalign{\smallskip}
S0409  & Metal-poor disk and bulge  &  11.8$-$15.6 & 16\,312 & 51\,823     &   B(1,0,0): 20, C(1,5,1): 1   & -- & 100 green \\ 
\noalign{\smallskip}
S0410  & PLATO (southern field)    &  10.7$-$14.8 & 2\,192  & 192\,609    & B(1,0,0): 21, C(1,1,650): 153266, C(1,5,1): 745   & -- & 100 green, 100 red (M dwarfs)\\ 
\noalign{\smallskip}
S0411  & K2                 &  9.1$-$15.0 & 2\,799  & 39\,614     & B(1,0,0): 43, C(1,1,650): 33800    & -- & 100 green \\ 
\noalign{\smallskip}
S0412  & TESS priority      &  7.4$-$12.2 & 11\,733 & 25\,751     & C(1,1,650): 21302    & -- & 100 green \\ 
\noalign{\smallskip}
\hline
\noalign{\smallskip}
Total &                     &             &         & 3\,638\,780 &     &     \\
\hline
\end{tabular}
\tablefoot{
(1)~Numbers correspond to catalogue export dated 17 April 2025. The selection functions of catalogues S0403 to S0412 follow the quality of auxiliary value-added data, such as global oscillation parameters or X-ray fluxes, and they are not based on photometric or geometric information (see text).
(2)~Column 3: Magnitude limit range provided for sub-surveys S0404 and above includes 95\% of the sample.
(3) Column 6: Cadence requests (number and spacing of multi-epoch spectroscopic observations) for types of stars in different \sd\ sub-surveys (see Sect.~\ref{sec:cadence} for details). These requests are expressed by the letter denoting the cadence use case (B, C, D), the encoded parameters (SINGLE\_VISIT, NREPEAT, DT\_MIN), and the number of targets for which the request is made.
}
\end{table*}
The complete input catalogue of the \sd\ survey comprises $3\,638\,780$ unique stars, distributed over 12 scientific sub-surveys (Table~\ref{tab:main}), as described below. The distribution of the full catalogue in the space of photometry  and in the $T_{\rm eff}$-$\log g$ plane is demonstrated in Fig. \ref{fig:stelparam}.

The largest fraction of targets, about 78\,\%, will be observed in the two main sub-surveys (S0401 and S0402), which aim to study the global structure and chemical evolution of the Galactic disk and bulge. Additionally, ten sub-surveys enhance the scientific value of the \sd\ programme by providing high-resolution spectroscopic follow-up of stars with quality photometric light-curves and/or X-ray fluxes observed by TESS, eROSITA, K2, and PLATO. Targets in these catalogues have a selection function that is entirely different from that of the two main sub-surveys, and it is set by constraints specific to these facilities (that is their own spatial footprints, magnitude ranges, and selection function) and the quality of time-domain data. Some of these catalogues (specifically, the deep bulge and Cepheids) have been developed in collaboration with the 4MIDABLE-LR survey \citep{Chiappini2019}. More information about the selection is provided below.
%
%
\begin{figure*}
\hbox{
\includegraphics[width=0.33\textwidth]{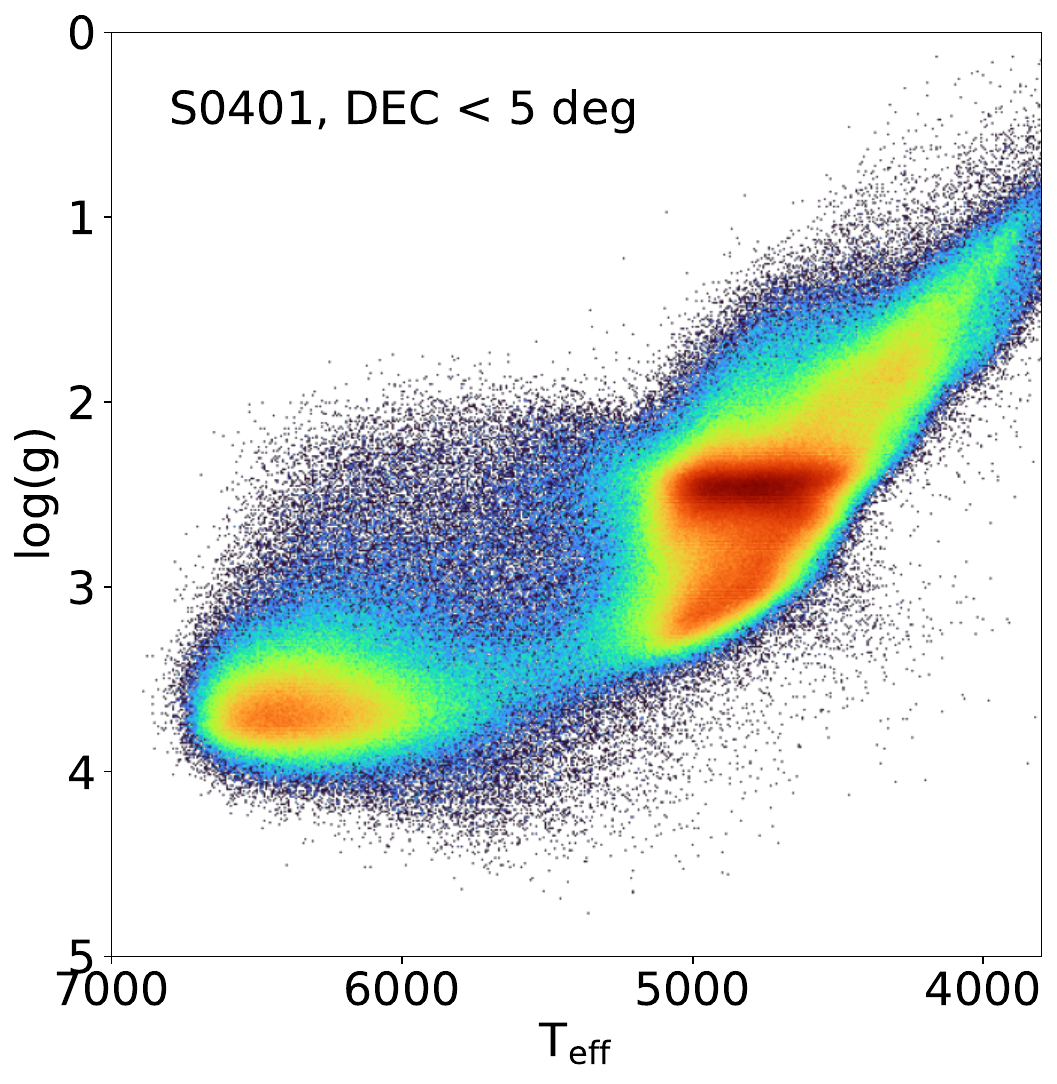}
\includegraphics[width=0.33\textwidth]{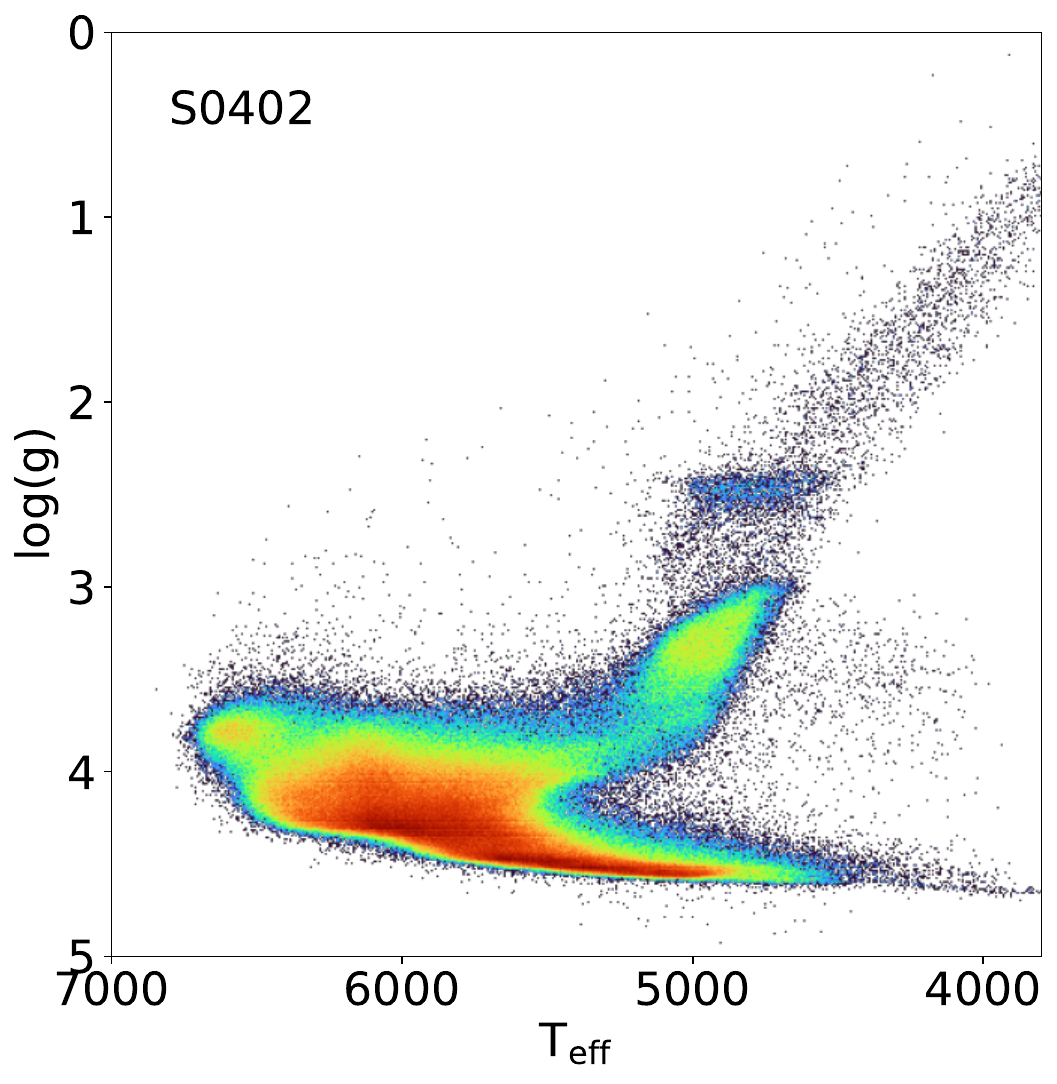}
\includegraphics[width=0.33\textwidth]{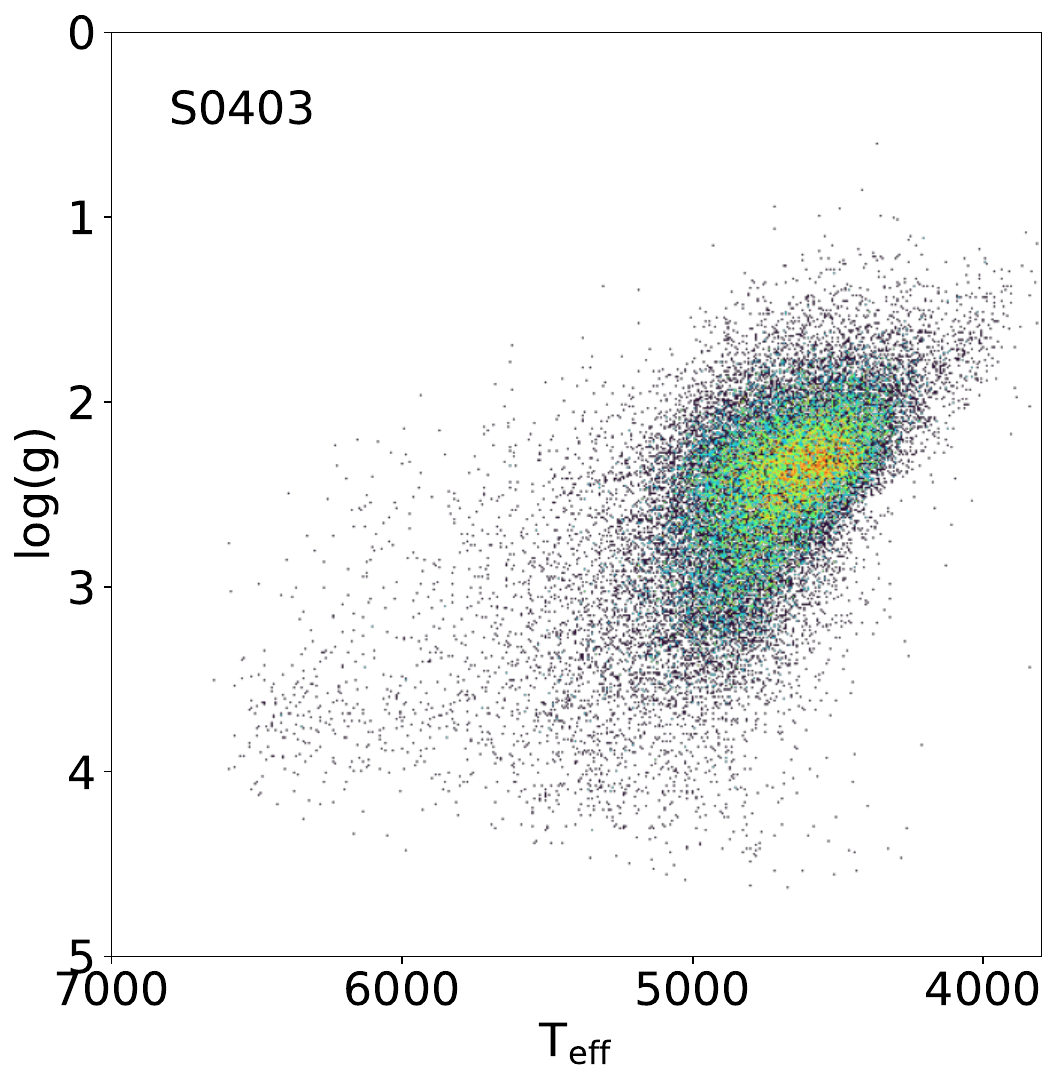}
}
\caption{Diagrams of $T_{\rm eff}-\log g$ for the three large sub-surveys of \sd: the main disk and bulge sub-survey (\textit{left}) , the Z-complete disk (\textit{middle}), and the deep bulge (\textit{right}).}
\label{fig:cmd1}
\end{figure*}
%
%
%
%
%
\subsection{Main disk and bulge (S0401)}
The prime scientific aim of this sub-survey is to map homogeneously and uniformly the entire Milky Way disk and bulge accessible within the 4MOST footprint in order to study comprehensively the structure, formation, and evolution of the disk and the bulge. The sample  selection of the sub-survey follows simple criteria, which are based on astrometric and spectro-photometric information derived from \textit{Gaia} DR3 catalogue \citep{GaiaDR3}. 

The S0401 sub-survey comprises $2\,002\,924$ stars with $13.5 < G < 16$, and spanning the full range of RA and Dec accessible to 4MOST. We also use a selection on absolute magnitudes ($M_{\rm G}< 2$), which means that the majority of our targets are red giant and red clump stars. This criterion was imposed to ensure that deeper regions towards the Galactic inner disk and bulge are probed well. The distribution of this sample in the $T_{\rm eff}$-$\log g$ plane is shown in Fig. \ref{fig:cmd1} (left panel); here the stellar parameters \citep{Andrae-2023} are used. The small contamination around $T_{\rm eff}$ $\gtrsim 6000$ K is due to the foreground main-sequence stars. Also we note that this figure does not show the high-latitude K2 fields (at Dec $> 5$ deg.), because these were included for a different reason (to ensure the observability of these fields in the nominal program) and their selection function (Sec. 5.6, sub-survey S0411) does not follow the S0401 criteria described in this section.

\textit{Gaia} RVS data were not taken into account when selecting the targets, and hence the sample is only limited by the quality of \textit{Gaia} DR3 astrometric and photometric information. Stars for the disk are taken with the following limitations: 
\begin{itemize}
    \item $13.5 < G < 16$ 
    \item $M_{\rm G} < 2$
    \item Dec $\leq$ 5.0
    \item $\rm -5 < X / kpc < 15$
    \item $\rm -15 < Y / kpc < 10$
    \item $\rm -10 < Z / kpc < 10$
    \item $E(B-V)< 0.8$
\end{itemize}
Stars are chosen randomly until a density of $400$ targets per square degree is reached. If the density is too low, that is, $400$ targets are not available in a square degree (which is typical for altitudes above $|b|\gtrsim 20^{\circ}$ from the Galactic mid-plane), then all targets are included. Stars with large extinction, which here is taken as is $E(B-V) > 0.8$, were filtered out. This primarily removes stars close to the Galactic plane and towards the Galactic Bulge. 

\subsection{The Z-complete disk (S0402)}
The Z-complete disk sub-survey aims to provide the first systematic large-scale investigation of the $s$-, $p$-, and $r$-process enrichment cycle in the Galaxy. We will do this by targeting $842\,159$ stars at a $S/N >250$ {\AA}$^{-1}$ in the blue window around $4315$\,\AA\,(see Sect.~\ref{sec:3})\footnote{According to our tests, the following range are rather clean and are usable for the continuum SNR estimate at the resolving power of the 4MOST HR spectrograph: 4207.5 - 4208.3, 4315.3 - 4316.5, 4335.3 - 4336.3.} to enable quantitative diagnostics of trans-Fe-group heavy elements up to gadolinium. 

The selection criteria for the Z-complete disk sub-survey are as follows:
\begin{itemize}
    \item $11.5 < G< 13.5$ 
    \item $2 < M_{\rm G} < 6$
    \item $E(B-V) < 0.6$
\end{itemize}

The distribution of this sample in the $T_{\rm eff}$-$\log g$ plane is shown in Fig. \ref{fig:cmd1} (middle panel); here the stellar parameters from the XGBoost catalogue \citep{Andrae-2023} are used. The apparent magnitude ranges ensure that the required $S/N$, as listed above, for robust detections of spectral lines of heavy elements is satisfied \citep[e.g.][]{Hansen2014,Storm2025b}. Extensive diagnostic tests were carried out in 2020, during the preparation of the ESO proposal, using the SIU Payne developed by the MPIA group \citep{Kovalev2019}\footnote{\url{nlte.mpia.de}}. The latest results for the accuracy and precision of abundance recovery based on 4MOSTified data (synthetic and archival) are presented in \citet{Storm2025b}, \citet{Ksoll2026} and in Fig. \ref{fig:precision}. We will determine abundances of Li, C, O, Mg, Al, Si, Ca, Ti, Cr, Mn, Ni, light $s$-process elements (Sr, Y), heavy $s$-process elements (Ba), and $r$-process elements (Gd), with a typical uncertainty of $0.1 - 0.16$\,dex.

The observed spatial distributions of $s$- and $r$-process elements relative to Fe-peak and $\alpha$-elements will  represent our key sample to constrain Galactic nucleosynthesis \citep[e.g.][]{Prantzos2018,Limongi2018,Kobayashi2020,Storm2025}. These distributions will also allow us to probe detailed physics and gas dynamics \citep[e.g.][]{Schoenrich2019,Tsujimoto2019} over ten billion years of Galactic evolution: radial migration, gas flows, the retention efficiency of elements in the interstellar matter, and the gas cooling times. By the end of the 2030s, this Z-complete disk legacy sample will represent the largest high-resolution and high $S/N$ sample in the optical of spectra for more than half a million Galactic stars.
\subsection{The deep bulge (S0403)}\label{sec:deepbulge}

Our Galactic deep bulge survey will probe the detailed chemical structure of stellar populations in the bulge. There are several intertwined structures and stellar components in the bulge region and our aim is to constrain the fractions associated with these structures, namely the  pseudo-bulge, the stellar halo and the thin and thick disks, and possibly also a classical bulge \citep{PerezVillegas2017,Bensby2017,Barbuy2018,Queiroz2020, Nepal2026}. We also aim to explore the long-standing question of whether the bulge is the cradle of extremely and ultra metal poor stars \citep{Springel2020,Howes2016}. 

The S0403 sub-survey contains $66\,468$ evolved stars selected to probe the inner bulge region, at $-11<b<-3$ and $-10<l<10$ (see panel~3 in Fig.~\ref{fig:targetsRADEC}). The field will receive up to eight hours total exposure time in order to enable the $S/N\approx100$ per \AA~ in the green. All deep bulge targets are red giant branch and red clump stars in the magnitude range $16<G<17.5$. The distribution of this sample in the $T_{\rm eff}$-$\log g$ plane is shown in Fig. \ref{fig:cmd1} (right panel); here the stellar parameters from the XGBoost catalogue \citep{Andrae-2023} are used. 

Additionally, we include about $1\,500$ red giant stars in the innermost bulge region, which overlaps with the planned footprint of the Roman Galactic Bulge Time-Domain Survey (RGBTDS) on the Nancy Grace Roman Space Telescope \citep{Penny2019,Huber2023,Wilson2023}\footnote{\url{https://roman.gsfc.nasa.gov/galactic_bulge_time_domain_survey.html}}. Our selection (priv. comm. Robert F. Wilson, Daniel Huber) covers the following spatial region \footnote{refined using the MAST website \url{ https://www.stsci.edu/roman/science-planning-toolbox/field-of-view-overlay}, \url{ https://gsss.stsci.edu/webservices/footprints/webform.aspx}} and the stellar target density was matched of the rest of our deep bulge field: 
\begin{itemize}
    \item $16.0 < G< 17.5$
    \item $-0.5 \leq l \leq 1.62$ degrees
    \item $-3 \leq b \leq -1.25$ degrees
\end{itemize}
The area around this pointing will be targeted by the Roman telescope to obtain time-series photometry with the expected precision of one milli-mag per 15-min cadence \citep{Huber2023}. The resulting dataset, including the asteroseismic values, will strengthen the science for our \sd\ bulge programme.

Combining photometry, spectra, and parallaxes, we will produce the largest high-resolution statistically significant catalogue of chemistry and phase-space in the Galactic bulge, fully homogeneous and consistent in terms of the analysis strategy with the other \sd\ sub-surveys spanning the inner and outer Galactic disk and stellar halo. 
\subsection{Cepheids (S0404)}
The sub-survey S0404 aims to study the distributions of chemical abundances in CCs across the thin disk and in the Milky Way warp \citep{lemasle2008,genovali2014,lemasle2022, trentin2024, Nunnari2025}, as well as T2Cs and ACs in the Galactic spheroid \citep{lemasle15, braga2018b, bono2020b, bono2024,  braga2020, ripepi2024}. This goal will be achieved by observing a sample of currently known $2\,643$ Cepheids with $G\leq16.5$\footnote{For most Cepheids in our catalogue, this is the average $G$ magnitude obtained by fitting the light-curve. For some variables, however, this is only the average magnitude from all measurements available, without light-curve fitting, which introduces a possible bias of less than $0.4$\,mag. Since the phase at which the observation will take place is unknown, the uncertainty of the true magnitude of the star at the time of observation is also of the order of $<0.4$\,mag.} distributed over the area of 3500\,deg.$^2$.

The current sample includes central helium burning variables: about 1420 CCs, young stars tracing the present-day ISM, about 90 ACs that trace intermediate-to.old age stellar populations. Moreover, the sample includes  about 860 T2Cs, old ($t>10$\,Gyr) stellar tracers in the AGB phase and about 270 dwarf Cepheids, that are intermediate-age stellar tracers in the central or shell hydrogen burning phase. ACs and dwarf Cepheids cover a very broad range in stellar ages (from several hundred Myrs to several Gyrs). However, the main difference is that the former group is only present in stellar systems hosting stellar populations more metal-poor than $\rm [Fe/H]\approx-1.5$, since only in this metallicity regime the helium burning sequence crosses the Cepheid instability strip \citep{fiorentinomonelli2012,monellifiorentino2022,bono2024,ripepi2024},  while the latter group covers a very broad range in iron abundances.

The cadence flag is set to $(1,1,9,60)$ to represent that these stars should be observed nine times with a cadence of at least $60$ days. Each time, observations should be completed within a single OB because the intrinsic variability of these targets does not allow stacking. For all of them, we require spectra at $S/N=100$ per {\AA} in order to determine abundances of $\alpha$-, Fe-peak and n-capture elements following the same approach adopted in previous investigations \citep{lemasle2008, genovali2014, lemasle2022, trentin2024, Nunnari2025, Trentin2025,Catanzaro2026}. The main reasons for this observing strategy is to  provide a good coverage of the pulsation cycle to improve the accuracy of abundance determination and to trace variations of atmospheric parameters (effective temperature, surface gravity, micro/macro turbulence). 

In this context, it is important to distinguish the complementary roles of \textit{Gaia} DR4 and 4MOST-HR. While \textit{Gaia} DR4 will provide accurate center-of-mass radial velocities for a significant fraction of known Galactic Cepheids ($G \le 16$ mag), its spectroscopic chemical analysis is more constrained. \textit{Gaia}’s abundances ([Fe/H] and [$\alpha$/Fe]) are limited to the brighter tail ($G \le 12$–$13$ mag) and are derived from RVS medium-resolution spectra ($R \approx 11,500$) within a narrow wavelength range ($\approx 845$–$872$ nm).
Because spectra of variable stars are collected at random phases, the accuracy of individual abundances depends heavily on the precision of the atmospheric parameters ($T_{\rm eff}$, $\log g$, $v_t$) estimated at those specific epochs. A primary objective of the 4MIDABLE-HR survey is to leverage high-resolution spectra ($R \approx 20,000$) and broader wavelength coverage to provide more robust atmospheric parameters. This, in turn, enables the measurement of accurate individual abundances for a wider range of species, including iron-peak, $\alpha$, and neutron-capture elements, that are not accessible with \textit{Gaia} RVS. This detailed chemical mapping is crucial for breaking intrinsic degeneracies in kinematic planes and tracing the origin of these stars across both space and time.
\subsection{eROSITA (S0405)}
The purpose of the eROSITA sub-survey is to study magnetic activity as a function of stellar type, age, and metallicity. This will be done via a dedicated spectroscopic follow-up of stellar targets detected in X-rays by eROSITA \citep{predehl2021, freund2024}. This sample will allow us to improve our understanding of the origin of stellar high-energy emission in the context of chromospheres and coronae \citep{linsky2020}. 
%
\begin{figure}
\hbox{
\includegraphics[width=\columnwidth]{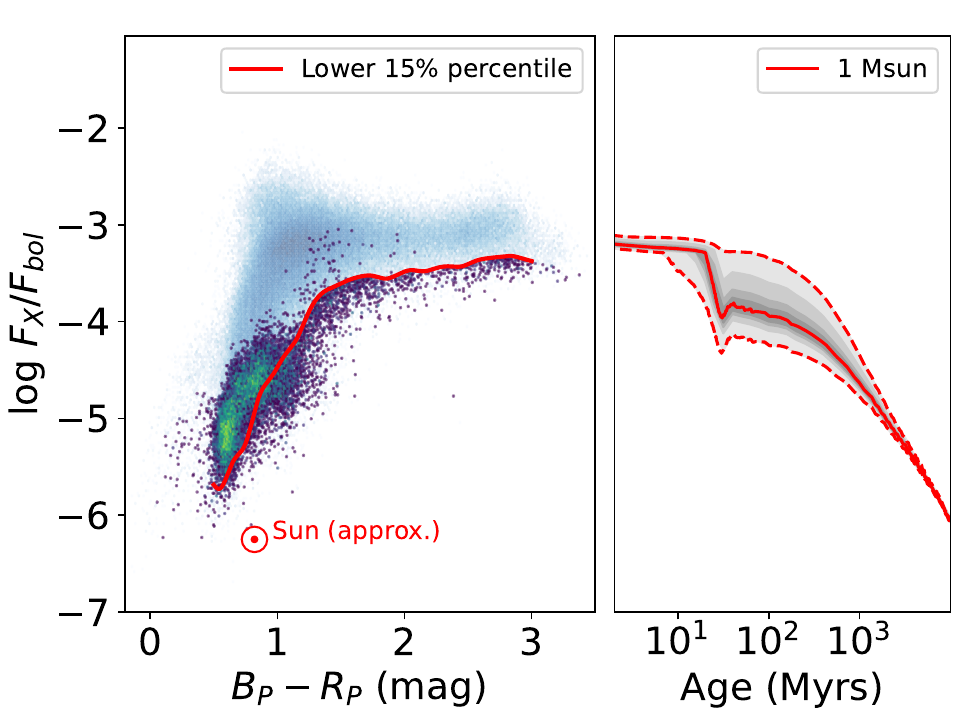}}
\caption{{\it Left}: Our S0405 sample of about 45\,000 F, G, and K-type main-sequence stars with weak X-ray emission from eROSITA. The shaded background represents the full stellar eROSITA sample, and the systems selected for 4MOST are highlighted (colour represents the density of targets). {\it Right}: Decay of magnetic activity, specifically the fractional X-ray flux, with stellar age. The curves shows synthetic models for the X-ray evolution of a 1\,$M_\odot$ star \citep[][]{Johnstone2021}; the dashed lines represent the expected 10\,\% and 90\,\% percentiles ranges. The sample will be used to systematically explore the link between fundamental parameters and the evolution of stellar activity.}
\label{fig:activity}
\end{figure}

Our eROSITA subsample contains $45\,696$ low-mass stars distributed over the area of sky $14\,000$ deg$^2$, which are comparably bright in the optical with a median $G$-magnitude of 10.6. The sample stars have been selected preferentially to have a weak but detectable X-ray emission in order to track the transition between saturation and decay as well as the decrease in magnetic activity. To this end, a colour-dependent selection probability was applied that scales inversely with fractional X-ray luminosity, that is, stars have a higher selection probability the less active they are.

Specifically, we use X-ray fluxes from the eROSITA All-Sky-Survey catalogue (eRASS) and used \texttt{eRASS:4.5}, that is, data accessible to the German community from all available all-sky scans (processing version \texttt{c030}). Targets were selected from the eROSITA source catalogue that have been identified as being stellar using the \texttt{HamStars} algorithm \citep{schneider2022,freund2024}. HamStar is a Bayesian framework that provides probabilities for each eROSITA source to be  stellar versus random chance alignments with, for example, active galactic nuclei. To correct for the detection bias of flux-limited X-ray surveys towards (highly) active stars, objects with  weak magnetic activity  were preferentially selected. Specifically, all stars below the colour-dependent 15\,\% percentile in $L_X/L_{\rm bol}$ are selected (the most inactive stars of their spectral type in eROSITA). To be representative, a small number of active stars was selected, too, and about $30$\,\% of the sample is composed of stars in the transition region towards the saturation limit. For this part, stars were selected with a probability that declines with increasing activity to counterbalance the eROSITA detection bias. This selection is shown in Fig.~\ref{fig:activity} and covers a broad range of activity levels; it naturally favors optically bright sources ($G< 16$), which  aligns well with the 4MOST limits for high-resolution spectra.

Our goal is to correlate the X-ray properties of these stars with their rotation and chromospheric properties, using the Ca H \& K and H$_{\alpha}$ activity indices, accessible in the 4MOST high-resolution blue and red arms. The sample will yield important constraints on the relative heating efficiencies between the chromosphere and the corona, and their relation with evolution of stellar rotation \citep{linsky2020, Kastner2022}. 
\subsection{Asteroseismic surveys (S0406, S0411, and S0412)}
\label{subsec:astroseismic_survey}
Radii, masses and ages of stars are fundamental quantities required to enable access to a range of science questions, including mixing and transport of angular momentum in stellar evolution \citep{Beck2012,Aerts2021}, mass loss \citep{Miglio2012,Handberg2017,Li2022}, effects of atomic diffusion and mixing \citep{Deal2020,Dumont2021}, and their influence on abundance ratios, and post-interaction chemical evolution in close binary systems, enabling studies of chemical abundance differences such as Li, in low-mass binaries with massive invisible companions, such as MS-NS or MS-BH systems \citep{ElBadry2023}, or the records of stellar mergers through the characterisation of odd populations \citep{Silva-Aguirre2018,Mass2025}.
The advent of asteroseismology \citep{Chaplin2011,Chaplin2013,Hekker2017}
has opened new avenues in this direction, enabling precise characterisation of stellar masses and ages \citep{Serenelli2017, Valentini2019, Pinsonneault2025} and thus also tests and calibrations of standard methods based on stellar evolution models \citep{Martig2016,vanSaders2016}.

\begin{figure}
\hbox{
\includegraphics[width=\columnwidth]{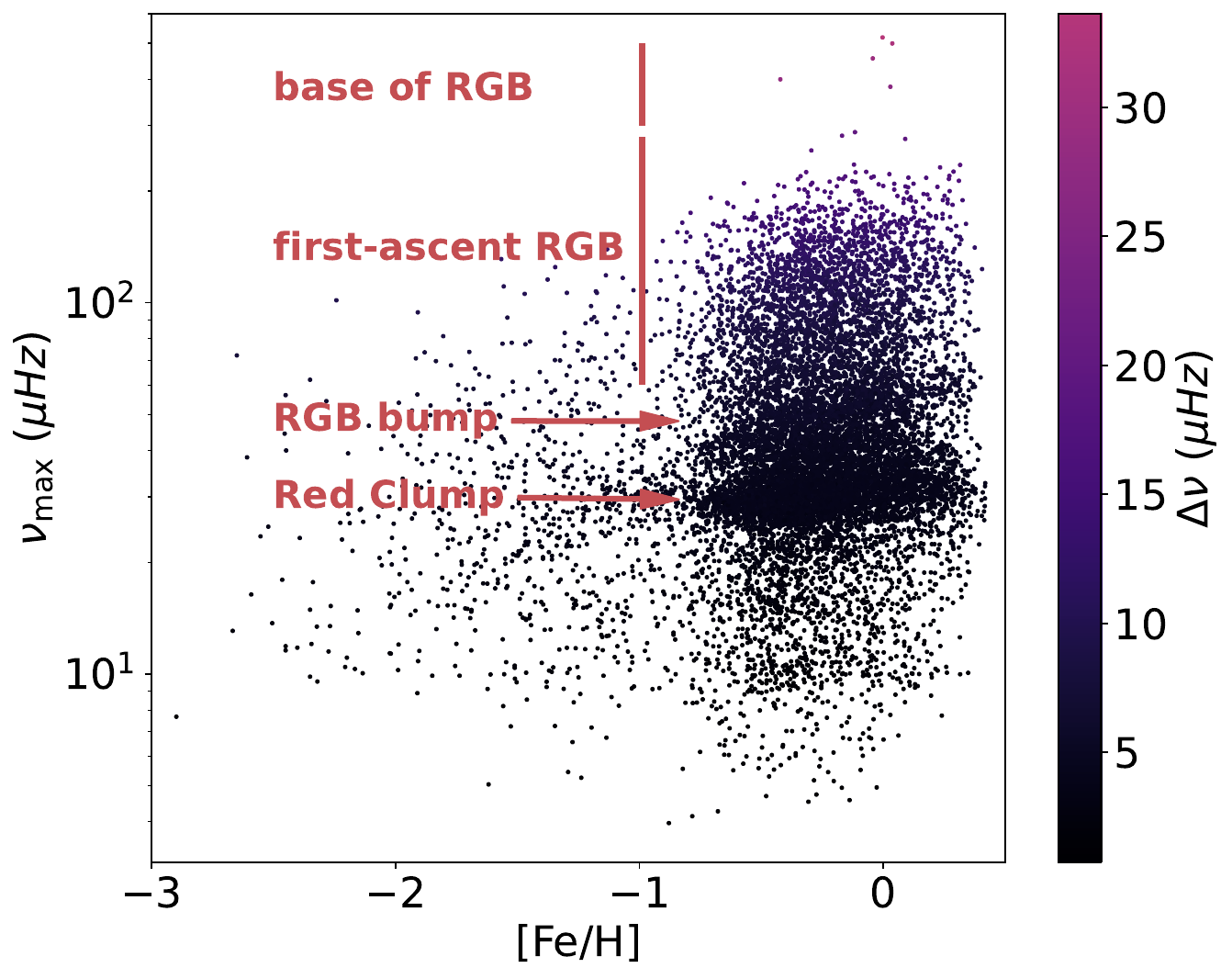}}
\caption{\sd\ asteroseismic sample in the space of stellar metallicity--frequency of maximum power $\nu_{\rm max}$. The colour-coding represents the large frequency separation $\Delta \nu$. The typical stellar evolutionary phases associated $\nu_{\rm max}$ ranges are indicated. The red giant bump and red clump are noticeable as quasi-discrete features. The absence of stars at the base of the RGB is due to $\nu_{\rm max}$  being larger than the Nyquist frequency of long-cadence K2 data, which is around 285$\mu$Hz.}
\label{fig:dnunumax}
\end{figure}

Three of our catalogues S0406, S04011, and S0412 are designed to address these questions by providing a large-scale comprehensive spectroscopic follow-up of subgiants and red giants with precise asteroseismic characterisation from the K2 \citep{Howell2014} and TESS \citep{Ricker2014} missions. Figure~\ref{fig:dnunumax} shows the distributions by cross-matching our original selection with catalogue for $\nu_{\rm max}$ from \citet{Hon2021,Mackereth2021} for TESS and \citet{Zinn2022} for K2. The vast majority of targets in these seismic sub-surveys have the \textit{Gaia} magnitudes within the following ranges: $10<G<12$ (S0406, seismic TESS main), $10<G<14.5$ (S0411, K2 fields), and $9<G<12$ (S0412, TESS priority). 

\begin{figure}
\includegraphics[width=\columnwidth]{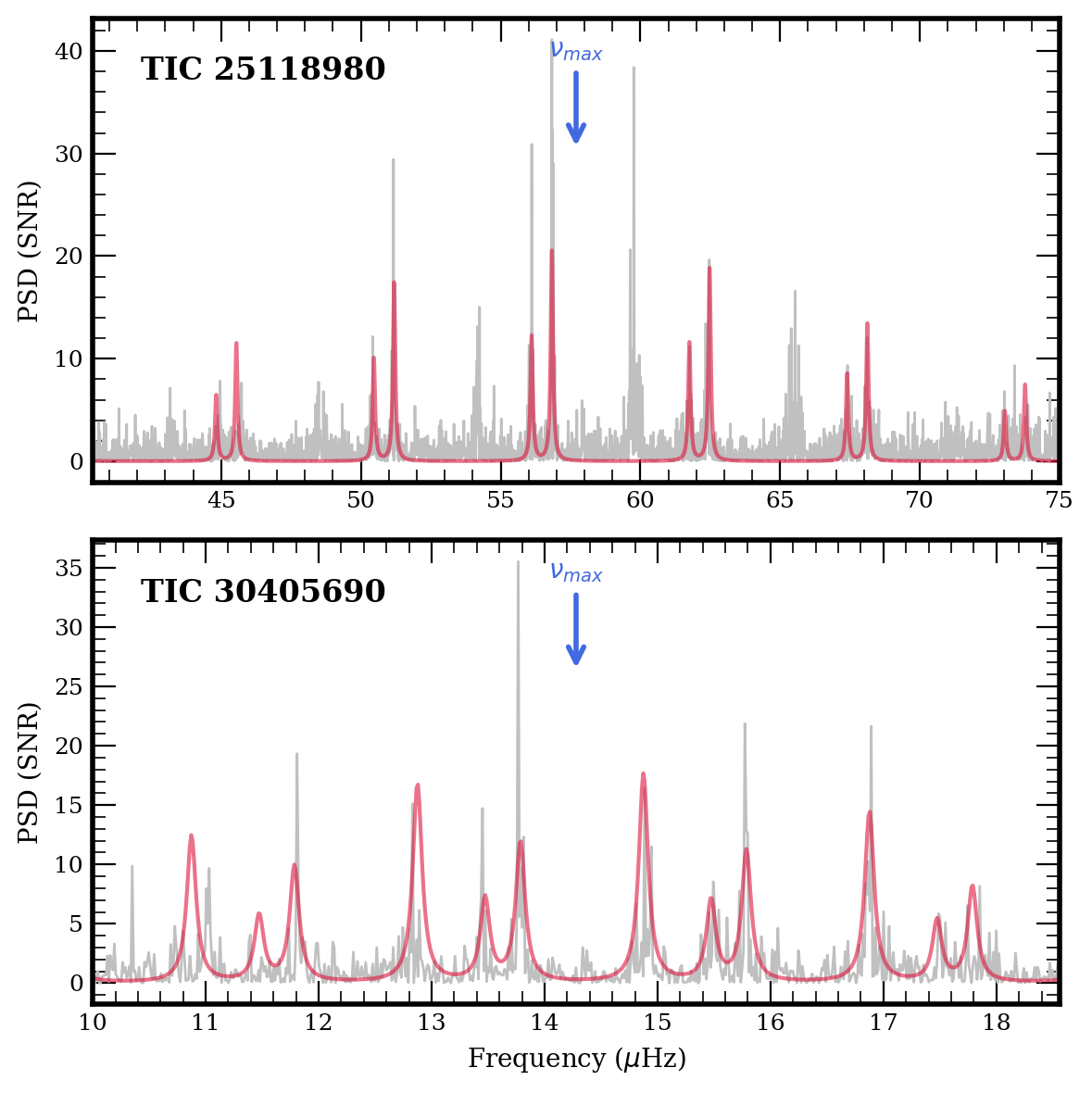}
\caption{Example of TESS power spectral density (PSD) of two targets in our \sd\ TESS surveys analysed with the \texttt{pyA2Z} pipeline. The grey curve shows the background-corrected power spectral density, the red curve represents the fitted asymptotic model, and $\nu_{\rm max}$ is indicated with the blue arrows. The top panel includes radial and quadrupole modes only, whereas the bottom panel additionally incorporates dipole modes.\label{fig:power_spectrum}}
\end{figure}

The division into three sub-catalogues follows from the inherent differences in the initial selection of targets. S0406 contains $364\,794$ stars in the TESS main field and is based on  results from \citet{Hon2021} with subsequently updated determination of asteroseismic parameters (D. Stello \& C. Reyes, private communication). The catalogue has been built by prioritising stars by the quality of their seismic parameters: 1) starting with stars for which both $\nu_{\rm max}$ and $\Delta \nu$ are very well measured; 2) stars for which $\nu_{\rm max}$ is very well measured but the quality is lower for $\Delta \nu$; 3) stars with only $\nu_{\rm max}$ measurements; 4) stars with detected seismic signal but measurement is reported (see discussion in \citealt{Hon2021}; 5) stars without reported seismic signal. For the latter, we limit the sample to stars with $G \leq 12$. Additionally, the sub-survey S0411 ($39\,614$ targets) contain red giant stars with reliable detection of oscillations from the K2 fields, specifically those in the K2GAP programme (\citealp{Zinn2022}\footnote{\url{http://www.physics.usyd.edu.au/k2gap/}}). Finally our S0412 sub-survey contains $25\,751$ TESS priority targets, also based on \citet{Hon2021} but with seismic detections and evolutionary state determination (D. Stello \& C. Reyes, private communication). The stars probe a wide range of masses and metallicities (Fig.~\ref{fig:dnunumax}), potentially enabling strong constraints for stellar physics as described before. Also, it opens up the possibility to test asteroseismic methods as well, in particular the dependency of global oscillation parameters and scaling relationships on metallicity \citep{Guggenberger2016,Viani2017,Huber2024}.

Using stars in S0406 and S0412 with more than ten TESS sectors, low-resolution \textit{Gaia} DR3 Spectra \citep{GaiaDR3}, and TESS magnitudes brighter than 9.5, a high-quality TESS calibration sample comprising about 10\,000 red giant stars has already been constructed. The seismic properties are characterised following the methodology of Liagre et al. (in preparation), using the Quick Look Pipeline \citep{Kunimoto2022} calibrated with the \texttt{pytadacs} pipeline \citep{Garcia2024}. The calibration sample will enable an improved determination of global seismic parameters for all stars in the S0406 and S0412 sub-surveys. In addition, the science enabled by the \sd\ S0406 and S0412 catalogues will extend beyond that related to global seismic parameters, as the high-quality of seismic data allows for the determination of precise frequencies of individual modes for many stars in the catalogues. Figure~\ref{fig:power_spectrum} illustrates this for two stars in the calibration sample, one close to the RGB bump and the second one higher along the RGB.

\subsection{Open clusters (S0407)}

\begin{figure*}
\centering
\resizebox{0.95\hsize}{!}{\includegraphics{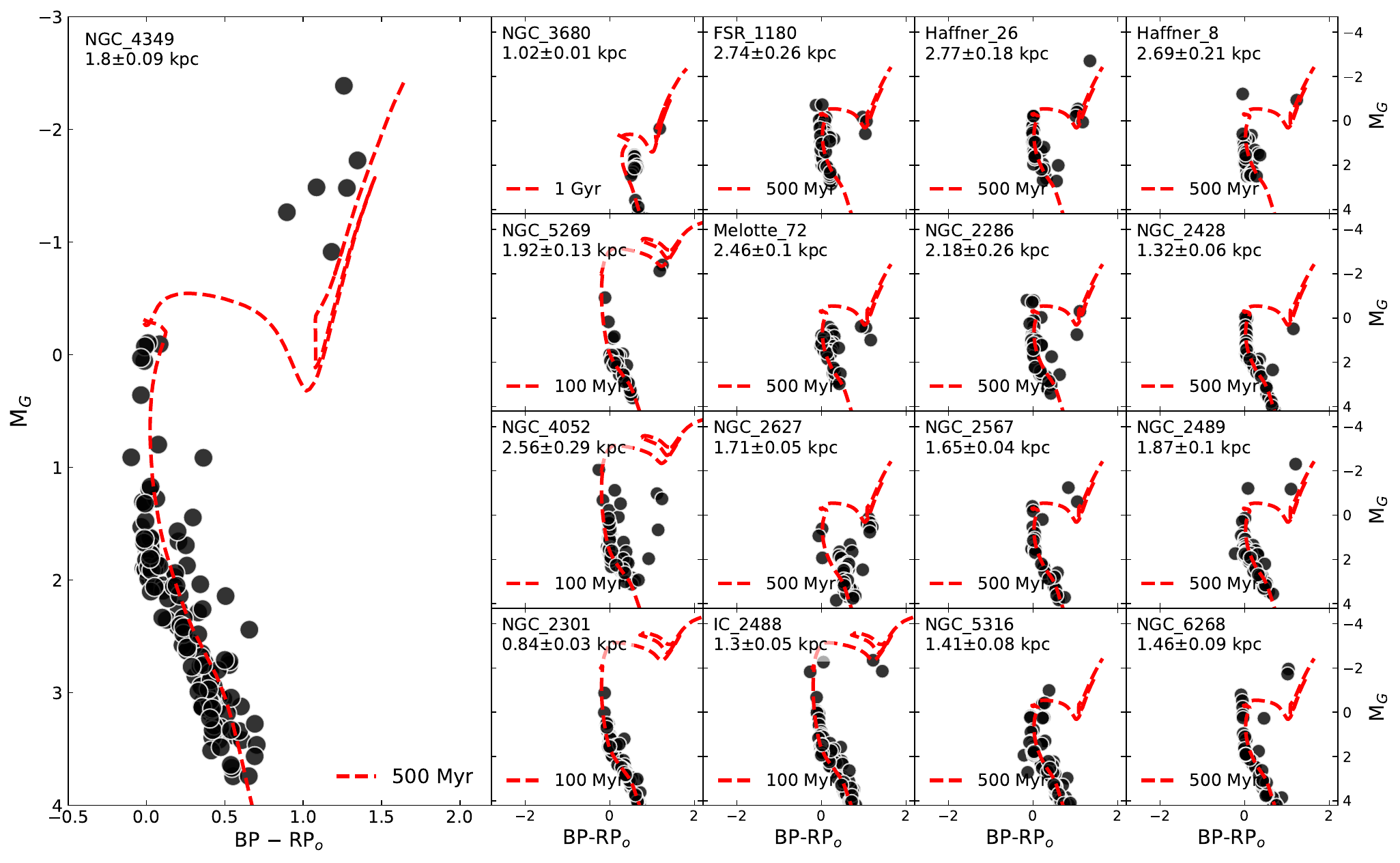}}
\caption{Our S0407 sample of 17 Galactic open clusters. The sample will be used to study atomic diffusion and its effect on surface chemical abundances of stars. The absolute G magnitude was computed using the zero-point corrected parallax \citep{Lindegren2021}, converted to distance, and extinction computed as described in \cite{Ernandes2024}. Here we show the isochrones from the PARSEC grid \citep{Nguyen2025}.}
\label{fig:clusters}
\end{figure*}

To quantify the effects of stellar evolution on observed abundance [X/Fe] ratios in our disk and bulge samples, it is essential to have a self-consistent and statistically representative samples of stars within a range of masses from $\sim 0.6$ to $\gtrsim 2$ M$_{\rm Sun}$, yet formed in the same star-forming environment. Our S0407 sub-survey is designed exactly for this purpose: to establish a clean and homogeneous set of cluster members to enable precise test of the variation of chemical abundance patterns with stellar evolutionary stage.

\begin{table}[t]
\centering
\caption{Age, metallicity, and number of selected stars of each open cluster.
\label{tab:OCs}
}
\begin{tabular}{lccc}
\hline
\hline
Name & Age (Gyr) & [Fe/H] & N$_{\rm stars}$ \\
\hline
NGC 2286 &    &0.00  & 38     \\
NGC 2301 &1.14&0.05  & 121    \\
NGC 2428 &    &0.00  & 42     \\
NGC 2489 &0.20&0.05  & 64     \\
NGC 2567 &0.30&0.02  & 38     \\
NGC 2627 &    &0.00  & 73     \\
NGC 3680 &1.8&-0.10  & 17     \\
NGC 4052 &    &0.00  & 70     \\
NGC 4349 &0.33&-0.06 & 146    \\
NGC 5269 &    &0.00  & 38     \\
NGC 5316 &0.16&0.13  & 83     \\
NGC 6268 &0.17&0.14  & 44     \\
FSR 1180 &    &0.00  & 53     \\
Haffner 26 &  &0.00  & 38     \\
Haffner 8  &   &0.00  & 30     \\
IC 2488    &0.19 &0.08  & 109    \\
Melotte 72 &  &0.00  & 35    \\  
\hline
\hline
\end{tabular}
\end{table}

In \sd\, we will achieve this goal by observing 17 well-studied open clusters  (Fig.~\ref{fig:clusters}) selected using \textit{Gaia} DR3 astrometry from \citet{Cantat-Gaudin2020}. Table \ref{tab:OCs} lists the clusters, along with their  metallicities and ages adopted from \citet{Netopil2016}. All clusters are sampled by at least a few tens of stars representing different evolutionary phases, including the upper and lower main-sequence, subgiant, and red giant stars. As this sample will be for science calibrations, we require $S/N =250$ per \AA\ in the blue to obtain the highest precision of abundance ratios, but also to constrain well their stellar parameters and to determine the projected rotation velocities v$\sin i$ and map activity via the Ca H and K resonance lines at $3933.66$ and $3968.47$\,\AA\, respectively.

The S0407 initial targets were selected from \textit{Gaia} DR3 in a cone with $4 \times r_{50}$, where $r_{50}$ is the half-light radius and is defined around the central RA and Dec coordinates of the clusters, where the latter values were adopted from \citet{Cantat-Gaudin2020}. We restrict the selection to within $2 \times r_{50}$, applying a $2 \sigma$ membership filter in proper motion and parallax of the open cluster, and prioritising the one-third brightest stars in $G$ – A($G$) within the central region ($r < r_{50}$). The resulting sub-survey contains $1\,052$ targets distributed over 3\,kpc in heliocentric distance. This sample will be used to determine masses of stars, aided with cluster ages based on CMD fits \citet{Randich2022} and to test stellar evolution models computed including the effects of atomic diffusion and turbulent mixing \citep{Souto2019,Semenova2020,Nordlander2024}.  

Since all these clusters are also observed by TESS, we expect that for a fraction of the red giant stars, especially in nearby open clusters, global asteroseismic parameters based on precise lightcurves will be available, boosting the quality of $\log g$ estimates and proving superior and less model-dependent (compared to CMD fits) constraints on masses and ages of stars.
\subsection{PLATO and exoplanet hosts (S0408,S0410)}
With our catalogues S0408 and S0410 we aim to provide a homogeneous chemical characterisation of all confirmed planet hosts in the 4MOST footprint and of PLATO stars in its southern long-pointing field. Both catalogues focus on main-sequence stars and subgiants. 

The S0408 targets were selected from the Exoplanet database\footnote{\url{exoplanet.eu}} and include the confirmed planet detections from the ground (e.g. EPIC, HATS, WASP, and the Korea Microlensing Telescope Network) and from the space-based facilities (TESS, K2, COROT, and OGLE). Additionally, we include all TESS Objects of Interest (TOI), which are confirmed or candidate exoplanet hosts, adopted from the online database released by the TESS collaboration\footnote{\url{http://tess.mit.edu/toi-releases/}}. Our S0408 catalogue thus contains $3\,216$ host stars of planets. The S0410 sub-survey represents the LOPS2 field of the PLATO PIC \texttt{v.2.1.0.1} \footnote{This catalogue has not been published yet.} and it will be observed in \sd\ as part of the 4MOST-PLATO Memorandum of Understanding (MoU). The sub-survey contains $192\,609$ stars. For the description of the PIC and PLATO  motivation, we refer the reader to \citet{Nascimbeni2022, Nascimbeni2025, Rauer2025}.

Our sample of $3\,216$ star-planet systems covers the entire range of planet types, from Earth-like to massive Jupiters ($0.1 M_{\mathrm{Earth}} \lesssim M_{\mathrm{planet}} \lesssim 10^{3} M_{\mathrm{Earth}}$) spanning a broad range of orbital periods from hours to years. The request that $SNR =250$ per \AA\ is achieved for all targets in these two catalogues, and a cadence request for one additional visit after two years from the first is adopted for S0410 in order to reveal their possible binary nature.

This reference planet-star sample forms an important legacy of the project. It will allow us to establish robust relationships between planetary parameters, properties of their host stars, and properties of analogous field stars, which do not host planets (especially, in combination with our Z-complete disk S0402 catalogue). We note that chemical characterisation for $\sim 2\,000$ exoplanet host stars is already available from \textit{Gaia} RVS \citep{deLaverny2025}; however, it is primarily limited to $\alpha$-elements, whereas our survey will provide  abundances of ice-forming (C, O) and rock-forming (Mg, Si, Fe, Ni) species. First, the data will enable a rigorous test of relations for planet occurrence rates and their dependence on metallicity, critically going beyond the current assumption of iron-based scaling relationships \citep[e.g.][]{Johnson2010, Boley2024}. Second, the data will establish a solid reference for planet formation models \citep[e.g.][]{Thiabaud2014, Dorn2017, Bitsch2020, Hatalova2025, ODonovan2026}, enabling direct tests of these models and their predictions in terms of population statistics. Finally, our sample will provide the hitherto best constraints on the true metallicity limit for planet formation \citep[e.g.][]{Li2021, Andama2024, Xenos2025} and its dependence of planetary (terrestrial vs. gas giants) and stellar (dwarfs vs. giants over a range of masses and ages) types, enabling the first systematic test of predictions of core accretion and gravitational instability scenarios on large stellar samples, where especially the data for low metallicity targets is missing.

\subsection{Very metal poor stars (S0409)}
The prime motivation for observing this sample of very metal poor and ultra metal poor disk and bulge targets is to probe the earliest epochs of the formation of the disk and bulge, and in particular the potential presence or remnants of the Galactic primordial disk \citep{Sotillo-Ramos2023, Semenov2025,GonzalezRivera2024}, evidence of which has been growing with the availability of data from the \textit{Gaia}-ESO survey \citep{Gent2024} and novel analyses of \textit{Gaia} RVS \citep{Nepal2024}. 

Our S0409 sub-survey targets $51\,823$ low-metallicity ($\rm [M/H]< -1.5$) stars in the disk and bulge in order to search for extremely metal-poor stars, characterise the metal-poor tail of the MDF, and identify and characterise low-metallicity substructures. We did not pre-select in this catalogue by magnitude. As a basis for the selection, we use the \textit{Gaia} XP catalogue \citep{Andrae-2023}, which reports  metallicities derived using the XGBOOST analysis algorithm.\footnote{We use their table table\_2\_catwise.fits.gz, which can be downloaded from \url{https://zenodo.org/records/7945154}} A detailed description of the method can be found in \cite{Rix2022}, which provides a similar catalogue for targets towards the Galactic centre. In addition, we use the \textit{Gaia} XP catalogue of $200\,000$ metal-poor turn-off and red giant stars from \cite{Yao2024}\footnote{Data can be downloaded from \url{https://zenodo.org/records/8360958}}. Also in this work, the XGBOOST algorithm was used to derive [M/H], their algorithm was tuned to improve the  search and characterisation of metal-poor stars. Thirdly, we use the Pristine Survey catalogue from \cite{Martin2024}\footnote{Data can be downloaded from \url{https://seafile.unistra.fr/d/ee0c0f05719d4368bcbb/}}, who derived [M/H] for around $50$ million stars using the narrow-band photometry around the Ca H \& K spectral region \citep{Starkenburg2017}. In \cite{Martin2024}, the observed magnitudes were derived from the \textit{Gaia} XP spectro-photometry, and synthetic magnitudes are based on  stellar model atmospheres. The Pristine Survey calibration is used, which performs well in the metal-poor regime. 

\section{Expected science outcome}\label{sec:1}


\subsection{Galactic structure and formation}

The science driver of the \sd\ survey is the origin and evolution of the Galactic disk and bulge. Therefore the main question that will be addressed by the survey is: what is the formation and morphology of these populations, and which physical mechanisms underpin the remarkable complexity of their chemical and dynamical sub-structure, where much of the information is gained by exploiting the concept of chemical tagging \citep{Freeman2002} alongside with detailed quantitative analysis of trends and dispersions in the distributions of phase-space, orbits, and ages of stars \citep[e.g.][]{Gilmore1989,Bland-Hawthorn2016, Barbuy2018}. 

Among the key chemical tracers there is the canonical  bimodality in the $\alpha$-abundance space \citep{Fuhrmann1998, Bensby2014,Hayden2015,Gent2024}, which has been associated with a variety of scenarios including radial migration \citep{Schoenrich2009,Minchev2014,Kordopatis2015}, hiatus in star formation between the disk components \citep{Chiappini1997,Spitoni2019,Khoperskov2021-bimodality},  turbulent instabilities in disks  \citep[e.g.][]{Bournaud2009,Clarke2019} and gas-rich accretion \citep{Grand2018,Buck2020}. As observed in the galaxy formation simulations, merging satellites contribute significantly to the disk morphology \citep{Brook2012, Belokurov2018, Helmi2018}. Mergers with dwarf galaxies \citep{Grand2018} may trigger or quench star formation depending on time, duration, and mass of infalling satellites \citep{Buck2020, Agertz2021, Renaud2021, Minchev2026, Bernaldez2026}. Such mergers may deposit metal-poor stars directly, but they also bring in cold (pristine) metal-poor gas \citep{Buck2020,Grand2018,Agertz2021,Orkney2026} from intergalactic medium  and leading to progressive formation of metal-poor stars in cold disk-like orbits.  

The debris of one of these mergers might currently be hiding within the stellar populations of the inner halo and bulge. Evidence of its existence has been provided using the dynamics of Milky Way globular clusters \citep{massari2019}, using cosmological simulations \citep{kruijssen2020}, and the chemistry of stars \citep{horta2021}. Identifying the many stellar members of this merger currently buried within the bulge \citep{Massari2026} region will enable characterisation of it with much finer detail, thus adding a crucial piece to the puzzle of the Milky Way assembly.

Predictions of models that rely on different disk and bulge formation scenarios are typically very similar in the metallicity and [$\alpha$/Fe] space (e.g. compare results in \citealt{Schoenrich2009}, Fig. 3, and \citealt{Buck2020}, Fig 3). Adding ages allows this degeneracy to be broken \citep{Minchev2019, Gent2024}, thereby allowing deviations from the canonical monotonic inside-out formation scenario \citep{Bergemann2014,Xiang2022}, caused by gas inflows, accretion, radial migration, and mergers, to be traced. To do this, it is essential to distinguish between, for example, young metal-poor stars formed from recently accreted pristine gas brought to the Galactic disk by a gas-rich merger (e.g. \citealt{Buck2023, Agertz2021}), and old metal-poor stars which might have formed in situ in the primordial disk \citep{Sotillo-Ramos2023}, even if such early fragile disks may long have been destroyed by subsequent mergers  \citep{Semenov2025}. Conversely, it is important to distinguish old metal-rich stars, which might have migrated from the inner galaxy \citep{Grenon1989,  Minchev2013, Silva-Aguirre2018, Haywood2024,Kordopatis2025}, from young metal-rich populations formed only recently in the well-mixed ISM in the solar neighbourhood \citep{Weinberg2017,Johnson2025}. Also, the reconstruction of stellar birth radii \citep{Minchev2018, Feltzing2020, Lu2024} requires precise stellar ages and metallicities, as demonstrated using the APOGEE survey data \citep{Ratcliffe2023} and tested extensively on cosmological simulations, including NIHAO-UHD, Hestia, and large TNG50 samples of Milky Way and M31 analogues \citep{Ratcliffe2024}.

Another outstanding feature of such galaxy interactions is the presence of stellar overdensities at high Galactic latitudes, which are thought to be oscillations of the disk itself \citep[e.g.][]{Xu2015,Sheffield2018,Laporte2022}. These overdensities, for example, Monoceros, ACS and TriAnd, are detectable as arcs in photometric and kinematic space, and have disk-like chemical abundances \citep{Bergemann2018,Hayes2018,Sheffield2018}. They are thought to be associated with the multiple passages of the Sagittarius dwarf Spheroidal (Sgr dSph, also sometimes referred to as dwarf Elliptical) galaxy \citep{Ibata1994}, perturbing the Milky Way's dark matter halo and thereby triggering oscillations of the disk \citep{Gomez2013, Laporte2022}. Such relaxation towards equilibrium following an external perturbation is also visible in the phase-space spiral, discovered in \textit{Gaia} DR3 \citep{antoja2018}.  Other kinds of systematic dynamical signatures in the Galactic disk were recently  interpreted in the framework of a major merger also known as Gaia-Enceladus \citep{Belokurov2018, Helmi2018}. Ultimately, stellar streams and the prominent features in the kinematic space could be due to resonant dragging by a slowing bar, dispersed open clusters, or transient spiral arms \citep{Masset1997, Monari2016, Khoperskov2020-bar, Chiba2021a, Sellwood2019,Li2024}.

All these effects produce weak signatures in stellar populations: their age-elemental abundance correlations \citep{Minchev2019} and their variability in phase-space, such as the asymmetries about the disk plane, small azimuthal variations in the disk \citep{Khoperskov2020-spiral,Khoperskov2023-spiral}, and vertical distributions influenced by the interaction of a spiral arm with the Galactic warp. Chemical abundance variation in the disk are very small, not exceeding $\sim 0.05 - 0.1$\,dex in abundance ratios as a function of Galactic azimuth, as seen in data and in models \citep{Poggio2022, Spitoni2023, Hackshaw2024}. Detailed analysis of these effects requires abundances of the elements that trace the differences in star formation history of galactic building blocks, such as O, Na, Ti, Ba, and Eu \citep[e.g.][]{Bergemann2018, Hayes2018}. The \sd\ survey will yield sufficient precision, depth, and multi-dimensionality in the elemental abundance space to quantify the detailed chemical sub-structure in the Galactic disk.
\subsection{The bulge and bar}

Stellar kinematics and star count maps imply that the Galactic bulge is a rotating, barred, and boxy/peanut-shaped bulge \citep{Shen2010, Wegg2013, Sanders2019, Clarke2019b, Johnson2022} that transits into a longer, thinner bar at around 2-3\,kpc along its major axis \citep{Wegg2015, Portail2017, Drimmel2023, Khoperskov2025b}. Boxy/peanut-shaped bulges embedded in thin bars are common in barred galaxies \citep{Luetticke2000, Bureau2006, Erwin2013} and also form in N-body models of unstable disks  \citep{Combes1990, Athanassoula2005, Athanassoula2015, Romero-Gomez2011}.

A majority of the stars in the bulge are old, 8-12\,Gyr, with younger and more intermediate age stars (4-8 Gyr) at higher metallicities and near the Galactic plane \citep{Clarkson2008, Bensby2013, Barbuy2018, Grady2020, Anders2023}. Stars in the bulge and bar follow a sequence from high-$\alpha$ at low-and intermediate metallicity to low-$\alpha$ metal-rich stars, with some bimodality seen at high metallicities \citep{Rojas-Arriagada2017, Barbuy2018, Queiroz2021, Imig2023}.

At metallicities $\lesssim -1.3$, a slowly rotating, pre-spin-up population is seen in the bulge and bar, similarly as previously in the Galactic disk \citep{Belokurov2022, Rix2022, Ardern-Arentsen2024, Pandey2025}. The main population of high-$\alpha$ stars around $\rm [Fe/H] \approx -0.4$ appears to be part of the bulge and bar with cylindrical rotation velocities similar to the low-$\alpha$, high-metallicity stars but with a thicker bar and weaker boxy/peanut-shape \citep{Ness2013,Wylie2021,Khoperskov2025b, Pandey2025}, whereas in the transition region with $\rm -1 \lesssim [Fe/H] \lesssim -0.6$ the structure becomes more spheroidal with the precise dynamical properties not yet clear. The low-$\alpha$ population is strongly barred; however, at the highest metallicities a more centrally concentrated distribution, the so-called knot has recently been discovered \citep{Rix2024, Horta2025}.

Recent work places the formation of the Galactic bar at $8-10$\,Gyr ago \citep{Haywood2024, Sanders2024}, likely later than the time when the high-$\alpha$ stars formed. Models in which a bar and boxy/peanut shaped bulge forms in a disk containing both a cold and thin, and a hotter, thicker disk component, indeed explain a number of the main chemo-dynamical properties and differences between metal-rich and metal-intermediate stars in the bulge through kinematic fractionation
\citep{diMatteo2015, Fragkoudi2017, Debattista2017}. After its formation, the pattern speed of the Galactic bulge and bar has likely slowed significantly, causing resonances such as co-rotation to grow while moving outwards. Stars captured along the way then radially migrate outwards, changing the population structure in the disk \citep{Chiba2021a, Chiba2021b, Haywood2024, Zhang2025}. 
%
%
\begin{figure}[!ht]
\centering
\includegraphics[width=0.9\columnwidth]{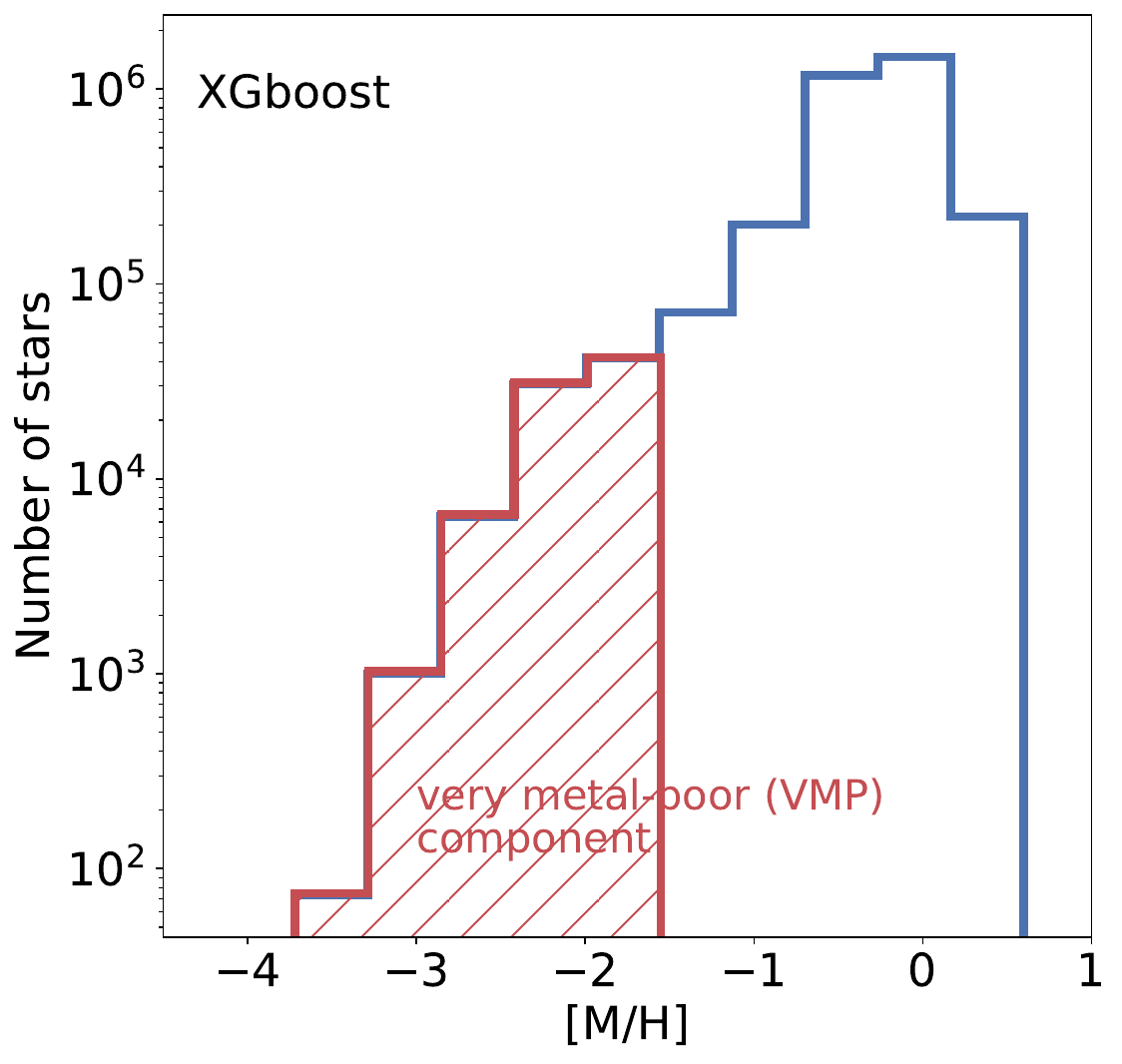}
\caption{Metallicity distribution of \sd\ stars in all sub-surveys. Metallicity was adopted from the XGBoost catalogue by \cite{Andrae-2023}.}
\label{fig:metallicity}
\end{figure}

The \sd\ sub-surveys as described below, specifically its deep bulge sub-survey (Sect.~\ref{sec:deepbulge}) will obtain accurate stellar abundances and kinematics for large ($\sim 66\,000$) samples of bulge, bar and inner disk stars. This will be the largest optical sample to date: the \textit{Gaia}-ESO survey \citep[e.g.][]{Rojas-Arriagada2017, Recio-Blanco2017} explored about $2\,500$ red clump stars in the inner regions of the bulge (11 small fields at $-10 < b < -4$ deg. and $-10 < l < 8$ deg.), whereas the APOGEE survey released abundances for about $10\,000$ stars that can be associated with the bulge and the bar by phase-space \citep[e.g.][]{Queiroz2021, Liu2025, Sales-Silva2026}\footnote{We remind that these numbers vary from study to study, even if based on the same APOGEE data release, as the population membership of stars requires a non-trivial analysis of their motions and orbits.}. For a fraction of our \sd\ stars, primarily red giants and red clump stars (Fig~.\ref{fig:cmd1}, for more details see below), also ages determined by asteroseismology (through the value-added Roman field), and spectroscopic calibrations will be available. These data will allow us to better study and separate the different inner Galaxy populations, with improved age resolution and precise chemical characteristics, including stars accreted onto the inner Milky Way and stars passing through. With these data, combined with distances and stellar kinematics, \sd\ will obtain a deeper understanding of the evolution of stellar populations in the bulge region, of the main evolutionary phases, and the transitions between them, from the early history to the present time.
\subsection{Chemical evolution of the Galactic disk}

The second major science theme driving the design of the \sd\ survey is the chemical evolution of the Galaxy \citep{Kobayashi2020, Arcones2023}. While major constraints on the evolution of $\alpha$-chain elements up to the iron-group have become available from recent surveys (see Introduction), the formation and origin of heavier elements, including the iron-peak \citep{Palla2021, Eitner2023, Storm2025}, p-capture elements \citep{Arnould2003, Travaglio2018}, slow- and rapid neutron-capture process \citep{Mashonkina2001, Magrini2018, Guiglion2024b}, remains the 'holy grail' of the complete theory of stellar nucleosynthesis. Whereas large samples have just become available from surveys like GALAH \citep{Buder2025}, most detailed constraints on Galactic chemical evolution are currently based on studies which are limited both in Galactic volume and in metallicity to $\rm [Fe/H] \gtrsim -1$ \citep{Lian2023}. Also the scatter due to observational errors in large samples is still unfortunately too large to provide strong quantitative constraints to s-, p-, and r-process nucleosynthesis.

The \sd\ programme was designed to answer the central question: Which events are the main sources of heavy elements in the Milky Way and how did they shape the chemical history of the Galactic disk. This goal will be achieved via two dedicated sub-surveys (S0402 and S0409; see Table~\ref{tab:main}) that aim to obtain high-resolution and high-SNR (in the blue) spectra of more than $800\,000$ main-sequence, turn-off and subgiant stars, with a very broad metallicity coverage, $-4 \lesssim$ [Fe/H]  $\lesssim 0.5$ (Fig.\ref{fig:metallicity}). Among our top-tier science questions is the role of extreme events in the formation of the neutron-rich nuclei away from the valley of stability. Such systems include mergers of neutron stars and black holes (NS-NS, NS-BH; \citealt{Korobkin2012,Cote2019}), rapidly rotating and highly magnetised massive stars \citep{Winteler2012,Siegel2019}, core-collapse supernova (SN) explosions driven by quark-hadron transitions \citep{Nishimura2011}, neutrino–driven winds of normal core-collapse SNe \citep{Farouqi2010,Arcones2007}, and collapsars \citep{Bauswein2013}, in addition to well-known production of heavy elements close to the nuclear stability valley in asymptotic giant branch (AGB; \citealt{Karakas2016}) and in rotating massive stars \citep{Limongi2018, Prantzos2018}.

One other key element to be studied by the \sd\ survey is Li, which can be traced via the diagnostic $^{6}$Li-$^{7}$Li doublet at $6707$\,\AA. The GALAH survey has released a sample of $\sim 1262$ Li-rich red giants \citep{Martell2024}, and $\sim 229$ thousand stars have precise Li abundance measurements \citep{Wang2024}. However, at [Fe/H] $\lesssim -1$, the GALAH Li sample is characterised by only a few hundred stars \citep[][their Fig. 5]{Wang2024}. Our sample will provide a ten-fold statistics, especially at low metallicities and greater Galactic distances, reaching the innermost disk and into the outer disk at $R_{\rm GC} \approx 20$ kpc (see Fig. \ref{fig:surveys}). According to our tests on 4MOSTified archival spectra \citep[see Figs. 5, 6][]{Storm2025b}, Li abundances will be precise to better than $\approx 0.05$ dex through the entire metallicity range $-4 \lesssim $[Fe/H] $\lesssim 0.0$, which will enable the nucleosynthesis history of Li to be traced from the Spite plateau \citep{Spite1982} to the super-solar regime  \citep{Bensby2018,Guiglion2019}. For instance, there is still a large uncertainty on the Li stellar yields, including the role of novae \citep[e.g.][]{Cescutti2019}, and state-of-the-art chemical evolution models will be used to address this question \citep[e.g.][]{Romano2021}.

Constraining GCE models requires multi-dimensional abundance distributions of abundance ratios [X/Fe] that; (a) probing different enrichment timescales (for example prompt versus delayed sources), (b) being inherently sensitive to the initial mass or metallicity of the progenitor (which leads to time-dependence in the observed GCE tracks), and/or (c) the explosion mechanism (that is, the internal structure of the object), are available. In \sd\ we will measure abundances of at least 18 chemical elements for stars spanning the entire range of ages and metallicities of the Galactic disk and bulge, which will allow us to disentangle the role of these astrophysical scenarios in the evolution of the Milky Way. 
\subsection{Stellar physics}
Stellar chemical composition is, along with mass and initial rotation, one of the main fundamental parameters of stars. Along with the pre-main-sequence mass, it defines all subsequent stellar evolution and its endpoints, also shaping stellar mass loss, driving convection and pulsations, and resulting nucleosynthesis, due to the profound influence of radiative opacities associated with detailed abundance rations on the energy balance, surface and sub-surface dynamics of stars \citep[e.g.][]{Maeder2009}.

Whereas the basic stellar evolution theory is generally rather well understood for the Sun (however, see \citealt{Acharya2025} for the key role of nuclear reactions that remain to be constrained), the effects of chemical composition in the context of stellar evolution are poorly understood. It has been established that C/N abundance ratios changes substantially during stellar evolution on the red giant branch \citep{ElEid1994,Boothroyd1999,Cao2026}, thus requiring evolutionary corrections to be applied to measured abundances in order to obtain pristine, that is, at-birth, abundances of C and N of a star, which are the quantities of an major interest to Galactic science. This relationship is also at the basis of chemical ages \citep{Salaris2015,Martig2016}. However, we are just beginning to explore quantitatively the detailed role of other chemical elements. The effects of atomic diffusion \citep{Thoul1994,Deal2018} include gravitational settling and radiative acceleration, as well as internal mixing due to rotation and turbulence, and thermohaline instabilities \citep{Charbonnel2007,Charbonnel2010,Deal2020}. Li is a critical tracer of physical processes that take place in stellar interiors. The `hot' and `cool bottom burning' mechanisms are responsible for the high lithium content observed in some red giant branch stars \citep{Sackmann1992, Sackmann1999, Smiljanic2018}. The abundance of Li can also be used a proxy for stellar age \citep[e.g.][]{Jeffries2023}. Finally, the determination of stellar ages, by asteroseismology \citep{Martig2015,Pinsonneault2025} or by classical isochrone fitting \citep{Pont2004,Ruchti2013} relies critically on the absolute precision of elemental abundances, especially those of $\alpha$-chain elements, such as Mg, Si, and Ca \citep{Salaris1993,VandenBerg2000}. 

In the context of \sd\ goals, it is central to quantify the effects of stellar evolution on surface elemental abundance, because studies of GCE generally rely on the assumption that chemical abundances measured in stellar atmospheres represent the pre-birth values \citep{Prantzos2018, Kobayashi2020}, that is, those of the ISM from which stars formed. Thus, any deviation from this assumption would have major consequences for understanding the measured 4MOST chemical abundance patterns and their gradients as a function of age, distance, and Galactic position.
\subsection{Star-planet connection}
%
\begin{figure}[t!]
\centering
\includegraphics[width=0.9\columnwidth]{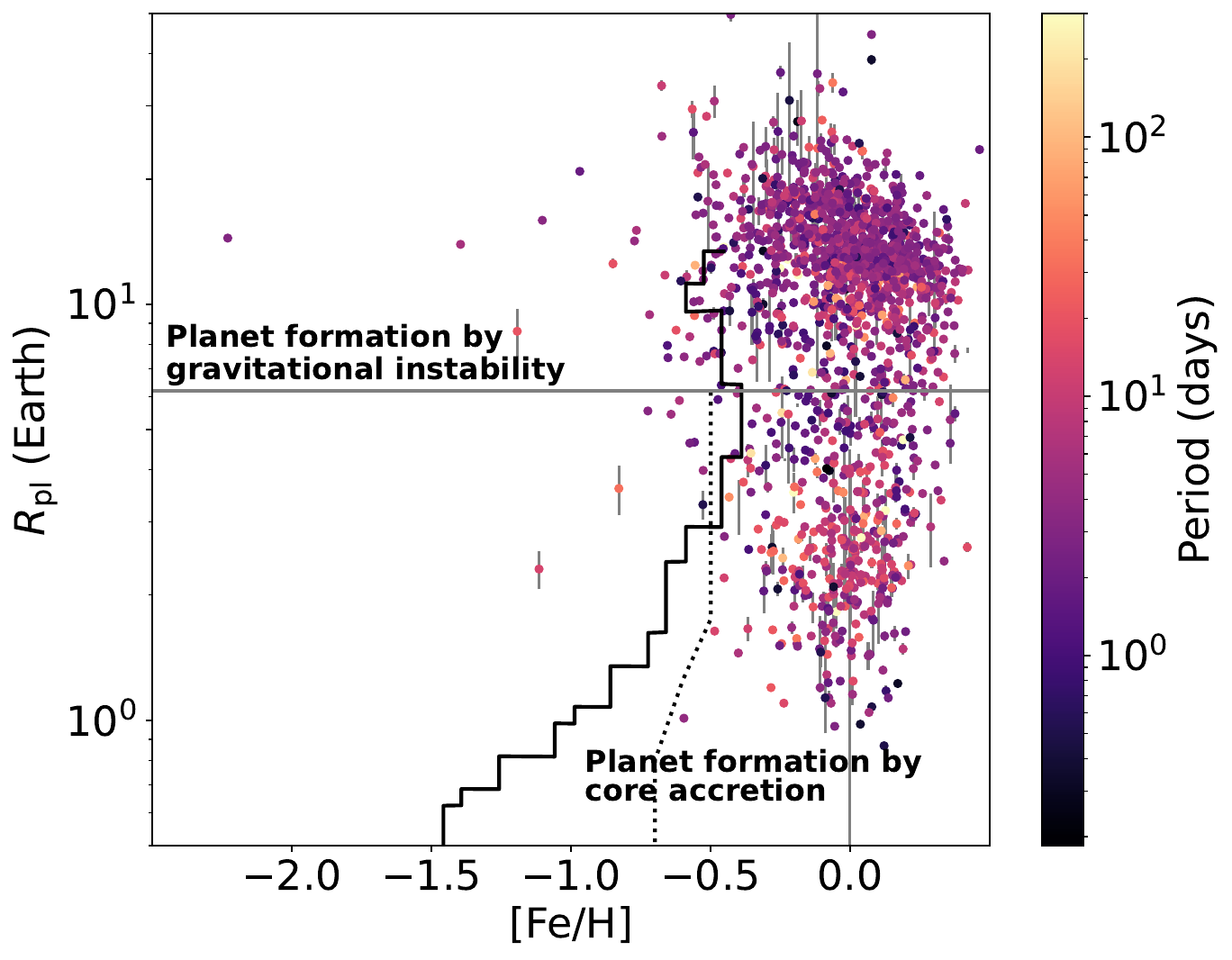}
\caption{Properties of planets orbiting a sub-sample of \sd\ survey stars. The colour-coding represents planetary orbital periods. Metallicities ($x$-axis) are adopted from the XGboost catalogue \citep{Andrae-2023}. The planet formation limits as a function of metallicity are adopted from \citet{Nielsen2023} (solid) and \citet{Xenos2025} (dotted) as well as from \citet{Boss2023} for the gravitational instability.}
\label{fig:exohosts}
\end{figure}
%
Another major opportunity arising from the \sd\ data is the study of chemical composition of exoplanet host stars. A key prediction of planet formation models and observations is the metallicity threshold above which planet formation becomes efficient \citep[e.g.][]{Santos2004, Fischer2005, Johnson2010, Boley2024, Andama2024}. For a given scenario, for example, the core accretion (CA), or the gravitational (disk) instability (GI), or extensions thereof, the demographics of planet populations is different \citep[e.g.][]{Ida2004, Mordasini2012, Nayakshin2017, Boss2023, Drazkowska2023}. As shown Fig.~\ref{fig:exohosts}, leading CA-based models \citep{Nielsen2023, Xenos2025} do not predict any small (Earth-like) planets below [Fe/H] $\lesssim -1.5$, with the population steeply increasing towards higher metallicity. However, the available planet detections suggest that these models are in contradiction with the observed data. The GI-based models, however, enable the formation of gas giants even at very low [Fe/H] $\lesssim -1.5$, which is also seen in the observed statistics of hot Jupiters  (Fig.~\ref{fig:exohosts}). All these distributions are expected to show a different scaling with metal mass fraction (Perdomo et al., in prep.), which sets the dust-to-gas ratio, which requires accurate abundances of ice- (C, N, and O) and rock-forming (Mg, Si, and Fe) elements \citep[e.g.][]{Fegley1987, Lodder2003, Lodders2025}. Furthermore, the detailed chemical abundances set the disk's initial composition which in turn determine the composition of forming planets \citep[e.g.][]{Bond2010, Bitsch2020, Cabral2023, Nielsen2023, Hatalova2025}, with hints to it influence the formation efficiency of planets themselves \citep[e.g.][]{Xenos2025}. Additionally, weak dependence of exoplanet host distributions on Galactic population membership may be present and requires a sufficiently robust statistics to confirm the trends \citep[e.g.][]{Rampalli2025}. 

Presently, a critical missing component is a complete uniform set of chemical abundances and stellar parameters, as well as their ages, and kinematics, for the entire sample of stars, which are known to host planets including Earth-like, hot Jupiters, sub-Neptunes and other types. Critically, any systematic dependence of planet populations and occurrence rates on stellar irradiation, mass, metallicity, or abundances of light or rocky elements would have major consequences for planet formation and evolution as well as their ability to retain atmospheres and liquid water and therefore for exploring the uniqueness of the Solar System in the context of Galactic habitability and extra-terrestrial life \citep{Prantzos2008, Scherf2024}. 

\subsection{Activity and magnetism}

The \sd\ survey, owing to its homogeneous all-sky selection, broad wavelength coverage, offers another unique opportunity to study stellar evolution - the regime of stellar magnetism and pulsations, taking advantage of synergies with multi-wavelength observations with high-energy facilities, like eROSITA.

Stellar magnetic activity causes a plethora of physical phenomena, including starspots, flares, and stellar winds. Coronae and chromospheres of low-mass stars ($M_\star \lesssim 2M_\odot$) are a direct consequence of magnetic activity. The stellar high-energy emission, X-ray to UV emission, is due to magnetic activity in low-mass stars. Magnetic activity is also very important for all studies that aim to search for and characterise planets and for studies of planetary habitability \citep{morris2018,rackham2023}, as starspots bias transit depth measurements and can cause line-profile variations that can be confused with a radial velocity signal originating from exoplanets \citep{Queloz2001}.  Stellar high-energy emission from magnetic activity powers planetary mass-loss \citep{owen2024}, and drives atmospheric chemistry of planets out of equilibrium \citep{france2020,locci2024}.

Magnetic activity is also relevant for analyses of stellar ages, as activity evolves significantly with the age of stars. This evolution is due to the dynamo origin of magnetic activity. Rapid rotation leads to strong magnetic fields but also strong rotational braking due to magnetised stellar winds. There is a limit in rotation rate after which the activity does not increase anymore. Many young solar-like (and lower mass) stars reside in this saturation regime, showing a pronounced X-ray emission, with the  fractional X-ray luminosity $L_{\rm X}$ of $10^{-3}\,L_{\rm bol}$ \citep{guedel2004,Wright2011}. When the magnetised winds slow the rotation below a stellar structure dependent threshold, magnetic activity decays approximately with $\sim {\tau}^{-0.5}$ \citep[][]{skumanich1972}; this $P_{\rm rot}$ threshold is typically reached after the first few tens of megayears, depending on stellar mass \citep{Kastner2022}. In the subsequent decay-phase, activity (in particular the stellar X-ray luminosity) decays rapidly with stellar age \citep{Getman_2022}. For instance, the Sun is a rather weak X-ray source \citep[with $\log L_{\rm X}/L_{bol}\sim-6.4\pm0.5$, ][]{Orlando2004,Tu2015}. 

Therefore, studying the evolution of magnetic activity requires a broad stellar sample spanning a range of activity levels (that is, ages), specifically stars near the transition between saturation and decay as well as stars well in the decay part. Such a sample will be uniquely available with \sd\ observations, providing a unique opportunity to study the relationships between magnetism, evolution, and detailed abundances of stars.
\subsection{Variable stars}

Pulsating stars such as Cepheid variables play a key role in stellar and galactic astronomy, since they obey to solid pulsation relations connecting the period of the oscillation  - minimally affected by systematic errors and independent of uncertainties affecting distance and extinction estimates - with stellar physical parameters, namely luminosity, effective temperature, and mass. In general, the Cepheid class includes several subclasses: 
a) Classical Cepheids (CCs) are young ($\rm age \sim 50-300$\,Myrs) pulsators, present in the disk of spiral galaxies and in dwarf irregular galaxies with ongoing star formation activity.
b) Type II Cepheids (T2Cs) are old ($\rm age\gtrsim 10$\,Gyrs) pulsators present in the Galactic spheroid (disk, bulge, and halo) and in globular clusters.  
c) Similar spatial distribution is shown by Anomalous Cepheids (ACs), which are intermediate-old stars ($\rm age \gtrsim 1-2$\,Gyrs) and are typically hosted in dwarf Spheroidal galaxies.  
d) Dwarf Cepheids \footnote{These targets have also been referred to, in the literature, as dwarf Cepheids, $\delta$ Scuti, or SX Phoenicis according to their chemical composition and/or Galactic location.} were recently introduced in the Cepheid sample as they trace quite well the Galactic thin disk, but are a few billion years older than CCs \citep{mcnamara11, fiorentino2015, coppola15}.

All these Cepheid classes are solid distance indicators thanks to the period-luminosity and period-Wesenheit\footnote{The Wesenheit magnitudes \citep[][]{Madore1982} provide a reddening-free magnitude by assuming that the adopted extinction law is universal.} \citep[ also known as Leavitt Law,][]{Leavitt1912} relations. This intrinsic feature also makes them invaluable tools for the extragalactic distance scale \citep[especially CCs, e.g.][]{Freedman2011,Riess2016}.

For the \sd\ survey, the prime role of Cepheids is to constrain the metallicity distribution and the abundance gradients in the Milky Way disk by using both young (CCs) and intermediate-age (ACs, dwarf Cepheids) stellar tracers. We also plan to provide independent estimates of chemical abundance ratios using T2Cs and ACs data for the bulge and the thick disk-halo interface. In addition, our data will be used to constrain the metallicity dependence of the period-luminosity and period-Wesenheit relations in order to refine the estimate of the Hubble constant \citep{Riess2021,hao2022,breuval2025,ripepi2025,madore2026}.
%
%
%
\begin{figure*}
\centering
\resizebox{\hsize}{!}{
\includegraphics[width=0.32\hsize]{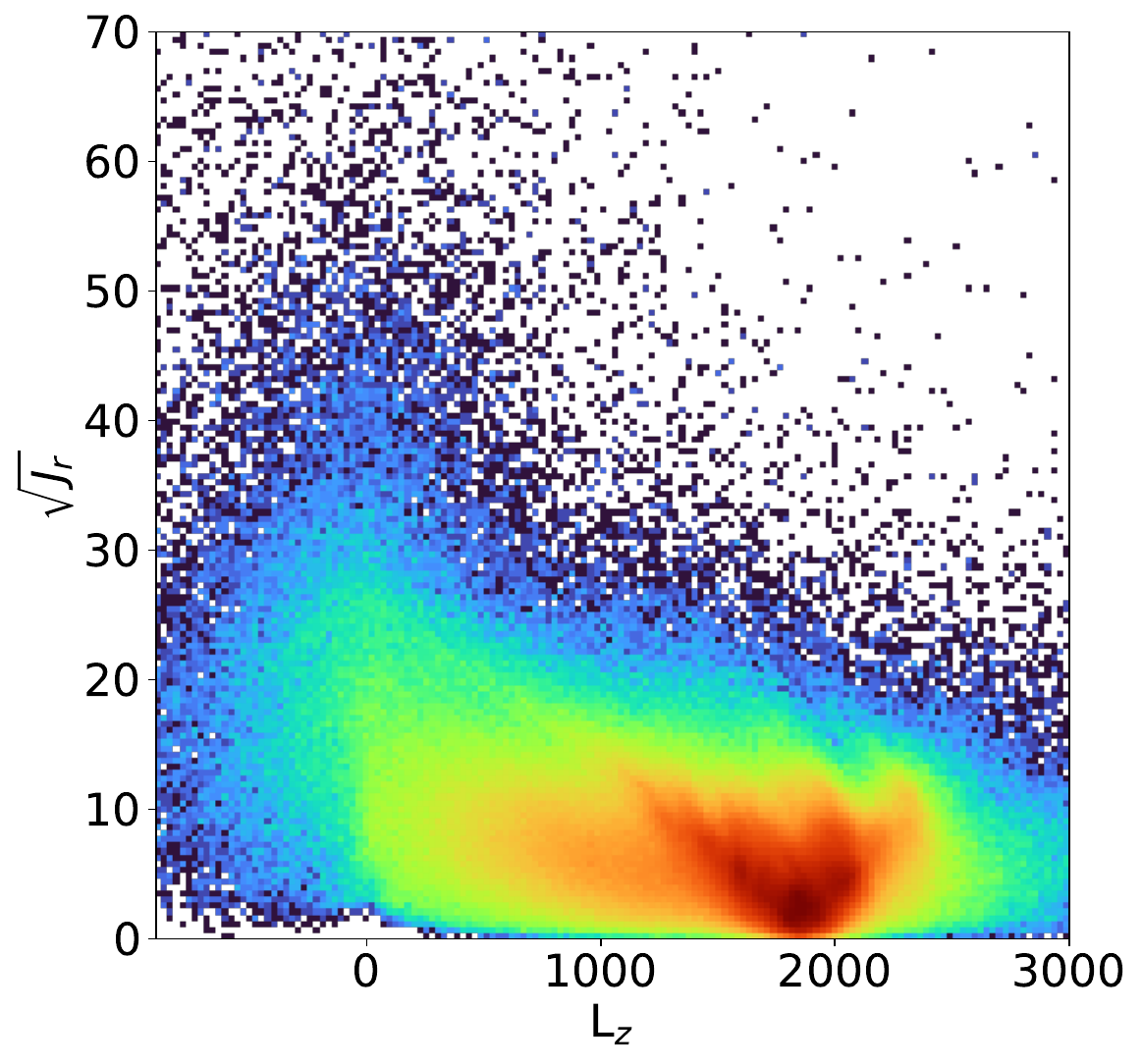}
\includegraphics[width=0.34\hsize]{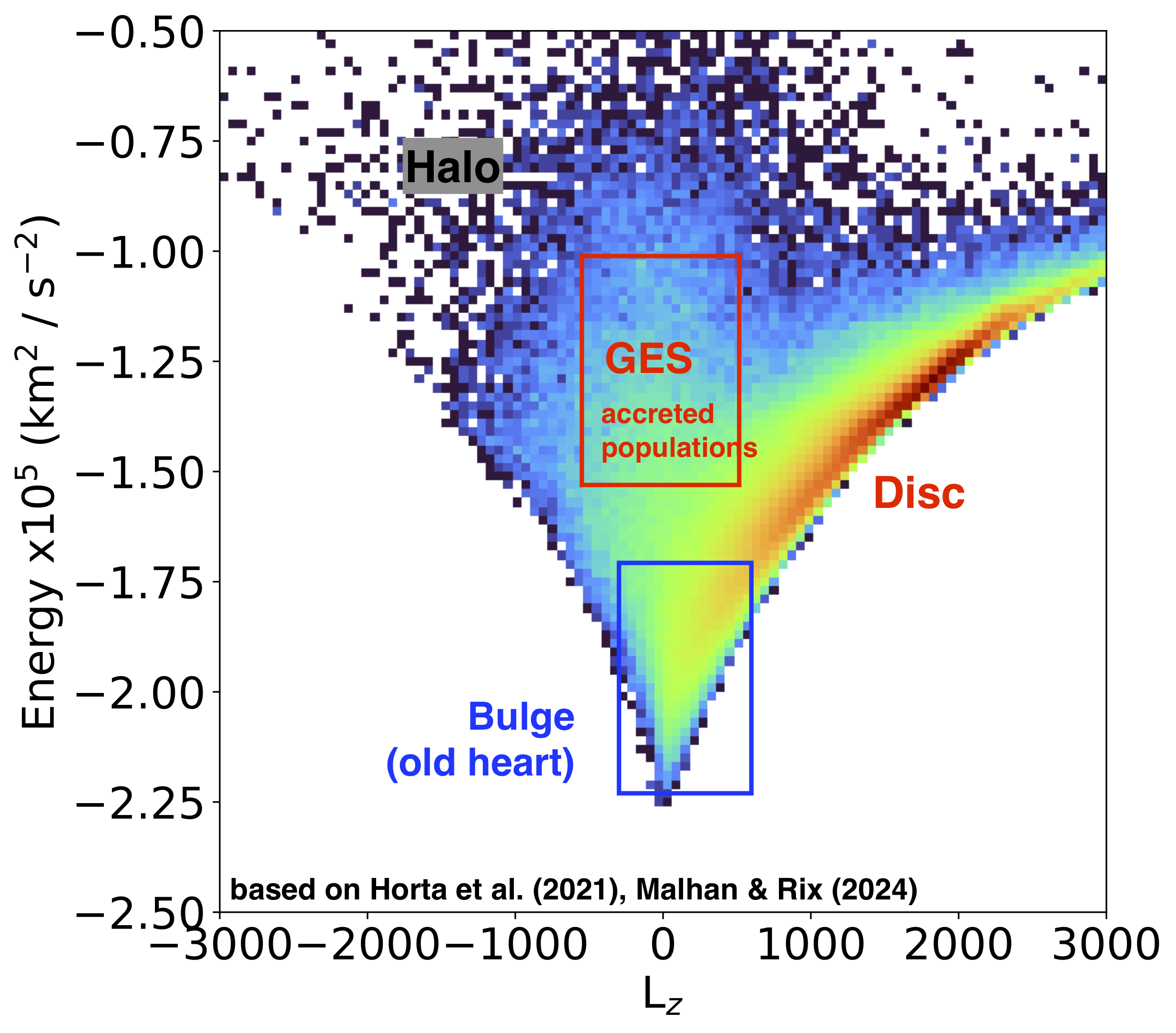}
\includegraphics[width=0.32\hsize]{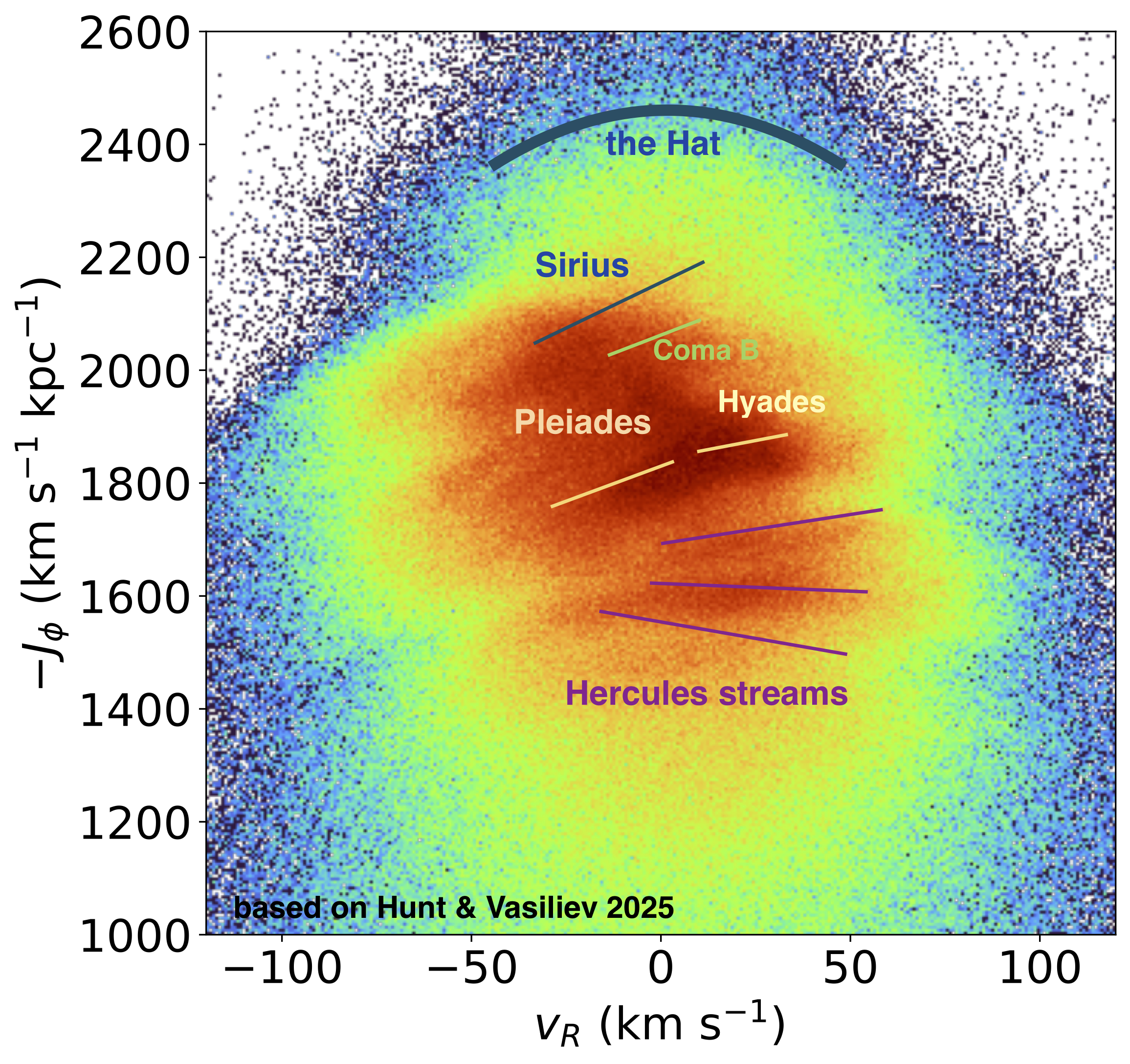}
}
\caption{Distribution of targets in the kinematic space, where L$_z$ is the orbital angular momentum and J$_r$ is the radial action.}
\label{fig:targetsup}
\end{figure*}
%
%
%
In this context, we point out that a recent investigation \citep{WangXu2024} found, by using accurate proper motions provided by \textit{Gaia} DR3, that 43 Galactic CCs are candidate members of 38 open clusters (OC). These OC~CCs can also be used constrain the efficiency of the first dredge-up event that ignites quiescently core helium burning. In the stellar mass regime $M/M_\odot\approx 2-7$, the first dredge up becomes significantly less efficient when moving from solar metallicity to more metal-poor stars \citep{bono2024}. Also, CCs obey the period-age and period-age-colour relations, enabling independent estimates of stellar ages, complementary to ages based on the method of isochrone fitting \citep{bono2005,desomma2025} and asteroseismology.  

\section{Auxiliary programs}\label{sec:5}

\subsection{Multiplicity and cadence} \label{sec:cadence}
The \sd\ dataset will also be used to detect and characterise a diversity of binary and higher-order multiple stellar system, including triples and quadruples. About 15\,\% of stars in a random sample in the field will appear as a spectroscopic binary (SB), either through the presence of multiple spectral components or through varying RV across multi-epoch spectra \citep{Moe2017}. Such serendipitous discoveries should allow statistical studies if properly identified and characterised. On the other hand, multiple star science is one of the main drivers for implementation of cadence functionality into the 4MOST observational strategy. In the expectation of the 4MOST cadence functionality, despite it being realised on a best effort basis, a scheme of cadence flags\footnote{Here we note that the flag (A) has been assigned to 'no cadence' by convention in the 4MOST nomenclature.} was devised to facilitate three use cases: B) variable stars (requiring optimal exposures in a single visit); C) deep/repeat field stars (requiring 1 $+$ NREPEAT visits at specific time delay DT\_MIN); D) active binary search (requiring specific time delay DT\_MIN between necessary multiple exposures). The different use cases are encoded by three parameters and listed in this notation also in Table~\ref{tab:main}: (SINGLE\_VISIT, NREPEAT, DT\_MIN). Use case B is always encoded as (1,0,0), C as (1, NREPEAT, DT\_MIN), and D as (0, 0, DT\_MIN).

\begin{figure}
\centering
\resizebox{1\hsize}{!}{\includegraphics{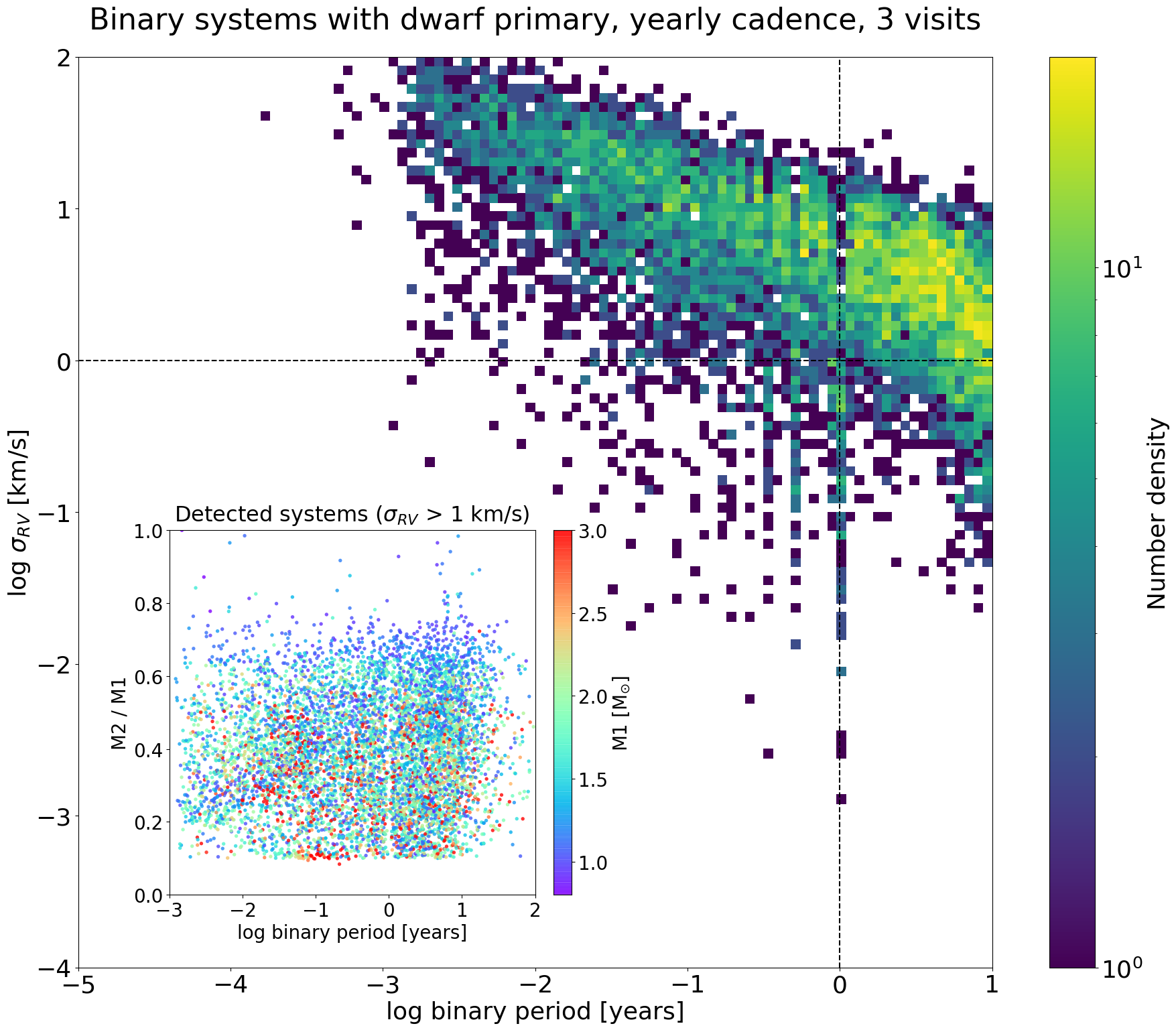}}
\caption{Simulated detection of SB1 systems from multiple visits. For systems with dwarf primary stars more massive than the Sun, about 27\% can be detected through velocity scatter greater than 1 km/s (required precision for 4MOST high-resolution spectra), assuming three visits separated by about 1 year. The detected systems cover well the lower part of the binary period distribution and a large range of mass ratios. Simulations were carried out by Rebecca Forsberg and are based on the code and parameters as described in \cite{Stonkute2018} and in line with S4 targeted population of stars. We excluded binary systems that are resolved and those with components that have a similar brightness, which can be detected as SB2s (see the lack of higher mass ratios in the inset).}
\label{fig:optimal_detection}
\end{figure}

All \sd\ sub-surveys except S0407 and S0408 use the cadence flags of the 3 use cases for optimising observation of variable stars and mostly for improved detection of multiplicity (see Table \ref{tab:main}). For S0401 and S0402, for stars that will necessarily have multiple exposures, we set the active binary search flags such that there is an optimal time delay of one year between exposures (see Fig~\ref{fig:optimal_detection}, but excluding fainter objects where such repeats could then drag beyond the five-year 4MOST programme.

\begin{figure}
\centering
\resizebox{0.8\hsize}{!}{\includegraphics{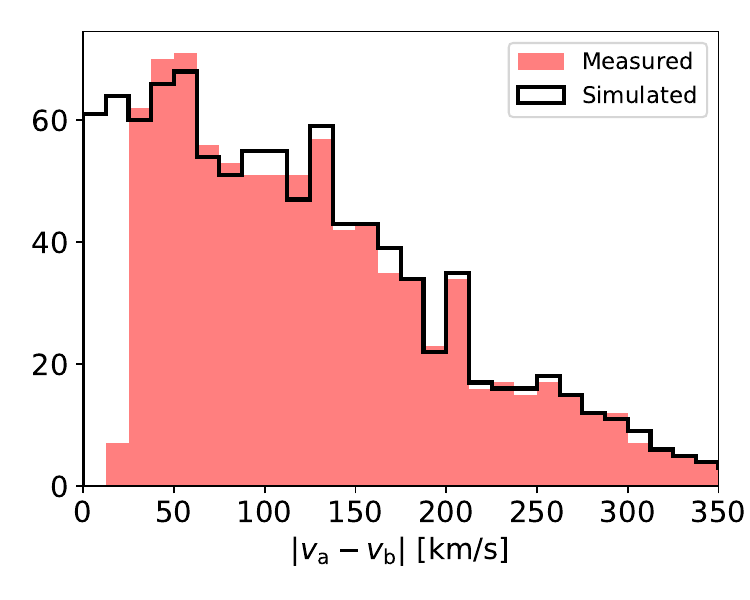}}
\caption{Distribution of the 876 SB2 RV differences between components as detected and measured by the 4GP DOE module (in red) compared to the simulated sample of 1000 SB2 (in black) from a synthetic detection experiment. 4MOST SB2 can be detected if the RV difference between components a and b is larger than approximately 25 km/s.}
\label{fig:doe_detection}
\end{figure}

\begin{figure}
\centering
\resizebox{1\hsize}{!}{\includegraphics{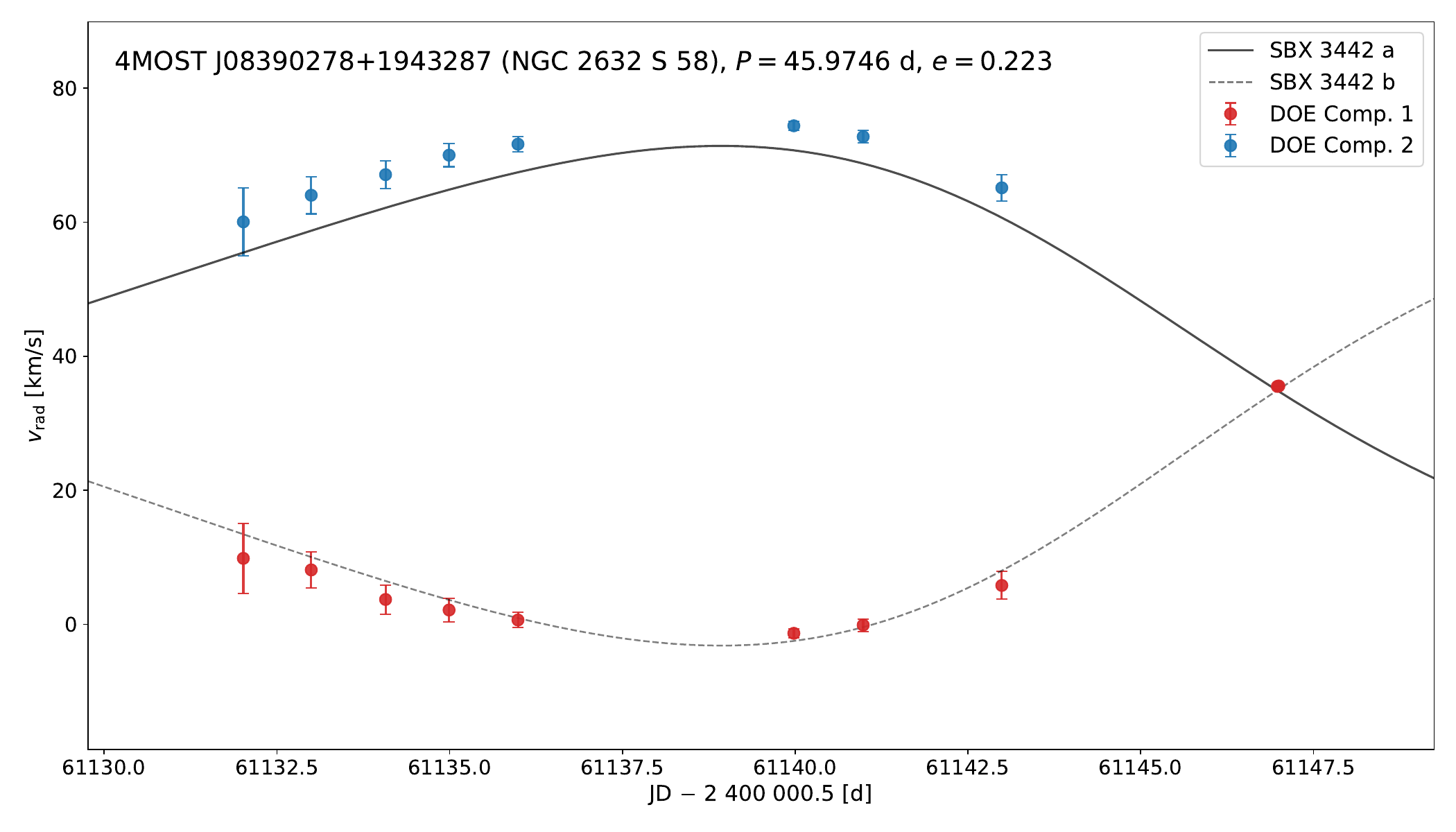}}
\caption{RV measurements from the 4GP DOE module of an SB2 in a cluster. The orbital solution from the SBX catalogue \citep{Merle2026} is displayed for comparison. The RV measurements were obtained from nine epochs of SPV observations covering almost one half of the orbital period. This kind of characterisation is what we aim for with specific cadence requests for selected targets.}
\label{fig:doe_orbit}
\end{figure}

\begin{figure}
\centering
\includegraphics[width=1\columnwidth]{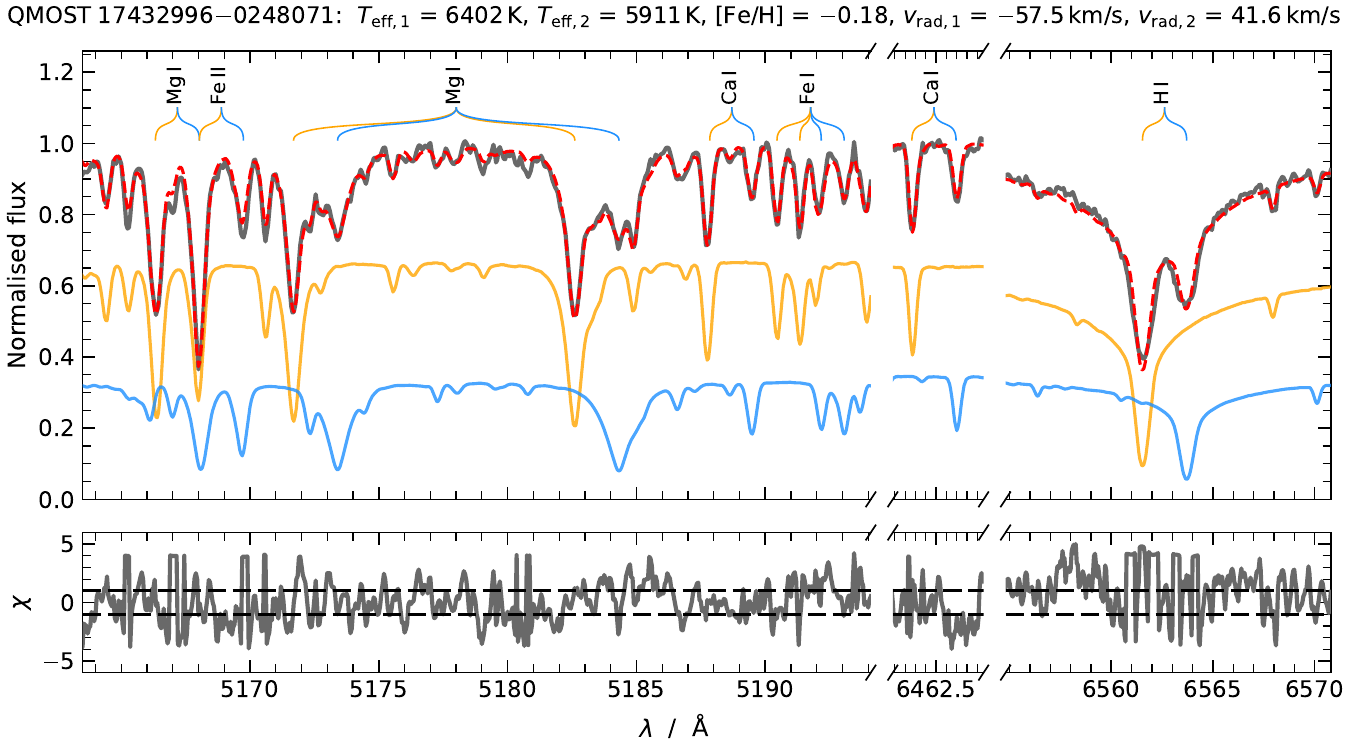}\\
\includegraphics[width=1\columnwidth]{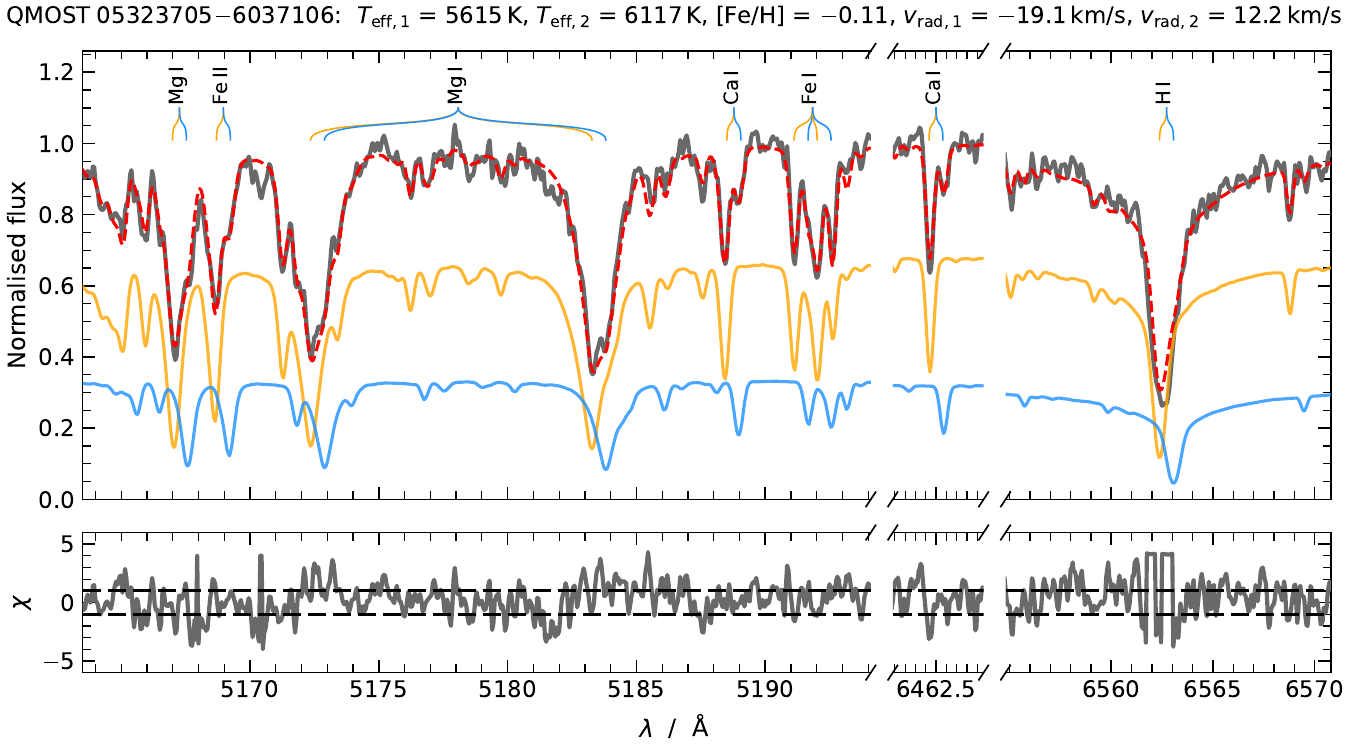}
\caption{Example HRS spectral regions of two newly discovered SB2 systems from the 4GP MULTIPLE\_STARS module. The colours are as follows: observed spectrum (grey), best-fit Payne model (red), primary component (orange), and secondary component (blue). The bottom panel for each SB2 shows the uncertainty-scaled residuals, $\chi$.}
\label{fig:multifit}
\end{figure}

For composite spectra, where light of more than one star enters a 4MOST optical fibre (aperture), an automated tool implemented in the 4MOST Galactic Pipeline (4GP) will allow for the detection of two or more components that are significantly contributing to the combined flux, the DOE module \citep{merle2017, Traven2020} (see Fig.~\ref{fig:doe_detection}). In turn this detection will enable the MULTIPLE\_STARS module to be triggered, which has the ability to simultaneously detect and characterise SB$n$ ($n\ge2$) spectra by creating a multiple-component model based on single-star spectra emulator The Payne \citep{Ting2019}, and also constrain orbits of these systems and deliver for example mass ratios based on multiple exposures. Additionally, accurate multi-epoch radial velocities enable the characterisation of component orbits and masses of single-lined spectroscopic binaries (SB1s), which can for example lead to the detection of the presence of unseen neutron-star or black-hole companions (see Fig.~\ref{fig:doe_orbit} and  Fig.~\ref{fig:multifit} for example results on survey programme validation (SPV) data from both 4GP modules).

Our sample, combined with the cadence strategy, will yield the largest harvest of binary stars from a ground-based spectroscopic survey and allow us to effectively characterise distinct populations of multiple stars that formed in different star-forming conditions, with the potential to address, for example, the question of multiplicity fractions and their recently suggested dependence on metallicity \citep[e.g.][]{Moe2019, Merle2020, Price-Whelan2020, Traven2020}. Another specific science goal of \sd\ is the `quiet' symbiotic stars, which are potentially important as SN\,Ia progenitors and Li producers (see Sect.~\ref{sec:SSP} and \ref{sec:SPV}).

\subsection{4MOST supplementary science programs aligned with \sd} \label{sec:SSP}

Several 4MOST supplementary science projects (SSPs) will be carried out with 4MOST, lead by different 4MOST science teams members. Some of these programs will be carried out in collaboration with our survey because their science scope is aligned with the \sd\ plans. These surveys are described below.

\paragraph{Improving ML performances in the direct solar neighbourhood (PI: G. Guiglion):} In the era of ML applications to large spectroscopic surveys, one challenge comes from building a training sample that covers a homogeneous parameter space, especially in the solar neighbourhood. For instance, incomplete training samples can lead to inaccurate metallicity distribution function predictions. To remedy that, we proposed to observe $\sim70,000$ nearby dwarf stars ($\pi > 4$ milli-arcsec) from the catalogue of \citet{Guiglion2024a}, mostly in the range $10 < G < 13$.

\paragraph{Li-rich giants (PIs: R.~Singh, N.~Vernekar):} Understanding the excess of Li discovered in the red giant phase of a fraction of low mass stars is a major puzzle in stellar astrophysics. Detailed chemical abundances supplemented by other properties of stars such as evolutionary phase, stellar activity, rotation velocity can constrain physical mechanisms responsible for Li enrichment. The 4MOST Supplementary survey of Li-rich giants proposes the spectroscopic observation of $30\,521$ red giant stars in the magnitude range 10$<$G$<$19 selected from \textit{Gaia} catalogue comprises binary stars, stars in TESS and GALEX survey.

\paragraph{4MOST: The benchmark elemental abundance census of nearby RR Lyrae with 4MOST (PI: Prudil)} BEACON at 4MOST will obtain a homogeneous, high-resolution spectroscopic census of $\sim10^{4}$ nearby RR Lyrae stars, providing a gold-standard abundance benchmark for time-domain standard candles. The survey will anchor PLZ and period–Fourier–metallicity relations using precise abundances tied to \textit{Gaia} parallaxes. This calibration set will strengthen spectroscopic pipelines and enable metallicity inference for large photometric surveys such as LSST. The resulting dataset will enable chemo-dynamical studies of the Milky Way and constrain the formation history of its oldest populations.

\paragraph{A search for GSE members using 4MOST (PI: S. Antony):}
\textit{Gaia} Sausage/Enceladus (GSE) is the remnant of a massive galaxy accreted by the Milky Way about ten billion years ago. However, only a few thousand stars have been identified as GSE members so far. \citet{Mori2024AA} and \citet{Skuladottir2025ApJ} show that GSE members span a larger region of integrals of motion space than previously thought, implying that the strict kinematic and dynamical cuts used in earlier studies may have missed many members and biased our view of the progenitor galaxy. We therefore propose to identify GSE members among a larger sample of stars with halo kinematics, pre-selected only on the basis of \textit{Gaia} DR3 radial velocities, that covers the full integrals of motion space, enabling tighter constraints on the progenitor’s properties.

\paragraph{Hot subdwarfs in HR, HSD at high resolution - Binarity and advanced atmospheric analysis (PI: Dorsch , S. Geier):} Hot subdwarf (sdO/B) stars are compact, core-helium-burning objects formed predominantly through binary interactions, including stable Roche-lobe overflow and common-envelope evolution. These channels can produce both short-period (hours to days) sdB$+$dM/WD systems and long-period ($\sim 10^3$ days) sdB$+$FGK binaries. We propose HR spectroscopy for 8\,000 bright hot subdwarfs from the \cite{Culpan2022} catalogue to complement LR data. These observations will deliver precise abundances and radial velocities, constraining both on hot subdwarf formation and the underlying binary mass-transfer processes.

\paragraph{SMR-HR: Super-metal-rich stars (PIs: S. Feltzing, G. Kordopatis):}
The properties of the stellar populations are shaped by the chemical and dynamical evolution of the Galaxy. A system stops forming stars when the gas has been consumed. Thus, without any infalling gas, the super-metal-rich stars are the end-product. Larger, and more precise, datasets than currently available are needed to further our understanding of this chemical endgame.  The giant star catalogue includes some $678\,327$ objects (117\,891 with a match in the S4 catalogue) and  the dwarf catalogue includes some 904\,311 objects ($102\,127$ have a match in the S4 catalogue). The programme will address the ability of the 4MOST system to analyse super-metal-rich stars ($\rm [Fe/H]\gtrsim 0.2$) and provide a statistical analysis of the upper limit of metallicity in general. 

\paragraph{STITCH at 4MOST: Spectroscopic observations of Galactic Type II and anomalous Cepheids (PI: T. Sicignano, V. Ripepi):}

This project aims to evaluate the potential of T2Cs and ACs as robust distance indicators, serving as viable alternatives to CCs. Achieving this requires a comprehensive assessment of potential systematic uncertainties. A primary unresolved challenge in this context is the dependence of their PL and PW relations on metallicity. Beyond their utility as standard candles, T2Cs and ACs are powerful tracers of galactic formation and evolution. Their chemical and kinematic properties offer unique insights into the assembly history of the Milky Way and Magellanic Clouds \cite[e.g.][]{bono2024,ripepi2024}.

The 4MOST facility is already set to deliver spectra for extensive samples of T2Cs and ACs in the Magellanic clouds via the 1001MC survey \citep[4MOST Consortium Survey 9: One Thousand and One Magellanic Fields][]{Cioni2019}, though these will primarily be at low resolution. By complementing these data with high-resolution observations of Galactic counterparts, we will be able to do the following:

\begin{itemize}
\item Quantify the systematic impact of metallicity on PL and PW relations \citep[as defined in][]{sic2024,sic2025} and subsequent distance determinations, thereby resolving discrepancies currently reported in the literature;
\item Conduct a direct comparison of PL-metallicity relations across environments with diverse chemical enrichment and star-formation histories using homogeneous datasets;
\item Use the chemical signatures of ACs and T2Cs in both the Milky Way and the Magellanic clouds to help clarify their progenitor origins and evolutionary pathways.
\end{itemize}

\paragraph{Characterising the population of accreting-only symbiotic stars - SySt HR (HR) (PI: G. Traven, T. Merle)}
Symbiotic stars (SySt) have been proposed to be possible progenitors of type Ia supernovae, sources of the ultraviolet excess of elliptical galaxies, and prolific Lithium and dust producers \citep{Mukai-2016, Munari-2021, Merc-2025}. The known objects, however, are believed to be only the tip of the iceberg of the underlying population of these red giant + compact companion systems, as they are difficult to detect in their longer-lived quiescent phase of evolution.
We aim to characterise this hidden population, which can nevertheless be uncovered by looking for subtle signs of accretion of mass from the giant onto the compact companion (typically a white dwarf - WD). 

In this proposal we build on the work of \cite{Munari-2021} who used the GALAH survey to construct spectral templates of M giants and searched for residual flux in the H$\alpha$ region to identify accreting-only SySt candidates. While their newly confirmed 33 accreting-only SySt doubled the number of previously known systems, we aim to further increase the detection yield and push towards even lower accretion luminosity systems. This will eventually deliver a sufficient number of systems for a statistical study of SySt via population modelling and allow us to estimate their overall number in the Galaxy. Our immediate goals for each SySt candidate are to (i) evaluate the current level of activity, (ii) detect orbital motion via RV measurement, (iii) analyse changes in the H$\alpha$ emission profile, (iv) estimate the mass ratio between the giant and its companion for systems with multiple visits by 4MOST, and (v) search for signs of possible past nova events via abundance patterns and, for example, the Li line.

\paragraph{4MOST HR follow-up and characterisation of \textit{Gaia} eclipsing binaries (HR),  (PI: T. Merle, G. Traven):}
A large proportion of stars exist as binaries, and when viewed edge-on their eclipses become observable. Combining this geometry with spectroscopy enables accurate determination of fundamental stellar properties such as masses and radii. The latest \textit{Gaia} data release provides the largest catalogue to date, containing 86\,000  eclipsing binaries from the Non-Single Star (NSS) catalogue and 2.2 million eclipsing binary candidates, of which 70\,\% were previously unknown. Most of them ($>90$\,\%) do not have \textit{Gaia} spectroscopy, essential for getting accurate masses. Here, we propose a 4MOST high-resolution follow-up of $20\,000$ \textit{Gaia} eclipsing binary (5\% from NSS and 95\,\% from candidates) where both components are spectroscopically resolved, allowing us (i) to confirm the binarity and determine the component atmospheric parameters in most cases with a single visit, (ii) to measure mass ratios and centre-of-mass velocities with two visits, (iii) and to characterise orbits and derive accurate stellar masses and radii with multiple visits.

\paragraph{4MSOS stellar orbits survey (PI: S. Khoperskov):}
The 4MSOS supplementary programme targets 548\,,096 Milky Way disk stars selected from \textit{Gaia} astrometry that currently lack radial-velocity measurements and therefore fall outside the \sd\ core sample. By delivering precise radial velocities and detailed chemical abundances for this astrometrically exceptional population, 4MSOS will provide the complete 6D phase-space and chemical information required for a fully self-consistent Schwarzschild orbit-superposition model of the Galaxy. The combined 4MSOS, \sd\, and \textit{Gaia} dataset will enable the simultaneous optimisation of orbit weights and the Galactic gravitational potential, yielding strong constraints on the 3D distribution of stellar and dark matter. This synergy will reveal fine-grained dynamical substructure and reduce degeneracies between Galactic components, orbital families, and stellar populations.

\paragraph{Asymptotic giant branch stars and their companions (4AGB) (PIs: \'{A}. Sk\'{u}lad\'{o}ttir, A. Ardern-Arentsen):}  The 4MOST supplementary survey of AGB stars and their companions (4AGB) aims to study AGB stars and binary companions that experienced AGB mass-transfer (Ba/CH/CEMP-s stars) with the 4MOST HRS in collaboration with \sd\ and the Milky Way Halo HR Survey. The 4AGB catalogue contains about $7300$ targets ($G<16$) and the target selection is based on existing catalogues and includes carbon-rich candidates with $\rm[Fe/H]<-0.5$ from \textit{Gaia} XP low-resolution spectrophotometry \citep{Ardern-Arentsen2025}, S-stars ($\rm C/O\leq 1$; \citealt{Stephenson1984,Chen2023a}), as well as C-stars ($\rm C/O>1$) identified with \textit{Gaia} \citep{Abia2020, Abia2022}. 4AGB will obtain radial velocities, key abundances (C, N, O, s-, and r-process elements), and search for technetium (Tc), the hallmark of active AGB stars. The resulting observations will be used to calibrate \textit{Gaia} XP$/$RVS spectra, and provide a unique homogeneous resource for studying AGB nucleosynthesis across metallicities. 

\paragraph{Diffuse interstellar bands in the Milky Way (PI: J. Kos):}  The red HR window contains several strong diffuse interstellar bands (DIBs) at $6196$, $6202$, $6270$, $6379$, $6613$, and $6660$\,\AA. DIBs are resolved absorption lines produced by yet unidentified molecules, with several hundred known in the visible and near-infrared wavelength ranges \citep{Cox2017}. Stars serve as background light sources, in which spectra we can observe weak absorption lines of the ISM molecules. To separate ISM features from stellar spectral lines, we will employ stellar spectral templates produced by the S4 analysis pipelines. From the residual spectra, we will measure the parameters of at least six strong DIBs. For stars with $E(B-V)=0.2$, the strongest DIBs have depths of $\sim3$\% \citep{Fan2019}, making them detectable in more than 50\% of stars within $|b|<15^\circ$ of the Galactic plane.

Each of these DIBs is likely caused by a different carbon-based molecule. \citet{Kos2017} has shown that these molecules behave differently under the influence of the interstellar radiation field. For instance, DIBs at $6196$ and $6202$\,\AA\ show the smoothest distribution of any studied DIB, and DIBs at $6379$, $6270$, and $6660$\,\AA\ form finely structured clouds, as the majority of studied DIBs do. The DIB at $6613$\,\AA\ forms ISM clouds with an average lumpiness. Such a diverse selection of DIBs has not been observed with any large spectroscopic survey to date. We will produce 3D maps of DIBs, study scale heights of the gas disks, interaction of DIB clouds with the interstellar radiation field, and the nature of DIB molecules in many distinctive regions of the Milky Way, such as star-forming regions, radiation-shielded clouds, diffuse ISM far from the Galactic plane, and also DIBs in the Large Magellanic Cloud. The DIB at $6613$\,\AA\ has also been observed by the GALAH survey, which additionally targeted the two strongest and most extensively studied DIBs at $5780$ and $5797$\,\AA\  \citep{Buder2025}. A common DIB across surveys will allow us to place DIB measurements on a unified scale.

\subsection{Survey programme validation} \label{sec:SPV}
%
\begin{figure*}
\sidecaption
\includegraphics[width=12cm]{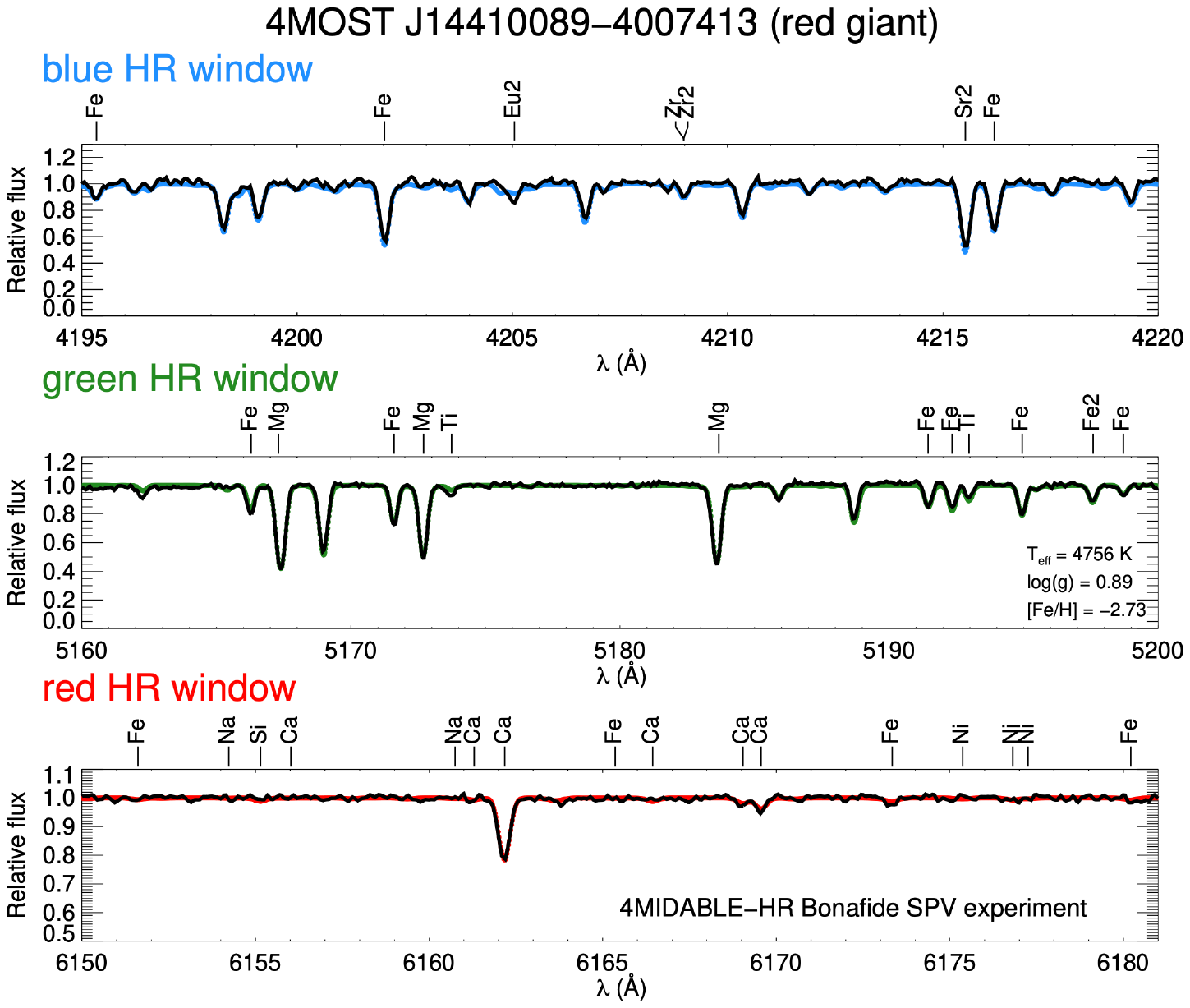}
\caption{Observed 4MOST high-resolution spectrum of a metal-poor red giant from the 4MIDABLE-HR SPV experiments. The spectrum was taken during the SPV phase, and it corresponds to \sd\ survey experiment SPV-SVG0401-HR (bonafide). The best-fit NLTE ANN model is overplotted, and the resulting astrophysical parameters are indicated in the legend. We note that these are small wavelength portions, and they are shown here for demonstration of the data quality. They do not represent the entire spectral coverage of the 4MOST HR spectrograph.}
\label{fig:spectrum1}
\end{figure*}

\begin{figure*}
\sidecaption
\includegraphics[width=12cm]{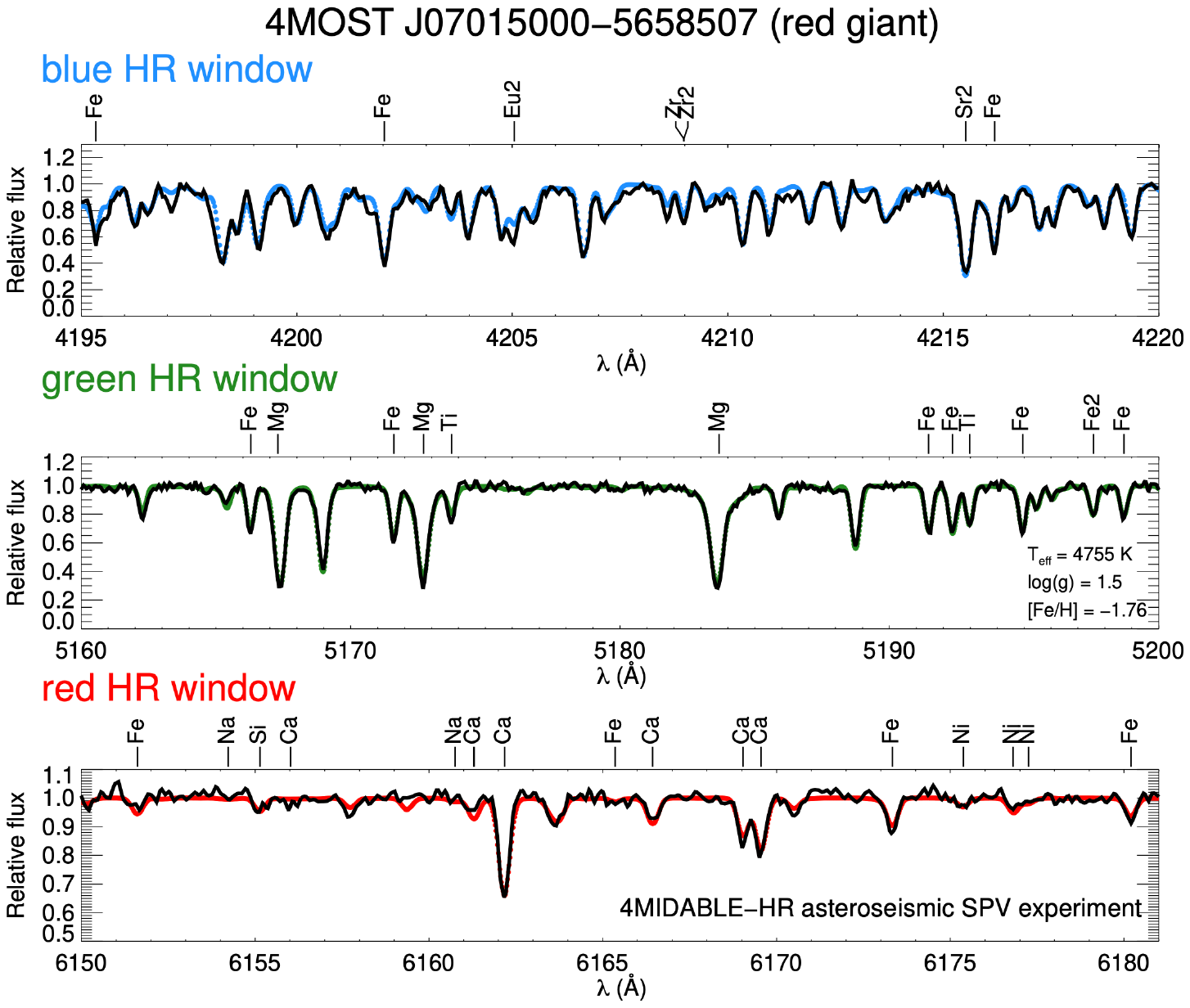}
\caption{Same as Fig. \ref{fig:spectrum1} but for a red giant star from the 4MIDABLE-HR SPV-SVG0409-HR (seismic) experiment.}
\label{fig:spectrum2}
\end{figure*}

\begin{figure*}
\sidecaption
\includegraphics[width=12cm]{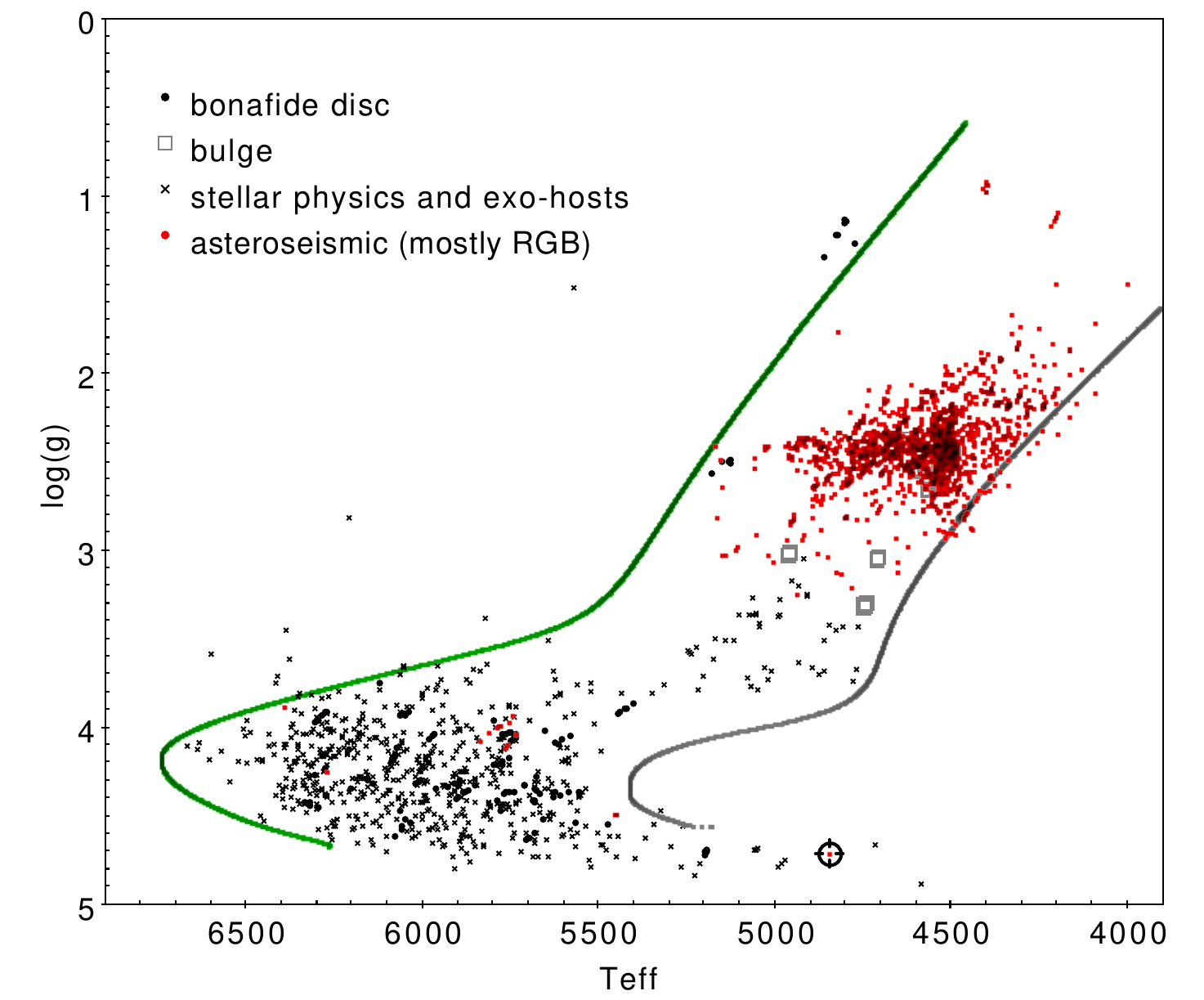}
\caption{Location of the S4 SPV observed targets in the T$_{\rm{eff}}$-log(g) plane. The 4MOST astrophysical parameters were determined from the SPV spectra, and they are based on the SAPP at 4GP analysis code \citep{Storm2025b}. Stellar evolutionary tracks for metallicities $+0.5$ and $-2.5$ are overplotted (masses 1 and 0.8 solar mass, respectively).}
\label{fig:SPV}
\end{figure*}

The first set of science-capable data to be collected after instrument commissioning will be for the Survey Programme Validation experiments. While the full 4MOST survey will run for five years, these experiments have been designed by the members of 4MOST public and community surveys a small, self-contained projects to demonstrate science readiness. SPV observations were carried out for a period of about seven weeks in March to May in $2026$, giving science teams an opportunity to verify their analysis methods. 

The SPV experiments for \sd\ cover the topics below. Figs. \ref{fig:spectrum1} and \ref{fig:spectrum2} show two examples of the HR spectra collected within our SPV programs in March-April 2026, and their best-fit NLTE ANN spectral models are overplotted. The derived NLTE stellar parameters are provided in figure inset. In Fig. \ref{fig:SPV}, we demonstrate  stellar parameters (T$_{\rm{eff}}$-log(g) plane) for other targets in our SPV experiments.
\paragraph{Bonafide disk stars:}
The experiment SPV-SVG0401 is our key spectroscopic analysis sample, which we will use for the validation of the \sd\ survey programme, especially its S0401 and S0402 sub-samples. The targets includes a selection of $1\,790$ \textit{Gaia}-ESO \citep{Gilmore2022, Randich2022} targets with high quality stellar parameters, chemical abundances \citep[similar selection to][]{Lian2023}, and several tens of \textit{Gaia}-ESO/4MOST/WEAVE/PLATO benchmark stars from $\rm [Fe/H] \approx 0.5$ down to the lowest $\rm [Fe/H] \approx -3$. These include mostly main-sequence and subgiant stars, and several red giants in the Galactic disk. These targets will allow for testing of the pipeline performance for stellar parameter and abundance determination compared to the \textit{Gaia}-ESO results.

\paragraph{Bulge sample:}
The bulge sample consists of 32 red giant stars observed as part of the \textit{Gaia}–ESO survey. All targets are located within similar sky position, allowing for tests of the derived stellar parameters and chemical abundances compared to the literature based on a single pointing field.

\paragraph{Multiplicity:}
Two experiments related to the science of multiple stars have been designed for the SPV phase. The first consists of confirming the binary nature of many spectroscopic binary candidates (SPV-SVG0406). Building up on the survey of surveys \citep{Tsantaki2022} which includes radial velocities and binarity flag from APOGEE, GALAH, \textit{Gaia}-ESO, RAVE, and LAMOST, we target a sizeable sample of SB candidates. 
With two epochs from 4MOST and at least one epoch from an external survey, statistical studies can be performed to estimate the period distributions. Some SB1 systems can be revealed as SB2 (double-lined). For SB2, centre-of-mass velocities and mass-ratio distribution can be obtained. 

The second experiment aims to reproduce the dependence of SB fraction on metallicity (SPV-SVG0407) for a random sample of filed stars. Here we select a quasi-random sample of stars from \textit{Gaia} that have XGBoost metallicity values \citep{Andrae-2023}. This enables us to select an approximately equal (but still random) number of stars per metallicity bin so that we get sufficient statistics across the metallicity range. We also limit the apparent magnitude between $11.5 <G<14.5$ and to distance of 2\,kpc based on \textit{Gaia} parallax. This provides a large enough sample to be available also as efficient fillers for other SPV experiments, while at the same time it provides high $S/N$ in short time and largely avoids giants, which is again beneficial for probing the binary fraction.
\paragraph{Exoplanet hosts:}
The goal of the SPV-SVG0408 experiment is to validate spectroscopic stellar parameters and chemical abundances for a sample of well-studied exoplanet hosts stars and for a sub-sample of PLATO targets in order to ensure a sufficient spectral quality for subsequent observations of exoplanet hosts and  candidates by 4MOST. 

The most important targets in the PLATO sub-survey are solar-type stars, main-sequence and subgiants, brighter than about $G\lesssim 12$ \citep{Montalto2021, Rauer2025}. These are objects that would enable robust analysis of global asteroseismic parameters, and for a sub-samples searches for Earth-analogs orbiting these stars will be carried out. These objects would be both important to the exoplanet and stellar objectives. 

In this experiment, we include $587$ known exoplanet host stars, carefully selected from TESS, K2, WASP, HAPRS, CoRoT, HATSouth survey, Backyard Worlds, etc, to probe a broad metallicity range ($\rm -1.4 \lesssim |Fe/H] \lesssim0.5$) and all stellar evolutionary stages, for which planet companions were reported in the literature. We also include a selected sub-sample of PLATO targets from the LOPS2 southern pointing field \citep{Nascimbeni2022, Nascimbeni2025}, following the collaborative agreement.\footnote{This sample is proprietary and cannot be disclosed}
\paragraph{Asteroseismology targets:}
The asteroseismology targets for the SPV-SVG0409 experiment were chosen from TESS input catalogue (From S0412) and we also include $12$ red giant stars with \textit{Gaia}-ESO UVES spectra and asteroseismic data obtained from CoRoT lightcurves \citep{Smiljanic2018}. These targets have seismic detections and evolutionary state determination (see also Sect.~\ref{subsec:astroseismic_survey}), allowing independent validation of the stellar parameters, masses, radii, and ages based on SAPP results.
\section{Conclusions}\label{sec:6}
This paper provides the first detailed description of the science goals, motivation, and input catalogue for \sd. The following topics will be addressed by \sd: Galactic structure and evolution, stellar physics, and star-planet connection as well as the chemical enrichment of the Milky Way. The survey has  $66$ members of $14$ institutes worldwide, and it will be carried out on the 4MOST telescope during the next five years of its operations in the survey mode.

The input catalogue of the \sd\ survey contains $3\,638\,780$ Galactic AFGKM-type stars, representing diverse evolutionary phases from main-sequence dwarf stars to tip red giant stars. We also cover core He-burning variables, Cepheids of Type I and II, anomalous Cepheids, stars in known binary and triple systems, and exoplanet-host stars. The stars are broadly distributed across the Galactic disk and bulge, spanning over 15\,kpc radially and 5\,kpc vertically from the Galactic mid-plane. Within $\sim 0.3$\,kpc of the mid-plane (thin disk), our sample probes the regions of the near end of the bar, Crux-Scutum, Carina Sagittarius, and the Outer spiral arms. For all of these targets, high-resolution ($R=20\,000$) optical spectra will be obtained with $SNR=100$ and 250 per {\AA}. In order to enable binary searches and characterisation, multi-epoch spectra will be taken for a sub-sample.

In total, our input catalogue comprises 12 dedicated sub-surveys addressing different science themes and enabling their coherent and staggered scientific exploitation, and it takes advantage of precise chemical abundances, kinematics, and ages. The selection function is simple, and for the main sub-surveys, the selection relies only on \textit{Gaia} DR3 astrometry and photometry. The entire southern sky is covered, with a total area of $28\,000$ sq.deg. and typical target densities of about $400$ targets/sq.deg. The stellar sample covers ten magnitudes in the \textit{Gaia} G-band, from the $8$th magnitude (brighter sub-surveys such as eROSITA, TESS prio.) to faint sub-surveys reaching to the 18th magnitude (the deep bulge and Roman deep field). For a fraction of our targets, value-added information from eROSITA (X-ray fluxes), TESS and K2 (global seismic oscillation parameters), OGLE (pulsation periods), and information on planet-mass companions (complementing confirmed exohost stars) are available. We will also perform a spectroscopic follow-up of the PLATO long-pointing southern field LOPS2, enabling uniform analyses of chemical abundances and fundamental parameters for the stars in this field. More generally, the combination of TESS and K2 asteroseismic data with effective temperatures and stellar abundances from S4 high-resolution spectra will produce the largest and most homogeneous dataset of its type.
 
In this paper, we have also presented the kinematic and dynamical classification of our catalogue using the \textit{Gaia} positions, astrometry, and photometry. We estimate that our sample comprises about $90-95\,\%$ disk and bulge stars, whereas a small fraction are kinematically halo-like. The fractions may fluctuate, depending on the kinematic and/or orbital characteristics used. By far the largest component of our stellar sample is the thin disk, with a fraction of over $70\,\%$, whereas the thick disk and the halo comprise $20\,\%$ and $\sim 5\,\%$, respectively. However, since the decomposition of populations requires chemical abundances and ages, we will revisit the membership once precise 4MOST radial velocities and stellar parameters are available. The preliminary chemical information from the \textit{Gaia} XGboost \citep{Andrae-2023} indicates that about $100\,000$ of our stars are very and ultra metal poor stars, with metallicities $\rm [Fe/H] \lesssim -2$.

We have also described the physical models and procedure planned for the analysis of the \sd\ high-resolution stellar spectra. Most of our targets will be processed through the NLTE SAPP and NLTE cINN codes \citep{Gent2022,Storm2025b, Ksoll2026}, with NLTE effects taken into account for 16 chemical elements (H, O, Na, Mg, Al, Si, Ca, Ti, Mn, Fe, Co, Ni, Sr, Y, Ba, Eu.). Auxiliary data, such as the asteroseismic parameters and X-ray fluxes, will enable better constraints on surface gravities, ages, and masses of stars. Binary and other multiple stellar systems are present in the \sd\ catalogue. Whereas we expect that most of them will not be resolved at the resolving power of the 4MOST HR spectrograph, we made efforts to detect and characterise these objects using a dedicated pipeline, and for selected sub-surveys in \sd\,, the observational cadence has been introduced in order to maximise the detection probability and improve the quality of reconstructed orbital parameters. For pulsating stars in our catalogue, such as Cepheids Type I and II, a dedicated NLTE approach \citep{Albarracin2026} will be used.

To summarise, the \sd\ Galactic survey will be the largest high-resolution optical Galactic survey in the coming decade in terms of the number of targets, photometric depth, and survey area. Its legacy value is its homogeneous target selection, completeness, and uniform spatial distribution covering the entire southern sky accessible from Paranal. The \sd\ survey also ideally complements other Galactic programs on 4MOST, such as the low- and high-resolution surveys of the disk, bulge, and halo \citep{Chiappini2019, Christlieb2019, Helmi2019}; Magellanic clouds \citep{Cioni2019}; and dwarf Spheroidals \citep{Skuladottir2023}. The \sd\ will have strong synergies with upcoming surveys on other astronomical facilities, including those on WEAVE, MOONS, and GALAH-2. In particular, MOONS on VLT \citep{Gonzalez2020, Cirasuolo2020}, due to its superb infrared spectroscopic capabilities (despite its small field of view), will ideally complement the \sd\ survey for about $500\,000$ stars as part of its REDdened Milky Way program. The data will constitute a unique legacy for all areas of astronomy and open the path for future observations and discoveries with planned next-generation astronomical facilities, such as HWO and WST.

\begin{acknowledgements}
We thank Matti Dorsch and Rebecca Forsberg for their kind assistance with some of the figures (Figs. 20, 23) in this paper.

MB, NS, RA, GG, and AP are supported from the European Research Council (ERC) under the European Union’s Horizon 2020 research and innovation programme (Grant agreement No. 949173).
T.B. acknowledges financial support by grants No. 2018-04857 and 2024-04990 from the Swedish Research Council, and the project grant "The New Milky Way" from the Knut and Alice Wallenberg Foundation.
A.S. was supported by grant PID2023-149918NB-I00 from the Spanish Ministerio de Ciencia, Innovaci\'on y Universidades (MCIU), the Spanish program Unidad de Excelencia Mar\'ia de Maeztu CEX2020-001058-M and by the MaX-CSIC Excellence Award MaX4-SOMMA-ICE.
G.G. acknowledges support by Deutsche Forschungs-gemeinschaft (DFG, German Research Foundation) – project-IDs: eBer-22-59652 (GU 2240/1-1 "Galactic Archaeology with Convolutional Neural-Networks: Realising the potential of \textit{Gaia} and 4MOST"). 
S.F. was supported by project grants from the Knut and Alice Wallenberg Foundation (KAW 2020.0061 Galactic Time Machine, PI: Feltzing, KAW 2013 The New Milky Way, PI: Feltzing).
G.T. acknowledges financial support of the Slovenian Research Agency (research core funding No. P1-0188) and the European Space Agency (Prodex Experiment Arrangement No. 4000143450).
R.A.G. acknowledges the support from the PLATO Centre National D'{\'{E}}tudes Spatiales grant.
S.M. acknowledges support from the Spanish Ministry of Science and Innovation with the grant no. PID2023-149439NB-C41.
K.L. acknowledges funds from the European Research Council (ERC) under the European Union’s Horizon 2020 research and innovation programme (Grant agreement No. 852977) and funds from the Knut and Alice Wallenberg Foundation.
S.K. acknowledges support by the Deutsche Forschungsgemeinschaft under the grant KH~500/2-1.
V.R. and T.S. have been supported through INAF GO-GTO grant 2023 “C-MetaLL - Cepheid metallicity in the Leavitt law” (P.I. V. Ripepi). 
T.S. thanks the Istituto Nazionale di Fisica Nucleare (INFN), Naples section, for support through specific initiatives QGSKY.
{A}.S. and R.Singh acknowledge funding from the European Research Council (ERC) under the European Union’s Horizon 2020 research and innovation programme (grant agreement No. 101117455).
V.K. acknowledges financial support by the Carl-Zeiss-Stiftung.
G.B. acknowledges partial support from the INFN InDARK project.
This research has used the SIMBAD database operated at CDS, Strasbourg, France ad data from the European Space Agency (ESA) mission \textit{Gaia}, processed by the \textit{Gaia} Data Processing and Analysis Consortium. Funding for the DPAC has been provided by national institutions, in particular, the institutions participating in the \textit{Gaia} Multilateral Agreement. Finally, this paper is based on work supported by COST Action CA21136, "Addressing Observational Tensions in Cosmology with Systematics and Fundamental Physics (CosmoVerse)," funded by COST (European Cooperation in Science and Technology).
\end{acknowledgements}

%
\bibliographystyle{aa} 
\bibliography{ref} 
%

\begin{appendix}

\section{Comparison with other ongoing surveys}

The \sd\ survey is highly complementary to other spectroscopic surveys, such as GALAH, APOGEE, \textit{Gaia}-ESO, and \textit{Gaia} RVS. Here we present a brief summary of the synergies and outline briefly where the data acquired by \sd\ can be transformative in terms of its science harvest. We note that this section is deliberately a part of the appendix, as the forecasts for \sd\ are based on theoretical estimates (of spectral quality, type of chemical elements accessible in the data, abundance precisions, Galactic volume sampled, etc.), whereas the data from other surveys are real samples with their true selection functions and realistic uncertainties encompassing. Therefore this comparison may not be fully fair, but it still provides an order of magnitude assessment of the value of \sd\ in the context of these programs.

For the analysis below, we use the following datasets from the literature. The GALAH DR4 data are adopted from \citet[][]{Buder2025}, and they are accessible at \url{https://www.galah-survey.org/dr4/overview/}. Specifically, we use the file \texttt{galah$\_$dr4$\_$allstar$\_$240705.fits}. The APOGEE DR17 data are adopted from \citet{Abdurrouf2022}, using the \texttt{allStar-dr17-synspec$\_$rev1.fits}; these are accessible at \url{https://www.sdss4.org/dr17/irspec/spectro_data/}. The \textit{Gaia} RVS data were obtained from \url{https://gea.esac.esa.int/archive/} using the following syntax \texttt{SELECT source$\_$id, ra, dec, phot$\_$g$\_$mean$\_$mag, parallax FROM gaiadr3.gaia$\_$source WHERE $has\_rvs = 'true'$}. Finally, the \textit{Gaia}-ESO final data release \citep{Gilmore2022,Randich2022} were adopted from the ESO public archive. In the subsequent discussion, we cross-match the catalogues by  \textit{Gaia} DR3/eDR3 ID, but we note that slightly different numbers in terms of overlap (of the order of $5,000$ to $10,000$ stars) can be obtained by using RA$/$Dec values instead.

\begin{figure*}[!ht]
\hbox{
\includegraphics[width=0.48\textwidth]{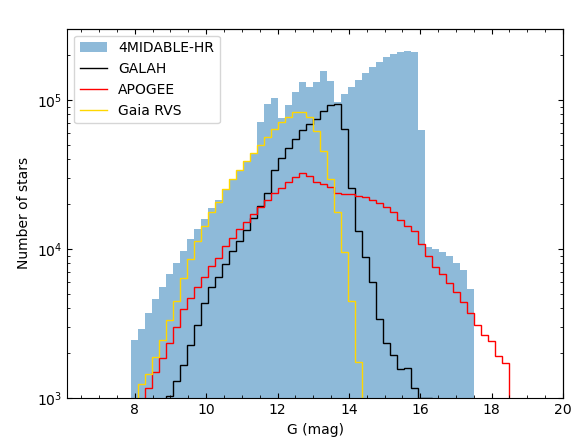}
\includegraphics[width=0.48\textwidth]{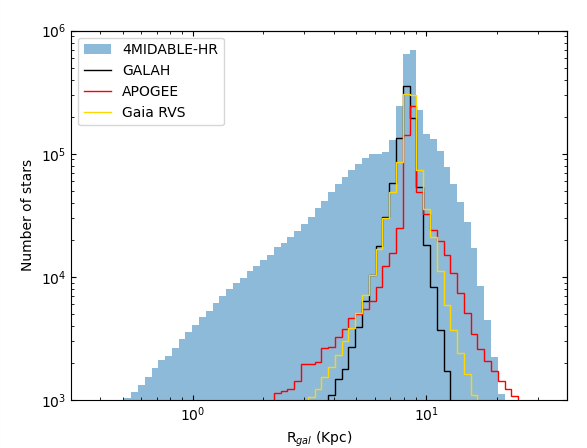}
}
\caption{Distribution of \sd\ targets in \textit{Gaia} G-mag (\textit{left}) and in galactocentric distance (\textit{right}), compared to other large spectroscopic surveys: GALAH DR4, APOGEE DR17, and \textit{Gaia}-RVS (see text). For simplicity and homogeneous comparison, Galactocentric distances for all these surveys were computed using \textit{Gaia} DR3 parallaxes.}
\label{fig:surveys}
\end{figure*}

Firstly, the \sd\ survey will have a substantially larger number statistics compared to other optical and near-infrared high-resolution spectroscopic surveys. Compared to GALAH, a similar high-resolution optical Galactic survey, the \sd\ survey improves the stellar sample size by a factor of three, and compared to the \textit{Gaia}-ESO survey by more than a factor of ten. In Fig. \ref{fig:surveys}, we show the distribution of the \sd\ survey catalogue in magnitude and in galactocentric distance, and we overplot the entire data from GALAH, APOGEE, and \textit{Gaia}-RVS (see the sources in the previous paragraph). The \textit{Gaia} RVS catalogue covers mostly brighter targets, $G$-mag $\lesssim 14$, rather similar to GALAH. The \sd\ survey, in contrast, provides a much improved statistics at fainter magnitudes, $G$-mag $\gtrsim 14$, partly overlapping with APOGEE.

The overlap between our survey and other surveys is as follows: there are $\sim 209\,476$ stars in common with GALAH, $70\,853$ with APOGEE, and $\sim 220\,347$ stars for \textit{Gaia} RVS, mostly limited to G-mag of $10$ to $14$. Also, we have $3\,916$ stars in common with the \textit{Gaia}-ESO survey. A greater overlap with APOGEE is expected at the faint end, G-mag $ \gtrsim 16$, although we note that the majority of our stars with $G$-mag $\lesssim 16$ are a part of the deep bulge sub-survey (S0403; see Table 1) and, due to the difficulty of observing the Galactic center, we cannot yet forecast the exact fraction of these to be observed with the required spectral success criteria.

The spatial coverage of the \sd\ survey is  much greater compared to that of APOGEE, GALAH, and \textit{Gaia}-RVS, which follows from the different magnitude ranges covered by the surveys. For this comparison, for simplicity and homogeneity, we use \textit{Gaia} parallaxes to compute Galactocentric distances for all stars (by using $d = 1/\pi$), in order to have a fair comparison of different survey catalogues, but we remind the reader that this approach is unreliable and may systematically over-estimate distances due to progressively increasing parallax uncertainties, limitations of inversion techniques, and magnitude-dependent biases  \citep{Schoenrich2014, Schoenrich2019, Lindegren2021, Bailer-Jones2021}. The \sd\ input catalogue covers the spatial depths out to about 10\,kpc from the Sun, including the innermost spheroid around R$_{\rm {GC}} \approx -3$ kpc, but also in the outer disk regions of $\sim 20$ kpc. Especially outside the solar volume, R$_{\rm{helio}} \gtrsim 1-2$ kpc, the \sd\ typically achieves a factor of $5$ larger sampling compared to GALAH, APOGEE, and \textit{Gaia}-RVS.

Finally, we compare the expected return of the \sd\ survey in terms of the number of elements and chemical abundance precision.  Here we use the predictions of our analysis described in Sect. \ref{sec:3}, along with the bias and scatter estimates demonstrated in Fig. \ref{fig:precision}. Comparing with GALAH DR4 \citep[][their Figs. 24, C2]{Buder2025}, our survey targets the same stellar spectral types and nearly the same chemical elements (except K I and Rb I, whose primary diagnostic features, $\sim 766-770$ nm and $\sim 780$ nm respectively, are outside the 4MOST HR windows), but our metallicity range will have a several orders of magnitude improved statistics down to [Fe/H] $\approx -4$. We expect that better precision will be achieved for abundances s- and r-process elements, because the 4MOST high-resolution spectra cover the near-UV spectral range, where most strong diagnostic lines of heavy elements are found (\citealt[][e.g. their Fig. 5 and Table 3]{Hansen2015}; see also \citealt{Kordopatis2023}). Specifically, in the blue-green-red window we will have access to the lines of the following  species: 
\begin{itemize}
  \item light elements: C (CH in the blue, parts of the C$_2$ Swan band in the green); 
  \item $\alpha$-capture: O (based on [O I] forbidden transitions), Mg (I), Si (I), Ca (I)
  \item odd-Z: Na (I), Al (I)
  \item Fe-peak: Sc (II), Ti (I, II), V (I), Cr (I), Mn (I), Fe (I,II), Co (I), Ni (I)
  \item s-process: 1-st peak Sr (II), Y (II), Zr (II); 2-nd peak: Ba (II), La (II), Ce (II), Nd (II); 3-d peak: Pb (I)
  \item r-process: Sm (II), Eu (II), Gd (II), Dy (II), Th (II, $^*$ unstable)
\end{itemize}
where we list species with the dominant s-process contributions at solar metallicity according to \citet{Bisterzo2014}; noting that contribution of r-process to Nd is nearly $50$ percent. Hence, compared to GALAH will be able to provide Pb, Gd, Dy, and Th for the first time, yielding the total combined dataset of 35 elemental abundances for over $200\,000$ stars  (GALAH and 4MOST). Furthermore, \sd\ will complement the APOGEE DR17 chemical dataset \citep[][Sect. 4.2.2]{Abdurrouf2022} with 12 trans-Fe group elements (only Ce was released in this study), thereby -- by combining the two samples (APOGEE and 4MOST) -- enabling a dataset of over $70\,000$ stars with robust abundances of over 30 chemical elements, including sulfur (S I) which is only measurable in the near-infrared. Finally, comparing with \textit{Gaia} RVS (GSP-spec; see \citealt[][their Fig.23, 'best sample']{Fouesneau2023}, \citealt{Recio-Blanco2023}), we will be able to add $21$ elements to the $12$ species accessible in the RVS window (N, Mg, Si, S, Ca, Ti, Cr, Fe, Ni, Zr, Ce, Nd). We note, however, that the 4MOST Galactic Pipeline incorporate different modules and subsequent improvements that may enable analyses of other elements is be anticipated.

\begin{figure*}
\hbox{
\includegraphics[width=0.65\columnwidth]{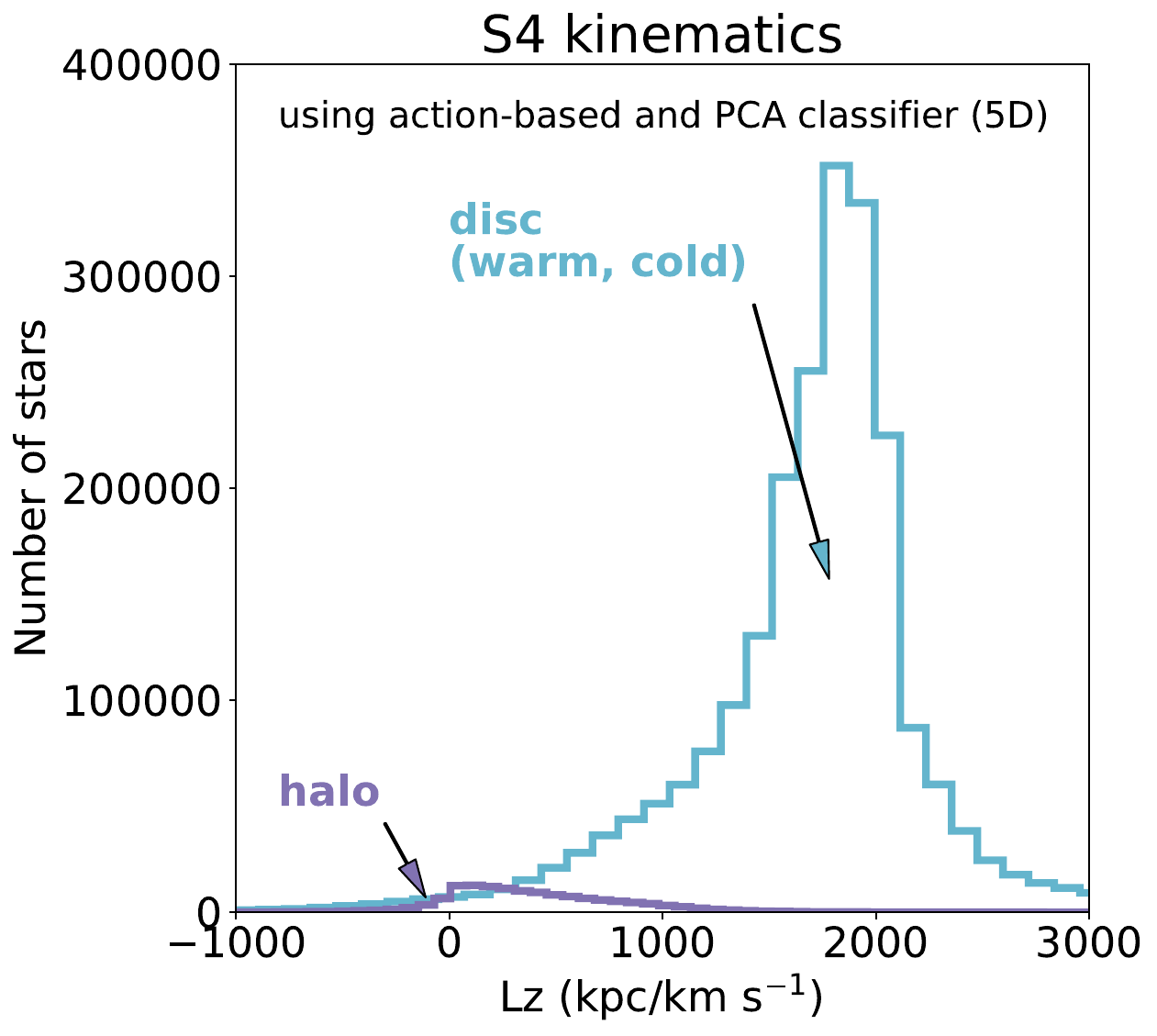}
\includegraphics[width=0.65\columnwidth]{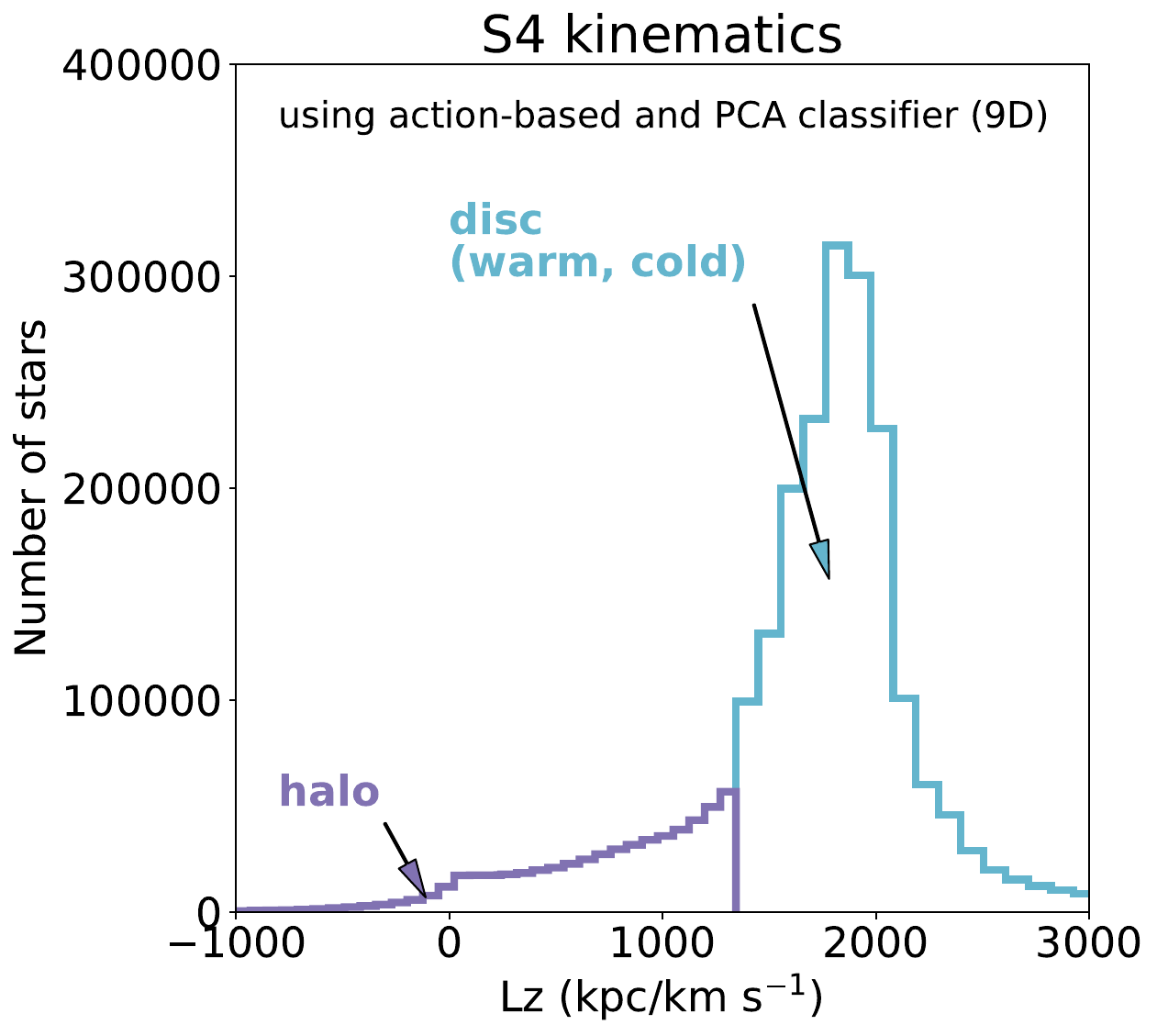}
\includegraphics[width=0.65\columnwidth]{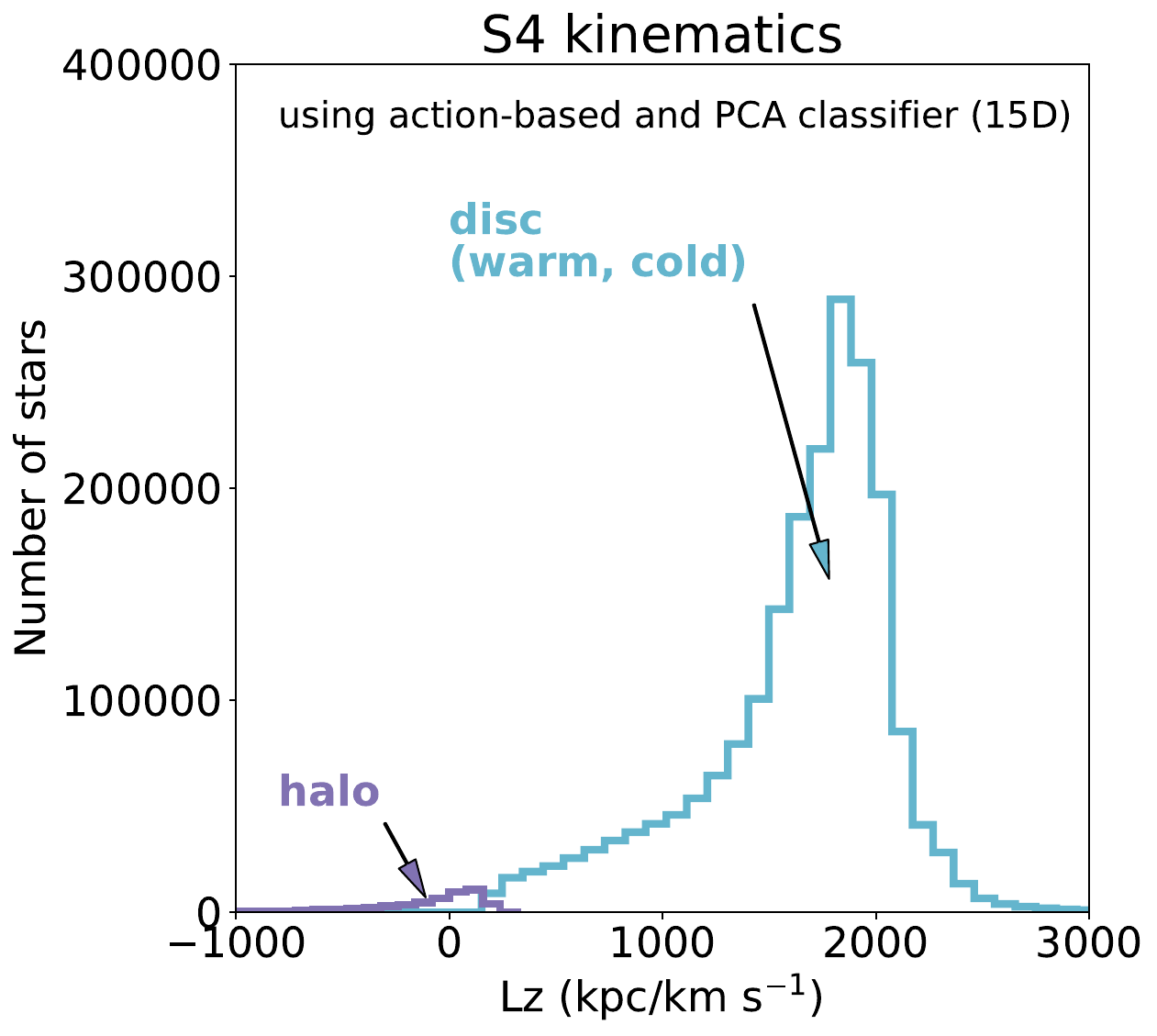}
}
\caption{Distribution of targets in S4 using the PCA analysis based on different number of input parameters. 5D: v$_{\rm tan}$, x, y, z, R. 9D: same as 5D, but also using L$_z$, v$_{\phi}$, v$_r$, v$_z$, that is, with RV but not assumptions on gravitational potential; 15D: same as 9D plus actions (J$_r$, J$_z$, J$_{\phi}$, R$_{\rm apocenter}$, eccentricity, z$_{\rm max}$}
\label{fig:a1}
\end{figure*}

\begin{figure}
\includegraphics[width=0.7\columnwidth]{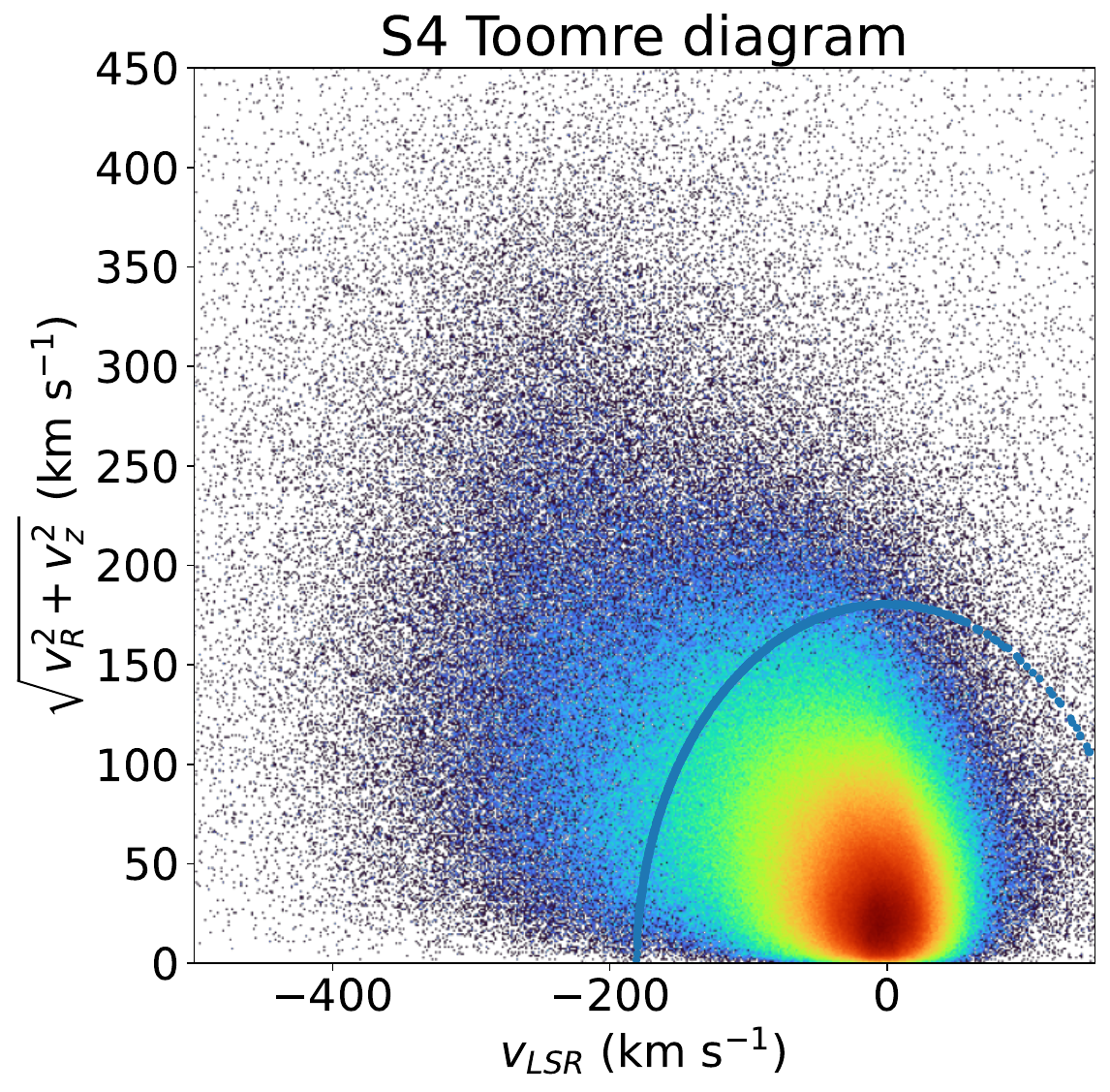}
\includegraphics[width=0.7\columnwidth]{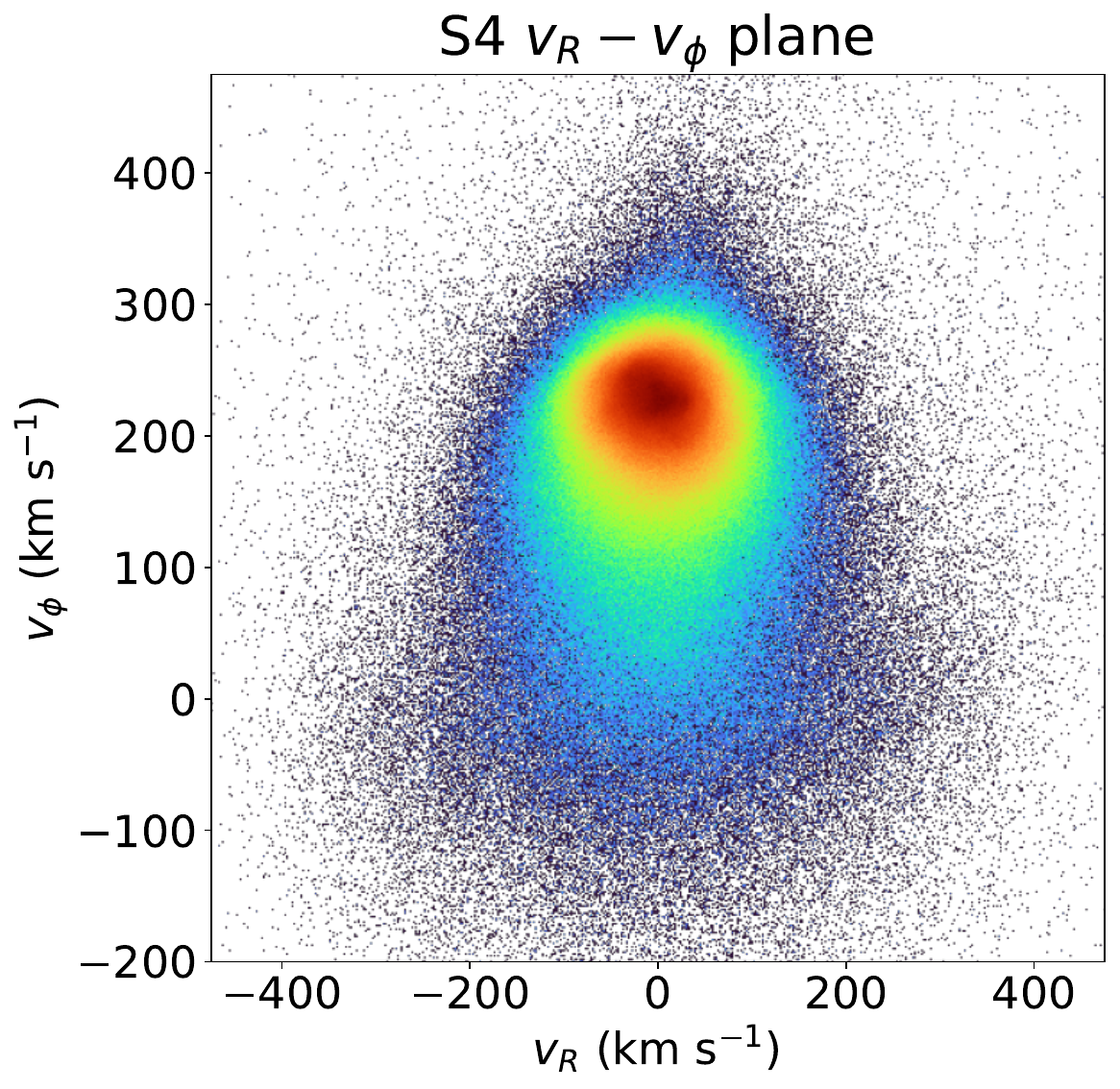}
\caption{Distribution of S4 targets in the space of kinematics.}
\label{fig:Toomre}
\end{figure}

\section{Astrometric classification of stellar populations}

The 4MOST facility has disk and halo surveys. Target stars for these surveys need to be selected carefully. Given \textit{Gaia}'s measurements of position, parallax and proper motions, this appendix shows how to classify if a star is likely to be in the Galactic disk(s), halo, or bulge. We note that this method is a balance between the current availability of observational data for the \sd\ targets and our ability to test the results (based on previous surveys), as described below.

\subsection{Class likelihoods}

As of \textit{Gaia}~DR3, most of stars visible to 4MOST have no radial velocity measurements yet. Therefore, we have to work from astrometry alone. The observational data are the right ascension $\alpha$, declination $\delta$, parallax $\pi$, and the two components of the proper motions $\mu_\alpha$ and $\mu_\delta$.

For each Galactic component, disk, bulge, halo, we have to adopt a gravitational potential and a phase-space distribution (3D space positions, 3D velocities). 
We then have to draw Monte Carlo samples from the phase-space distribution for each of these components, and convert from these Galactocentric coordinates to the heliocentric coordinate frame of \textit{Gaia} observables:
\begin{equation}
(x,y,z,V_x, V_y, V_z) \;\longrightarrow\; (\alpha, \delta, \pi, \mu_\alpha, \mu_\delta, v_{\rm rad})
.\end{equation}
These coordinate transformations were done using \texttt{astropy}. We make use of the \textit{Gaia} EDR3 mock catalogue\citep{Rybizki2020}, which has done exactly those steps. \textit{Gaia} EDR3 mock is based on the \texttt{galaxia} simulator \citep{2011ApJ...730....3S} and the exact definition of the gravitational potentials used is described in Sect.~2.1 in \citet{Rybizki2020}. Using the mock catalogue's population ID, we can select disk, bulge and halo samples.

In order to decide to which component a given star belongs, we evaluate the class likelihoods for each component and choose the class with the highest likelihood:
\begin{align*}
k = \arg\max_k \Bigl[
    &P_{\rm thin-disk} (\alpha, \delta, \pi,\mu_\alpha, \mu_\delta), \\
    &P_{\rm thick-disk} (\alpha, \delta, \pi,\mu_\alpha, \mu_\delta ), \\
    &P_{\rm halo} (\alpha, \delta, \pi,\mu_\alpha, \mu_\delta ), \\
    &P_{\rm bulge} (\alpha, \delta, \pi,\mu_\alpha, \mu_\delta )
\Bigr]
\end{align*}
where each component's likelihood is a marginalisation (i.e.\ integral) over the unknown true velocities and distance:
\begin{displaymath}
P_{\rm k} (\alpha, \delta, \pi,\mu_\alpha, \mu_\delta ) 
\end{displaymath}
\begin{displaymath}
=\int dr \int dV_x \int dV_y \int dV_z\;
p(\mu_\alpha, \mu_\delta, \pi | \alpha, \delta, r, V_x, V_y, V_z)
\end{displaymath}
\begin{equation}
\phantom{=}
\cdot p_{\rm k} (\alpha,\delta,r,V_x, V_y, V_z) 
.\end{equation}
The last term is just the phase-space density of the potential component. Note that the term $p(\mu_\alpha, \mu_\delta, \pi | \alpha, \delta, r, V_x, V_y, V_z)$ does \emph{not} depend on the population component $k$ because it is merely the likelihood of the measured \textit{Gaia} astrometry. Specifically, $p(\mu_\alpha, \mu_\delta, \pi | \alpha, \delta, r, V_x, V_y, V_z)$ is a 3D Gaussian whose mean values, standard deviations and correlations are given by the \textit{Gaia}~DR3 archive. Given that we have drawn $S$ Monte Carlo samples from $p_{\rm k} (\alpha,\delta,r,V_x, V_y, V_z)$, we can estimate the class likelihood as follows:
\begin{displaymath}    
P_{\rm k} (\alpha, \delta, \pi,\mu_\alpha, \mu_\delta) \approx \frac{1}{S} \sum_{s=1}^S p_{\rm k} (\mu_\alpha, \mu_\delta, \pi | {\alpha}_s, {\delta}_s, r_s, V_{x,s}, V_{y,s}, V_{z,s})
\end{displaymath}
\begin{equation}    
\propto \sum_{s=1}^S\exp\left[-\frac{1}{2}\left( \vec{x} - \vec{x_{\text{s}}} \right)^T \Sigma^{-1} \left( \vec{x} - \vec{x_{\text{s}}} \right)\right],
\end{equation}
where $\vec x=(\pi,\mu_\alpha,\mu_\delta)$ and $\Sigma$ is the $3\times 3$ covariance matrix from \textit{Gaia} astrometry whose elements are (names according to columns in \textit{Gaia} archive table \texttt{gaia\_source}):

\begin{flalign}
&\Sigma_{11}=\texttt{parallax\_error}^2 \\
&\Sigma_{12} = \Sigma_{21} = \texttt{parallax\_error}\cdot\texttt{pmra\_error}\cdot \\
    &\texttt{parallax\_pmra\_corr} \\
&\Sigma_{13} = \Sigma_{31} = \texttt{parallax\_error}\cdot\texttt{pmdec\_error}\cdot \\
    &\texttt{parallax\_pmdec\_corr} \\
&\Sigma_{22} = \texttt{pmra\_error}^2 \\
&\Sigma_{23} = \Sigma_{32} = \texttt{pmra\_error}\cdot\texttt{pmdec\_error}\cdot\texttt{pmra\_pmdec\_corr} \\
&\Sigma_{33} = \texttt{pmdec\_error}^2
\end{flalign}
Note that $\Sigma$ is not diagonal. We also neglect any uncertainties in \textit{Gaia}'s position measurement $(\alpha,\delta)$ and any correlations between position errors and parallax or proper-motion errors.
\subsection{Class probabilities versus\ class likelihoods}

The method laid out above gives us class likelihoods of the form
\begin{eqnarray}
P(\alpha, \delta, \pi,\mu_\alpha, \mu_\delta | k) = P_k(\alpha, \delta, \pi,\mu_\alpha, \mu_\delta)
.\end{eqnarray}
If we wanted class probabilities instead, i.e.\ class posteriors, we need to introduce class priors:
\begin{eqnarray}
P(k|\alpha, \delta, \pi,\mu_\alpha, \mu_\delta) \propto P(k)\cdot P(\alpha, \delta, \pi,\mu_\alpha, \mu_\delta | k)
.\end{eqnarray}
Such class priors essentially amount to the mass or number ratio of stars that we a-priori expect to be in the components (thin disk, thick disk, bulge, halo). For example, using the \textit{Gaia} EDR3 mock catalogue \citep{Rybizki2020} with a selection $G<17.5$, which is the approximate faint limit of the 4MOST disk sample (c.f.~Figs.~\ref{fig:targetsmag1},\ref{fig:targetsmag2}), we obtain the class priors given in Table~\ref{appedix-table-edr3-mock-priors}. Such priors should probably also take into account the \textit{Gaia} and 4MOST selection functions. At $G<17.5$, the \textit{Gaia} catalogue can be assumed to be complete \citep{2023A&A...669A..55C}, and therefore the 4MOST completeness can be obtained by straightforward ratioing against the \textit{Gaia} catalogue.

\begin{table}
\centering
\caption{Class priors from \textit{Gaia} EDR3 mock catalogue \citep{Rybizki2020} for $G<17.5$ according to Figs.~\ref{fig:targetsmag1},~\ref{fig:targetsmag2}.}
\label{appedix-table-edr3-mock-priors}
\begin{tabular}{c|c|c|c|c}
  & thin disk & thick disk & bulge   & halo  \\
\hline
stars & 187\,479\,029 & 36\,172\,437 & 8\,950\,829 & 86\,182  \\ 
fraction & 0.802557 & 0.154846 & 0.038317 & 0.000369  \\ 
\end{tabular}
\end{table}

\end{appendix}
\end{document}